\documentclass[twocolumn,showpacs,preprintnumbers,amsmath,amssymb,prb,floatfix]{revtex4}

\usepackage{graphicx}
\usepackage{dcolumn}
\usepackage[tight]{subfigure}
\usepackage{amsmath}
\usepackage{amsfonts}
\usepackage{bbm}
\usepackage{mathrsfs}
\usepackage{verbatim}
\usepackage{simplewick}
\usepackage{float}
\usepackage[usenames,dvipsnames]{color}
\usepackage{bm}
\usepackage{cancel}

\newcommand{\key}[1]{{#1}} % key points of the text
\newcommand{\hl}[1]{{#1}} % changes for referee
\newcommand{\HL}[1]{{#1}} % changes additional
\newcommand{\byhand}[1]{{#1}} % mark citations of footnotes inside footnotes BY HAND
\newcommand{\clr}[1]{{#1}}    % color macros below

\newcommand{\unit}{\clr{\mathbbm{1}}}
\renewcommand{\vec}[1]{\clr{\mathbf{#1}}}

\newcommand{\vecg}[1]{\clr{{\bm #1}}}
\newcommand{\im}{\clr{\textrm{Im}}}

\newcommand{\bra}[1]{\clr{\langle #1|}}
\newcommand{\ket}[1]{\clr{|#1 \rangle}}

\newcommand{\tot}{{\clr{\mathrm{tot}}}}
\newcommand{\R}{{\clr{\mathrm{R}}}}
\newcommand{\Cite}[1]{\clr{Ref.~\onlinecite{#1}}}
\newcommand{\Tab}[1]{\clr{Table~\ref{#1}}}
\newcommand{\Fig}[1]{\clr{Fig.~\ref{#1}}}
\newcommand{\Eq}[1]{\clr{Eq.~(\ref{#1})}}
\newcommand{\eq}[1]{\clr{(\ref{#1})}}
\newcommand{\App}[1]{\clr{App.~\ref{#1}}}
\newcommand{\Sec}[1]{\clr{Sec.~(\ref{#1})}}
\renewcommand{\sec}[1]{\clr{(\ref{#1})}}
\newcommand{\Tr}[1]{\clr{\underset{#1}{\mathsf{Tr}}}}
\newcommand{\tr}{\clr{\mathsf{Tr}}}
\newcommand{\sdagger}{{\clr{\dagger}}}
\begin{document}
\title{Fermionic superoperators for zero-temperature non-linear % quantum
  transport:
  \\
  real-time perturbation theory and renormalization group for Anderson quantum dots
}
\author{R. B. Saptsov$^{(1,2)}$}
\author{M. R. Wegewijs$^{(1,2,3)}$}

\affiliation{
  (1) Peter Gr{\"u}nberg Institut,
  Forschungszentrum J{\"u}lich, 52428 J{\"u}lich,  Germany \\
  (2) JARA- Fundamentals of Future Information Technology\\
  (3) Institute for Theory of Statistical Physics,
  RWTH Aachen University, 52056 Aachen,  Germany
}
\date{\today}

 \begin{abstract}
We study \HL{electron} quantum transport through a strongly interacting Anderson quantum dot
at finite bias voltage and magnetic field at \emph{zero temperature}
using the real-time renormalization group (RT-RG)
in the framework of a kinetic (generalized master) equation for the reduced density operator.
To this end, we further develop the \HL{general, finite-temperature} real-time transport formalism
by introducing \emph{field superoperators} that obey fermionic statistics.
This direct second quantization in Liouville-Fock space strongly simplifies the construction of operators
and superoperators \HL{that} transform irreducibly under the Anderson-model symmetry transformations.
The fermionic field superoperators naturally arise
from the univalence (fermion-parity) superselection rule \HL{of quantum mechanics} for the total system of quantum dot plus reservoirs.
Expressed in these field superoperators, the \emph{causal structure} of the 
 perturbation theory for the effective time-evolution superoperator kernel
becomes explicit.
Using the constraints of the causal structure,
we construct a  parametrization of the \emph{exact} effective time-evolution kernel for which we analytically find the eigenvectors and eigenvalues in terms of 
a minimal set of only 30 independent coefficients.
The causal structure also implies the existence of a \emph{fermion-parity protected eigenvector} of the exact Liouvillian,
explaining a recently reported result on adiabatic driving [L.D. Contreras-Pulido \emph{et al} Phys. Rev. B 85, 075301 (2012)] and generalizing it to arbitrary order in the tunnel coupling $\Gamma$.
Furthermore, in the wide-band limit 
the causal representation exponentially reduces the number of diagrams for the time-evolution kernel.
The remaining diagrams can be identified simply by their topology and are manifestly independent of the energy cut-off term-by-term.
By an exact reformulation of this series
we integrate out all \emph{infinite-temperature effects}, obtaining
an expansion targeting only the non-trivial, finite-temperature corrections,
and the exactly conserved transport current follows directly from the time-evolution kernel.
From this \HL{new} series the previously formulated RT-RG equations are obtained naturally.
We perform a complete one-plus-two-loop RG analysis at finite voltage and magnetic field,
while systematically accounting for the dependence of all renormalized quantities on both the quantum dot \emph{and} reservoir frequencies.
Using the second quantization in Liouville space and symmetry restrictions,
we obtain
analytical RT-RG equations\HL{, which can be solved numerically in an efficient way,}
%with an efficient numerical solution
 and we extensively study the model parameter space, excluding the Kondo regime where the one-plus-two-loop approach is obviously invalid.
The incorporated renormalization effects result in an enhancement of the inelastic cotunneling peak, 
even at a voltage $\sim$ magnetic field $\sim$ tunnel coupling $\Gamma$. Moreover, we find a \emph{tunnel-induced} non-linearity of the stability diagrams (Coulomb diamonds) at finite voltage, both in the 
single-electron tunneling and \hl{inelastic cotunneling} regime.
 \end{abstract}
 \maketitle
\section{Introduction\label{intro}}
\key{Non-linear transport spectroscopy of nanoscale systems is a key technique
in modern day physics.}
\HL{Although the linear (equilibrium) transport regime is
well-understood}~\cite{Glazman88,Ng88} (see \Cite{Bulla08} for a review),
 the theoretical description of non-equilibrium transport, especially at low
temperature,
 remains challenging.
Recent progress in this direction has led to an improved understanding of
quantum transport through strongly interacting systems (see \Cite{Eckel10}
for a recent comparative review).
 \HL{Several} fully numerical approaches \HL{have appeared},
such as
the scattering-state time-dependent numerical renormalization group (TD-NRG)
\cite{Anders}
relying on the discretization parameter approximation,~\cite{Rosch12}
time-dependent  density matrix \HL{renormalization group}
(TD-DMRG),~\cite{White04,Daley04,Schmitteckert}
iterative path integrals (IPSI),~\cite{Thorwart} numerically exact influence
functional path integrals (INFPI),\cite{Segal10,Segal11}
quantum Monte Carlo (QMC) in combination with the Nakajima-Zwanzig projection
technique \cite{Cohen11} or with \HL{an} imaginary-time formalism,\cite{Han} and
diagrammatic
Monte Carlo (diagMC).\cite{Komnik}
Partially analytical approaches involve the
non-crossing approximation (NCA),~\cite{Meir,Kroha}
equations of motion for Green's functions,~\cite{vanRoermund}
Bethe ansatz,\cite{Andrei,Konik}
and the flow-equation approach.~\cite{Kehrein}
Fully analytical approaches
include \HL{the} Keldysh perturbation theory to high orders in \HL{the Coulomb
interaction} $U$ (\Cite{Oguri} and \Cite{Aligia}) or \HL{the} dual-fermion superperturbation
theory.\cite{Kirchner}
Finally, perturbative renormalization group studies have mostly started from the
Kondo-model mapping of the Anderson model,
either working with Keldysh-Green functions~\cite{Paaske_Rosch}
or the reduced density operator approach.~\cite{Schoeller09a,Schoeller09b}

The density operator approach has a long history in various fields \HL{of}
physics and chemistry.
In the context of nanoscale transport, it is a natural starting point for the
description of systems with large interaction energies in the high temperature,
weak-coupling limit, $U\gg T \gg \Gamma$.
It can be systematically extended to include high-order tunneling processes
using the real-time diagrammatic~\cite{SchoellerSchoen} or Nakajima-Zwanzig
approach~\cite{Koller10} technique, in particular when combined with Liouville
space techniques~\cite{Leijnse08a}.
However, at low temperature this approach becomes problematic due to effect of
high-energy contributions \HL{that} renormalize the effective low-energy
physics.
\key{
In this paper, we show how this approach can be extended to this regime.
In particular, we show that much of the insightful structure of  generalized the
master / kinetic equation approach at high temperature is preserved
in this regime.} We formulate a real-time renormalization group approach that
naturally connects to the generalized quantum master or kinetic equation
approach.

For this purpose, we study the simplest possible benchmark model of an
interacting \hl{quantum dot} (QD) coupled to \HL{metallic} electrodes, the
Anderson model.
We explicitly set up a general approach to deal with the non-linear transport at
very low temperature \HL{using a renormalization group formulated in the
framework of the real-time perturbation theory.}
We proceed analytically as far as possible, making this a technically
challenging task.
In fact, the problem is not really manageable without a different physical approach
for dealing with superoperators.
The development of this \emph{Liouville-Fock-space approach} is a central topic
of this paper.
Our approach differs
\HL{from previous formulations~\cite{Prosen08,Schmutz,Kosov, Mukamel08},
both by its construction and by the scope of its application.}
\HL{The most closely related is that} of Prosen,~\cite{Prosen08} which was used
to calculate steady states of quadratic effective Liouvillians.
Here, we extend it to the reservoirs with continuous fields as well and further
develop it to simplify 
the microscopic \emph{derivation} of effective Liouvillians for non-quadratic
problems. 
The approach introduced by Schmutz~\cite{Schmutz} and used by other
authors~\cite{Kosov, Mukamel08} differs from our approach in principal details
\HL{that} are essential for our application.
\HL{See also the very recent~\Cite{Dzhioev}}.

A large part of the paper is devoted to the development of our Liouville-Fock space approach since it allows for
general physical insights into the
problem and is crucial for overcoming technical difficulties in setting up
\HL{the} real-time renormalization group (RT-RG). We will illustrate its potential in \HL{the} application
to \HL{the} RT-RG calculation of the non-linear transport at zero temperature.
As \HL{a} result, this paper by necessity is extensive. Further motivation for
\HL{its length} is that several general physical insights into the real-time
approach have not been pointed out, although this approach has been developed
for some time and \HL{has} found widespread use (see \Cite{Schoeller09a} for 
a review).
%To include these it is required
\HL{It is therefore required} that the approach is set up from scratch,
paying special attention to
(i) \HL{the} causal structure and the related Keldysh rotation,
(ii) \HL{the} Liouville space formulation, in particular \HL{the second}
quantization for \emph{superoperators},
(iii) \HL{the} spin- and charge-rotation symmetries,
\HL{and}
(iv) the infinite-temperature limit\HL{, which serves as a reference point, both
for} the \HL{second} quantization technique that we develop, as well as \HL{for}
the perturbation theory \HL{and RT-RG}.
Only by fully exploiting these does the application of RT-RG to the Anderson model
becomes \HL{feasible}.
Clearly, these developments are best presented coherently in the context of the
application to the RT-RG for which it is absolutely crucial.

To indicate the impact of these developments for \HL{our study of the Anderson
model},
we note that the simplest approximation \HL{that} includes the exact result for
the $U=0$ limit
requires an infinite series of diagrams in the standard \HL{real-time
perturbation theory}.
Using the RT-RG, this result is recovered only when performing a one- \emph{and
two-loop} analysis for the effective Liouvillian \emph{and} including one-loop
vertex corrections.
We emphasize that when applied to the interacting case, this incorporates
renormalization effects from strong tunneling,
while neglecting spin-fluctuation processes relevant only in the  Kondo regime,
which enter only in a three-loop RG analysis
(the latter has been addressed previously based on a Kondo-model
mapping~\cite{Schoeller09b}).
Naively formulating these \HL{RG} equations leads to hundreds of non-linear,
coupled integro-differential equations for frequency-dependent coupling
functions.
The central result of this paper is the derivation of 30 coupled differential
coupling functions, \HL{which
systematically incorporate} the leading frequency dependence
\HL{and describe} the $U=0$ limit exactly.
On the way, we \HL{derive} several exact results of general importance.
\HL{Altogether, this makes} an efficient numerical implementation possible
and allows experimentally relevant stability diagrams to be calculated from wide
ranges of parameters in the non-linear zero-temperature regime (excluding the
narrow Kondo regime).
Many of the results can be extended to generic models involving local
interactions (multiorbital Anderson-type models) with bilinear tunnel coupling
to reservoirs.
See \Cite{Andergassen11a} for a recent study of non-local interactions using the
RT-RG.

The paper is organized in \HL{three} main parts as follows.
In the first part, \Sec{sec:model},  we formulate the model and directly
\HL{revert} to a Liouville space description
and develop the kinetic equation approach for the stationary QD density
operator.
We formulate the perturbation series for the effective Liouvillian $L(z)$
appearing in this equation
using what we will call \emph{causal representation} of field \emph{superoperators} $G$ with
fermionic statistics.
We emphasize that this formulation of the perturbation theory, although
equivalent to previous formulations,~\cite{Koenig96a,Leijnse08a,Koller10}
leads to many simplifications beyond the application of interest here,
and therefore warrants a proper, extensive discussion.
Several of these results have already found application,~\cite{Reckermann13}
and even provide insights into, and generalization of, recent interesting
predictions.~\cite{Contreras12}
Moreover, a renormalized perturbation theory
\HL{that} takes the \emph{infinite-temperature limit} as a formal reference
point,
suggests itself.
It also connects in a natural way to the renormalization group approach
 while preserving much of its general perturbative structure.
We show how the calculation of the current requires little additional
calculation
and prove that in our non-linear approach the linear current vanishes at zero
bias,
\HL{a fact that} is not obvious from the general structure of the theory.

In the second main part of the paper, \Sec{sec:rg},
the explicit one- and two-loop RT-RG equations are derived,
accounting for the energy dependence of both the Liouvillian and \HL{the}
vertices due to the finite non-\HL{linear} transport voltage.
\HL{For the non-interacting case, $U=0$, the} \emph{current} is shown to arise
naturally as an exact result already in \HL{the} one-loop RT-RG, implying that all
two-loop corrections \emph{to this observable} arise from the strong local Coulomb
interaction.
We find, however, that for $U=0$ non-zero, two-loop terms exist,
which are relevant when one is interested in, e.g., the density matrix (and not
just the current).
The non-trivial frequency dependence of two-loop equations is systematically
accounted for
in powers of the renormalized dimensionless coupling superoperators $\bar{G}$,
resulting in an effective RG equation \emph{for \HL{an} effective Liouvillian
only} \HL{that} accounts for vertex renormalization corrections.

This simplification enables the detailed numerical study of the
zero-temperature, non-linear transport \HL{in the third part of the paper,
\Sec{sec:results}.}
\HL{This covers all regimes, except} for the Kondo regime of low applied voltage
and magnetic field.
The importance of accounting for both one- and two-loop corrections,
 as well as the \HL{non-equilibrium} Matsubara axes is demonstrated numerically.
Finally, we show that the \emph{tunnel-induced} renormalization effects
incorporated in our one- plus-two-loop approach enhance the inelastic cotunneling 
resonance at finite magnetic field and voltage and generate non-linearities of
the single-electron tunneling (SET) stability diagrams (Coulomb diamonds).

\section{Model and real-time transport theory  \label{sec:model}}
\subsection{Anderson model\label{sec:anderson-model}}

\key{In this section we introduce the model and our compact notation, which is crucial to the Liouville space formulation of the theory.}
The simplest model Hamiltonian of a QD that takes into account Coulomb interaction effects involves just a single orbital:
\begin{align}
  \label{dot_ham}
  H  =   \epsilon n + B S_z  + U  n_\uparrow  n_\downarrow
  .
\end{align}
\hl{where
$ n=\sum_\sigma  n_\sigma$ and
$ n_\sigma=  {d}^{\dagger}_\sigma  {d}_\sigma$
are the occupation operators.
Here, $\epsilon$ denotes the energy of the orbital, experimentally controlled by the gate voltage $V_g$ (we take $\epsilon= - V_g$), and $U$ is the Coulomb charging energy.}
The index $\sigma=\pm$ corresponds to spin up ($\uparrow$) and down ($\downarrow$)
and $S_z=\tfrac{1}{2}\sum_\sigma \sigma n_\sigma$ is the $z$ component of the spin vector operator
$\vec{S}=
\sum_{\sigma \sigma'}  \tfrac{1}{2} \vecg{\sigma}_{\sigma,\sigma'}  {d}^{\dagger}_{\sigma}{d}_{\sigma'}
$ along the external magnetic field $\vec{B}=B\vec{e}_z$ (in units where $g \mu_B =1$)
and
$\vecg{\sigma}$ is the vector of Pauli matrices.
The dot is attached to electrodes, treated as free electron reservoirs:
\begin{align}
    H^{R}=\sum_{\sigma,r,k} \epsilon_{r,k}  {a}^{\dagger}_{\sigma, r,k}  {a}_{\sigma, r,k}
    ,
\end{align}
where \HL{the index of the spin $\sigma=\pm$, quantized along the $z$-axis,
corresponds to $\uparrow,\downarrow$,
the reservoir index $r=\pm$ corresponds to $L,R$
and $k$ is the orbital index.}
The reservoir electron number and spin \HL{operator} can be decomposed \HL{as}
${n}^{R} = \sum_r {n}^r$
and
$\vec{s}^{R} = \sum_r \vec{s}^r$\HL{, respectively, into}
\begin{align}
  {n}^r &=\sum_{\sigma,k}  {a}^{\dagger}_{\sigma, r, k}  {a}_{\sigma,r, k}
  ,
  \\
  \vec{s}^r&=\sum_{\sigma,k}  \tfrac{1}{2} \vecg{\sigma}_{\sigma,\sigma'}{a}^{\dagger}_{\sigma, r, k}  {a}_{\sigma',r, k}
  .
\end{align}
In the continuum limit the reservoirs are described by the density of states
$
\nu_r (\omega)= \sum_k \delta (\omega - \epsilon_{r,k} +\mu_{r} )
$ and we go to the energy representation of the fermionic operators,
\begin{align}
  {a}_{\sigma, r} (\omega)
  = \frac{1}{\sqrt{\nu_{r} (\omega) } } \sum_k  {a}_{\sigma,r,k}\delta (\omega - \epsilon_{r,k} +\mu_{r} )
  ,
\end{align}
with the anticommutation relations
\begin{align}
  [ {a}_{\sigma,r} (\omega),  {a}_{\sigma',r'}^{\dagger} (\omega')]_{+}
  &=
   \delta_{\sigma,\sigma'} \delta_{r,r'} \delta (\omega - \omega')
   ,
  \\
  [ {a}_{\sigma,r} (\omega),  {a}_{\sigma',r'} (\omega')]_{+} &= 0
  ,
\end{align}
where we denote (anti)commutators by $[A,B]_{\pm} = A B \pm B A$.
Thus we have for the reservoir Hamiltonian:
\begin{align}
   H^{\R} = \sum_{\sigma,r} \int d\omega (\omega + \mu_r )  {a}^{\dagger}_{\sigma,r} (\omega)  {a}_{\sigma,r} (\omega)
   .
\end{align}
In contrast to $\epsilon_{r,k}$ the energy $\omega$ is the electron energy relative to $\mu_r$, i.e., the reference energy depends \HL{on} which reservoir \HL{$r$} is considered.

The junctions connecting the dot and reservoirs are modeled by the tunneling Hamiltonian
\begin{align}
   {V} &=  \sum_{r} V^r
   \label{Vdef}
   ,
  \\
   {V^r} &=  \sum_{\sigma} \int d\omega \sqrt{\nu_r (\omega)}
   \left( t_r (\omega) {a}^{\dagger}_{\sigma,r} (\omega)  {d}_{\sigma}+ \mathrm{h.c.} \right)
   \label{Vrdef}
   .
\end{align}
The Hamiltonian of the total system is denoted by
\begin{align}
  \label{tot_Ham}
   H^{\tot}= H  +  H^{\R} +  {V}
   .
\end{align}
We assume $t_r (\omega)$ to be real and introduce the spectral density
\begin{align}
  \Gamma_r (\omega) =2\pi \nu_r (\omega) |t_r(\omega)|^2
  \label{spectdens}
\end{align}
and rescaled field operators
\begin{align}
   {b}_{\sigma,r} (\omega) =\sqrt{\frac{\Gamma_r (\omega)}{2\pi}}  {a}_{\sigma,r} (\omega)
   .
\end{align}
To make the notation more compact we introduce an additional particle-hole 
index \HL{$\eta$:}\footnote{\HL{To treat the reservoir and dot field operators in the same way our $\eta$ sign convention is opposite to that of \Cite{Schoeller09a}}}
\begin{align}
  {b}_{\eta,\sigma,r} (\omega)
  & =
  \begin{cases}
    {b}^{\dagger}_{\sigma,r} (\omega) &  \eta=+ \\
    {b}_{\sigma,r} (\omega)         &  \eta=-
  \end{cases}
  ,
  \\
  {d}_{\eta,\sigma}
  & =
  \begin{cases}
    {d}^{\dagger}_{\sigma} &  \eta=+ \\
    {d}_{\sigma}         &  \eta=-
  \end{cases}  
  .
  \label{etadef}
\end{align}
Throughout the paper we will denote the inverse value of a two-valued index with a bar\HL{, e.g.,}
\begin{align}
  \bar{\eta} = -\eta
  .
\end{align}
We combine all indices into a multiindex variable written as a number:
\begin{align}
  1 = \eta,\sigma,r,\omega,
  &&
  \bar{1}= \bar{\eta},\sigma,r, \omega
  .
\label{bar1}
\end{align}
\HL{By} way of exception, the bar denotes inversion of the particle-hole index \HL{$\eta$} only.
\HL{With} ${b}_{1}= {b}_{\eta,\sigma,r} (\omega)$ and $ {b}_{\bar{1}}= {b}_{-\eta, \sigma,r} (\omega)$
the various independent anticommutation relations \HL{can be} compactly summarized \HL{as}
\begin{align}
  [ d_1, d_2 ]_{+} & =  \delta_{1\bar{2}}
  \label{danticomm}
  ,
  \\
  [ b_1, b_2 ]_{+} & = \frac{\Gamma_1}{2\pi} \delta_{1\bar{2}}
  \label{banticomm}
  ,
\end{align}
where $\Gamma_1=\Gamma_r(\omega)$. The interaction then simply reads as
\begin{align}
  V &=    b_{\bar{1}} d_{{1}}
  \label{Vshort}
  ,
\end{align}
where we implicitly assume summation over all discrete parts of the multi-index $1$ (i.e., $\eta,\sigma,r$) and integration over its continuous 
part ($\omega$).
If we have more than one multi-index, we distinguish their components by corresponding subscripts: $1=\eta_1,\sigma_1,r_1,\omega_1$, $2=\eta_2,\sigma_2,r_2,\omega_2$.
We \HL{will often omit these subscripts if only one multi-index appears}.
\hl{Importantly, it can be shown}~\footnote{This gives the correct grand-canonical average over the reservoirs, see Appendix A in \Cite{Schoeller09a} for a proof.} \hl{that operators (and, below also superoperators) of the dot and the reservoirs can be treated \emph{as if} they \emph{commute}
(rather than anticommute), i.e., $[d_1,b_2]_{-}=0$ for all multi-indices $1$, $2$.}

The reservoirs are assumed to be \HL{in} thermal equilibrium,
each described by its own grand-canonical density operator
\HL{with temperature $T$ and electrochemical potential $\mu^r$}:
\begin{align}
  \label{grand_canon}
  \rho^{\R}=\prod_{r}\rho^{r},
  &&
  \rho^r = \frac{1}{z^r} e^{-\frac{1}{T} ( H^{r}-\mu^r {n}^r)},
\end{align}
where
$z^r = \mathrm{Tr}_{r}~e^{-\frac{1}{T} ( H^{r}-\mu^r {n}^r)}$.
We assume that a symmetric bias is applied to the electrodes, i.e.,  $\mu_{L,R}=\pm V/2$.
We note the key property
\begin{align}
  b_1 \rho^{\R} = e^{\eta_1 \omega_1/T} \rho^{\R} b_1
  \label{rhores-property}
  .
\end{align}

In \Eq{spectdens}, the density of states varies on the energy scale of the bandwidth $D$,
 which we assume to be much larger than any other energy scale in the problem.
In this wide-band limit, we can assume $\Gamma_r(\omega)$ to be energy independent
and cut off all reservoir energy integrals ($\omega$) at the scale $D$.
The detailed energy dependence of $\Gamma_r(\omega)$ at high energies is not crucially important for the results. \cite{Schoeller09a}
In the actual applications in \Sec{sec:results}, we will assume for simplicity that the tunnel couplings are symmetric, i.e., $\Gamma_L=\Gamma_R=\Gamma$,
and consider the low-temperature limit, i.e., $T \ll U ,  V,  \Gamma_r$, by setting $T=0$.
The results of this section and much of \Sec{sec:rg}, however, do \emph{not} depend on these assumptions unless explicitly indicated.

For a non-zero magnetic field $B$ and finite Coulomb interaction $U$, the total system possesses two locally and globally conserved quantities
that will play an important role.
\HL{The charge and the spin components along the magnetic field are conserved individually
in the dot and in the reservoirs},
\begin{alignat}{2}
  [H ,n]_{-}  &= 0
  ,
  &
  ~~~~~~~~
  [H ,S_z]_{-} &= 0
  \label{nconserv}
  ,
  \\
  [ H^{\R} , n^{\R}]_{-}  &= 0
  ,
  &
  [ H^{\R}, S_z^{\R}]_{-}  &= 0 
  \label{nresconserv}
  .
\end{alignat}
These conservation laws extend to the total charge, $N^{\tot}=n+ {n}^{\R}$, and spin,
$S_z^{\tot}=S_z+ S_z^{\R}$, since the interaction $V$ commutes with these operators:
\begin{align}
  \label{chgcons}
  [H^{\tot},  N^{\tot} ]_{-}  &= 0
  ,
  \\
  \label{spincons}
  [  H^{\tot}, S_z^{\tot} ]_{-} &= 0
  .
\end{align}

\subsection{Density operator and diagram rules}
\key{The purpose of this section is twofold.
First, we briefly review the real-time approach to the calculation of the stationary reduced density operator,
introducing the central quantities $\Sigma(z)$, the self-energy superoperator, and $L(z)$, the effective Liouvillian,
and their perturbative expansions in vertex superoperators $G$.
Second, we introduce a ``causal'' representation of the perturbation theory, which allows for a compact formulation and derivation of the diagrammatic rules for the self-energy $\Sigma$.
Moreover, many general physical insights become explicitly clear in this representation. In particular, \HL{this new formulation naturally suggests} the possibility of a two-stage, real-time renormalization group (RT-RG), \HL{which} will be set up in \Sec{sec:discrete} and \Sec{sec:rg}.
}

\subsubsection{Stationary density operator}
In order to find the QD stationary state, we need to consider the evolution of the total system density operator.
It evolves according to the Liouville-von Neumann equation,
\begin{align}
  \partial_t \rho^{\tot} (t)
  = -i \left[  H^{\tot} ,  \rho^{\tot} (t) \right]_{-}
  = -i L^{\tot} \rho^{\tot} (t)
  \label{von Neumann}
  ,
\end{align}
with the superoperator Liouvillian
$L^{\tot}=[H^{\tot} , \bullet]_{-}$.
Superoperators are linear 
transformations of operators
and throughout the paper we let the solid bullet ($\bullet$) indicate the operator on which a superoperator acts (if needed).
Explicit matrix representations of superoperators are only required for the QD part and will be discussed later on in \Sec{sec:sym_Liv}.

The initial state of the total system at the initial time $t_0$ is assumed to be the direct product of the dot density matrix
 and the equilibrium density matrices \eq{grand_canon} of the electrodes:
\begin{align}
  \rho^{\tot}(t_0) = \rho (t_0) \rho^{\R}
  .
\label{initialstate}
\end{align}
We will discuss some properties of  $\rho(t_0)$ further in the following.
The formal solution of \Eq{von Neumann} is:
\begin{align}
\rho^{\tot} (t) &= e^{-i H^\tot (t-t_0)} \rho^{\tot}(t_0) e^{i H^\tot (t-t_0)}
\\
& = e^{-i L^\tot (t-t_0)} \rho^{\tot} (t_0)
.
\end{align}
\HL{The} reduced density matrix of the dot is obtained by integrating out \HL{the reservoir} degrees of freedom:
\begin{align}
  \rho (t) = 
\hl{  \Tr{\R} \rho^{\tot} (t)} =\Tr{\R}\left( e^{-i L^\tot (t-t_0)} \rho (t_0)\rho^{\R} \right)
  \label{rhot}
  .
\end{align}
We now decompose $L^{\tot}= L + L^{\R} +L^V$, \HL{where}
$L    =[H  ,\bullet]_{-}$ \HL{and}
$L^{\R}=[H^{\R},\bullet]_{-}$,
and set up the perturbation series in the tunnel coupling $L^V=[V,\bullet]_{-} \sim \sqrt{\Gamma}$.
It is then more convenient \cite{Schoeller09a} to use
the Laplace transform of the dot reduced density matrix for $\im  z > 0$:
\begin{align}
  \label{Laplace-rho}
  \rho (z)&=\int \limits_{t_0}^{\infty} dt e^{i z (t-t_0)} \rho (t)
  \\
  &=\Tr{\R} \left( \frac{i}{z-L - L^{\R}- L^V}\rho^{\tot}(t_0)\right)
  .
\end{align}
We will refer to $z$ as the dot frequency.
We expand the resolvent in $L^V$, resulting in a geometric series with terms of the form
\begin{align}
  \label{exp}
  \Tr{\R} \left(\frac{1}{z -L - L^{\R}} L^V ... L^V \frac{1}{z -L - L^{\R}} \rho^{\tot}(t_0) \right)
  .
\end{align}
The average over the reservoirs can now be calculated directly using a Wick theorem for field superoperators (see \Eq{wick} and \App{sec:Wick}).
Collecting irreducible contractions into the self-energy superoperator $\Sigma(z)$
(see below, \Sec{sec:perturb}),
the perturbation series can be resummed to
\begin{align}
  \label{kineq}
  \rho (z) =\frac{i}{z-L(z)} \rho (t_0)
  ,
\end{align}
where we have introduced the \HL{\emph{effective dot Liouvillian}}
\begin{align}
  \label{Leff}
  L(z) = L+\Sigma (z)
  .
\end{align}
To keep the notation to a minimum we distinguish this quantity from the ``bare'' dot Liouvillian $L=[H,\bullet]_{-}$ by simply appending the dependence on the frequency $z$.
$L(z)$ completely determines the time-evolution of the reduced density operator.
A key idea exploited both in the perturbation theory and in the renormalization group is that one is free to modify $L$ and $\Sigma(z)$ as long as their 
sum remains equal to $L(z)$.
The equation determining the stationary density matrix
$\rho 
 = \lim_{t - t_0 \rightarrow \infty} \rho(t)
 = \lim_{z \rightarrow i0} (-i z) \rho(z)$
is now obtained by multiplying \Eq{kineq} by $-i z (z-L(z))$ and taking $z \rightarrow i0$:
\begin{align}
  L(i0) \rho=0
  \label{zeroeigenvector}
  .
\end{align}
Before deriving the  perturbation series for $\Sigma(z)$ in \Sec{sec:perturb},
we first introduce a convenient representation of the field \emph{superoperators}.

\subsubsection{Causal representation of fermionic field superoperators\label{sec:causal}}

In the following, we integrate out explicitly the reservoir degrees of freedom while keeping track of those of the QD.
To facilitate this,
the tunnel coupling superoperator $L^V=[V, \bullet ]_{-}$ should be written as a convenient product of \HL{superoperators of} the dot 
and the reservoirs: inserting \Eq{Vshort}, we have
\begin{align}
  \label{lvp}
  L^V = p^{L^{N^\tot}} p \mathscr{J}^{p}_{\bar{1}} \mathscr{G}^{p}_1
  ,
\end{align}
\HL{where we defined the following ``naive'' field superoperators:}
\begin{align}
  \label{keld_j}
  \mathscr{J}^{p} \bullet  =
    \begin{cases}
      b_1 \bullet &  p=+\\
      \bullet b_1 &  p=-
    \end{cases}
    ,
  \\
  \label{keld_g}
  \mathscr{G}^{p} \bullet =
  \begin{cases}
    {d}_1 \bullet  &  p=+\\
    \bullet  {d}_1 & p=-
  \end{cases}
  .
\end{align}
The superscript $p=\pm$ keeps track of whether a field operator acts from the left or right
and is referred to as \HL{the} Keldysh index by analogy to the Green's function and functional integral techniques.
In \Eq{lvp}, we implicitly sum over $p$, in addition to the multi-index $1$.
A crucial difference to the formulation of \Cite{Schoeller09a} is that we introduced
a harmless additional superoperator
\begin{align}
  p^{L^{N^{\tot}}} \bullet =  p^{L^{n^{\R}}} p^{L^{n}} \bullet
  \label{harmless}
\end{align}
into \Eq{lvp}, where
\begin{align}
  L^{{N^\tot}} & =[  {N^{\tot}} , \bullet ]_{-} =   L^{{n}} +   L^{{n^{\R}}},
  \label{Lntot}
  \\
  L^{{n}} &= [ {n} , \bullet]_{-},
  \label{Ln}
  \\
  L^{{n^{\R}}} & = [ {n^{\R}} , \bullet ]_{-},
  \label{LnR}
\end{align}
are the superoperators associated with the total, QD, and reservoir electron numbers, respectively.
Clearly, for $p=+1$ this factor is trivially equal to 1.
However, it may seem at first sight that for $p=-1$ this is not the case:
when applied to a projector of states of the total system, this superoperator counts the relative parity of the fermion numbers $N$ and $N'$:
$(-1)^{L^{N^\tot}} |N\lambda\rangle\langle N'\lambda'| = \pm |N\lambda\rangle\langle N'\lambda'|$ for $N-N'= $ even / odd, where $\lambda$ and $\lambda'$ denote further quantum numbers.
However, in all calculations we can assume that the total-system \HL{density} operators on which it acts have even parity,
since odd fermion-parity components of states can neither be measured by any physical operator
nor be prepared using physical evolutions.
This is referred to as the \HL{\emph{fermion-parity superselection rule} of quantum mechanics}.~\cite{Wick52,Aharonov67}

Since the fermion parity plays an important role in what follows
but is often not mentioned or used explicitly in density operator approaches, it warrants some discussion,
in particular since odd-fermion-parity operators do appear in the renormalization group approach.
Physical Hamiltonians and observables (and their corresponding superoperators) always contain only products of even numbers of fermionic operators.
This implies that only the part of the density operator $\rho^{\tot}(t)$ with even fermion parity can enter into the calculation of any physical observable $\langle A \rangle(t) = \mathsf{Tr} A \rho^{\tot}(t)$. 
This even part of $\rho^{\tot}(t)$ is generated solely from the even parity part of $\rho^{\tot}(t_0)$ at earlier times
since by the same token the parity of the total fermion number is conserved during time evolution.
Therefore, only the even fermion-parity part of the initial state $\rho^{\tot}(t_0)$ can contribute to an observable and one may set any odd-fermion part of any density matrix equal to zero.
As a result, we can take in \Eq{lvp} $p^{L^{N^\tot}}\bullet =1\bullet$, even for $p=-1$.
For the factorized initial state \Eq{initialstate} that we assumed here, this implies that $\rho(t_0)$ must be assumed to be of even fermion parity, 
since $\rho^{\R}$ also has even fermion parity 
($L_{n^{\R}} \rho^{\R} = [ n^{\R} , \rho^{\R}]_{-} = 0$, which follows from \Eq{grand_canon}).

The useful implications of this fermion-parity conservation become clear when 
performing a linear transformation of the field superoperators with respect to their Keldysh indices $p$.
The naively chosen field superoperators \eq{keld_j} and \eq{keld_g} have the disadvantage that they commute or anticommute depending on the Keldysh index $p$:
\begin{align}
  \mathscr{G}^{p}_1\mathscr{G}^{p'}_2  + p p' \mathscr{G}^p_1\mathscr{G}^{p'}_2 &= \delta_{pp'} \delta_{1\bar{2}}
  \label{keld_g-comm}
  ,
  \\
  \mathscr{J}^{p}_1\mathscr{J}^{p'}_2  + p p' \mathscr{J}^p_1\mathscr{J}^{p'}_2 &= \delta_{pp'} \delta_{1\bar{2}}
  .
\end{align}
This complicates many calculations as noted, e.g., in~\Cite{Mukamel03}.
However, the factorization of the total fermion parity into a dot and reservoir factor in \eq{harmless}
naturally suggests a transformation of the field operators. By absorbing the fermion parity superoperators of each subsystem into new field superoperators, 
\begin{align}
  \mathcal{G}_1^p & =p^{L^{ n}} \mathscr{G}^{p}_1
   \label{semi-naive_g}
   ,
  \\
  \label{semi-naive_j}
  \mathcal{J}_1^p & =p^{L^{ {n}^{\R}}} \mathscr{J}^{p}_1
  ,
\end{align}
\HL{the latter} obey definite anticommutation relations
\begin{align}
  [ \mathcal{G}^{p}_1, \mathcal{G}^{p'}_2 ]_{+} & = p \, \delta_{pp'}\delta_{1\bar{2}}
  \label{semi-naive-comm}
  ,
  \\
  [ \mathcal{J}^{p}_1, \mathcal{J}^{p'}_2 ]_{+} & = \frac{\Gamma}{2\pi} p \, \delta_{pp'}\delta_{1\bar{2}}
  .
\end{align}
\key{This allows one to prove the Wick theorem directly for the operators $\mathcal{J}$ using simple algebra,~\cite{Reckermann13}
avoiding the need to carefully keep track of sign factors as done in~\Cite{Schoeller09a}.}
However, the non-vanishing anticommutators still depend on the Keldysh index $p$ on the right-hand side.
Moreover, the fields have no simple Hermitian superconjugation relation (see \App{sec:operG}).
This can be avoided by a rotation of the QD fields,
\begin{equation}
  \hl{ G^{q}_1 }  =\left\{
    \begin{alignedat}{7}
      \tilde{G}_1&=& \tfrac{1}{\sqrt{2}} \sum_p \hl{ \mathcal{G}^p_1 }
                 &=& \tfrac{1}{\sqrt{2}}  \sum_p p^{L^{ n}}  \mathscr{G}^{p}_1 
                 & ~~& q=+
      \\
      \bar{G}_1 &=& \tfrac{1}{\sqrt{2}} \sum_p p \hl{ \mathcal{G}^p_1 }
               & =& \tfrac{1}{\sqrt{2}} \sum_p p^{L^{ n}+1}  \mathscr{G}^{p}_1 
               & ~~& q=-
    \end{alignedat}
  \right.
  \label{kr_d}
  ,
\end{equation}
and a contravariant rotation (cf. \Sec{sec:perturb}) of the reservoir fields
\begin{equation}
  J^{q}_1 =\left\{
    \begin{alignedat}{7}
      \tilde{J}_1 &=& \tfrac{1}{\sqrt{2}} \sum_p p \hl{ \mathcal{J}^p_1 }
                  &=& \tfrac{1}{\sqrt{2}}\sum_p p^{L^{ {n}^{\R}}+1} \mathscr{J}^p_1  & ~~ & q=+
      \\
      \bar{J}_1   &=&\tfrac{1}{\sqrt{2}}\sum_p \hl{ \mathcal{J}^p_1 }
                  &=& \tfrac{1}{\sqrt{2}}\sum_p p^{L^{ {n}^{\R}}}  \mathscr{J}^p_1  & ~~ & q=-
    \end{alignedat}
  \right.
  \label{kr_el}
  .
\end{equation}
To facilitate later discussions, we introduced both an index $q=\pm$ as well as ``tilde'' and ``overbar'' symbols to distinguish the new field components.
Now, the anticommutation relations are completely analogous to those of the usual fermionic operators,
\begin{equation}
  \left[ G_1^{q} , {G}_{1'}^{q'} \right]_{+}  
  = \delta_{q,\bar{q}'}\delta_{1,\bar{1}'}
  ,
  ~~
  \left\{
    \begin{alignedat}{3}
      \label{commutG}
      \left[ \tilde{G}_1 , \bar{G}_{1'}\right]_{+}  &= \delta_{1,\bar{1}'}
      \\
      \left[ \tilde{G}_1 , \tilde{G}_{1'}\right]_{+} &= 0
      \\
      \left[ \bar{G}_1 , \bar{G}_{1'}\right]_{+} & = 0
    \end{alignedat}
  \right.
  ,
\end{equation}
whereas in the reservoirs we incorporate the coupling into the normalization factor:
\begin{equation}
  \left[ J_1^{q} , {J}_{1'}^{q'} \right]_{+}   = 
  \frac{\Gamma_1}{2\pi} \delta_{q,\bar{q}'}\delta_{1,\bar{1}'}
  ,
  ~~
  \left\{
    \begin{alignedat}{3}
      \label{commut}
      \left[ \tilde{J}_1 , \bar{J}_{1'}\right]_{+} &=  \frac{\Gamma_1}{{2}\pi}  \delta_{1,\bar{1}'}  \\
      \left[ \tilde{J}_1 , \tilde{J}_{1'}\right]_{+} &= 0
      \\
      \left[ \bar{J}_1 , \bar{J}_{1'}\right]_{+} & =0
    \end{alignedat}
  \right.
  .
\end{equation}
This second transformation is known as the Keldysh rotation~\cite{Larkin75}
and is, e.g.,
applied to fermionic fields represented by Grassmann numbers in functional integral theories~\cite{Kamenev09} or in \HL{the} Green's-function formalism.~\cite{Larkin75}
Here, we find that it also considerably simplifies the real-time transport theory
in which not all degrees of freedom can be integrated out,
in contrast to the cited approaches.
We note that a transformation similar to Eqs.~\eq{semi-naive_g} and \eq{semi-naive_j}, which also results in the usual anticommutation and conjugation
relations for the fermionic  superoperators,
was introduced in \Cite{Schmutz,Kosov,Mukamel08}, however, without performing the Keldysh rotation. 
This transformation, however, is less convenient since
it does not reveal a general structure of the fermionic superoperators \HL{that} is important for our applications. \HL{In the following, this will be related to causality
and we will} therefore refer to \eq{kr_d} and \eq{kr_el} as the \emph{causal representation}
and \HL{to} the index $q=\pm$ as the \emph{causal index}. 
Again, we denote its inverse by $\bar{q}=-q$.
\HL{The superoperator approach introduced by Prosen~\cite{Prosen08} is related to ours, but is constructed in a different way
by first introducing a Fock basis (see \Sec{sec:fock-liouville}). This construction does not allow one to easily see the relations
between Prosen's superoperators and our causal superoperators \eq{kr_d}.
See our detailed comparison of the existing approaches in \App{sec:Pros} and \ref{sec:Schmutz}.}
\key{For the \emph{dot} field superoperators ($G^q$)
this representation was already introduced in \Cite{Schoeller09a},
but it was not exploited in the context of the perturbation theory,
where it provides many useful additional simplifications that we now address.}

First, the property \HL{that for any multiindex 1}
\begin{align}
  \label{trG}
  \Tr{D} (\bar{G}_1 \bullet ) = 0
\end{align}
ensures the probability conservation on the dot during the time-evolution [see \Eq{trSigma}].
\HL{This property of $\bar{G}_1$} is extremely important for the formulation of the RT-RG in \Sec{sec:rg}
and is preserved during renormalization.
We note that probability conservation in the Liouville approach corresponds to the 
normalization conservation for the partition function in the path-integral approach~\cite{Kamenev09}.
Equation \eq{trG} follows directly from the definition of the causal representation:
using $\mathrm{Tr}_{D} L^{n} = 0$, and, using the cyclic property of the trace, \HL{we see that}
$\mathrm{Tr}_{D} \mathscr{G}^p_1 \bullet = \mathrm{Tr}_{D} d_1 \bullet$ is independent of $p=\pm$.
Therefore,
\begin{align}
  \Tr{D} (\bar{G} \bullet )
  = \sum_p  \Tr{D} p^{L^n+1} \mathscr{G}^p_1 \bullet 
  = \sum_p p \Tr{D} d_1 \bullet = 0
  .
\end{align}

Second, the fields $\bar{G}_1$ and $\tilde{G}_{\bar{1}}$ are related by Hermitian conjugation of superoperators:
\begin{align}
  \bar{G}_{1}=( \tilde{G}_{\bar{1}})^{\sdagger}
  .
  \label{Gconj}
\end{align}
For a superoperator $S$ the Hermitian conjugate $S^\sdagger$ is defined by $\mathrm{Tr}(A^\dagger S B) = \mathrm{Tr} ((S^\sdagger A)^\dagger B)$ where $A$ and $B$ 
are arbitrary \emph{operators} (see also \Sec{sec:braket} and \App{sec:operG}).
\key{ This indicates that the conjugate fields $\bar{G}_{1}$ and $\tilde{G}_{\bar{1}}$ are similar to usual creation and annihilation operators, respectively.
}
In \Sec{sec:sym_Liv}, we exploit this ``second quantization'' in Liouville space
to construct a convenient basis \HL{for this space}
that includes the left and right zero eigen-supervectors of both vertices.
This considerably simplifies the matrix representations of many superoperators required in \HL{the} perturbation theory and the RG.
An immediate consequence of the probability conservation property \hl{\eq{trG} and the property} \eq{Gconj} is that $\tilde{G}$ has the unit operator $\unit$ as a right zero eigen-supervector:
\begin{align}
  \tilde{G}_{1} \unit = 0
  \label{Gtildeunit}
  .
\end{align}

\key{The above properties of \HL{the} causal field superoperators provide explicit insight into important physical issues that are otherwise not obvious in the general form of the perturbation theory,
e.g., the wide-band limit, the infinite temperature limit, and \HL{the dependence on the} energy cutoff $D$, which we discuss in \Sec{sec:perturb}.}
Introducing the corresponding causal representation for the \emph{reservoir} field superoperators also yields several simplifications, which we now discuss.

\paragraph{Wick theorem.}
First, we note that due to the local interactions on the QD, \Eq{commutG} cannot be used to formulate a Wick theorem for the vertices $G$.
In contrast to this, for the reservoir field superoperators this is possible due to the relation
\begin{align}
  \label{FDT}
  \bar{J}_1\rho^{\R} = \tanh (\eta_1 \omega_1/2T) \tilde{J}_1\rho^{\R}
  .
\end{align}
This result follows from \Eq{rhores-property} by writing it in superoperator notation,
$
 \mathscr{J}^+_1 \rho^{\R} = e^{\eta_1 \omega_1/T}  \mathscr{J}_1^{-} \rho^{\R}
$, and then applying the transformations \eq{semi-naive_j} and \eq{kr_el}.
With \Eq{FDT} we can algebraically prove the Wick theorem for the superoperators $J^q_1$ in close analogy to the usual case of fermionic operators (see \App{sec:Wick}):
the average of a product equals the product of pair contractions summed over all contractions of pairs $\langle ik \rangle$,
\begin{align}
  \label{wick}
  \Tr{\R} \left (J^{q_1}_1 ... J^{q_{m}}_{m} \rho^{\R} \right)
  =\sum_{i<k} \prod_{ \langle ik \rangle} (-1)^{P} \langle J_i^{q_i} J_k^{q_k} \rangle_{\R}
  ,
\end{align}
 with the usual fermionic sign $(-1)^P$  of the permutation $P$ that disentangles the contractions.

\paragraph{Causal structure.}
The number of possible pair-contractions appearing on the right-hand side of \Eq{wick} is \HL{strongly limited by the causal structure}.
Applying $\Tr{\R}  \bar{J}_1 \bullet$ to \Eq{FDT} written for the operator $\bar{J}_2$ we obtain a relation between two of the possible four pair contractions:
\begin{align}
  \langle \bar{J}_1 \bar{J}_2 \rangle_{\R} = \tanh (\eta_2 \omega_2/2T) \langle \bar{J}_1 \tilde{J}_2\rangle_{\R}
  \label{contr-rel}
  .
\end{align}
This is actually a statement of the equilibrium fluctuation-dissipation theorem for each reservoir separately (see, e.g., \Cite{Esposito09rev}).
In close analogy to \Eq{trG}, one proves \HL{that}
\begin{align}
  \label{trJ}
  \Tr{\R} (\tilde{J}_1 \bullet ) =  0
  .
\end{align}
This implies, in particular,
$\langle \tilde{J}_1 \tilde{J}_2 \rangle_{\R}=0$,
which is related to the well-known fact that of the four reservoir Green's functions only three are independent.~\cite{Lifshitz_physkin}
This is analogous to the absence of \HL{the} so-called ``quantum-quantum'' contractions \HL{and of the} ``classical-classical'' term in the action in the Keldysh functional integral approach~\cite{Kamenev09} and has been referred to as the ``causal'' structure of \HL{the} Green functions.
However, \Eq{trJ} entails an additional simplification:
$ \langle \tilde{J}_1 \bar{J}_2 \rangle_{\R}=0$.
This is particular to our real-time superoperator approach, and is most explicitly related to causality.
To see this, note that we keep track of the left and right action of operators using superoperator notation
and that these superoperators inherit their ordering in the Laplace representation
from their \emph{forward} time-ordering in the time-evolution
 [cf. \Eq{exp}].
We never have to introduce a fictitious backward time-propagation as in the Keldysh technique.
Therefore, no advanced reservoir Green functions can appear in our theory,
which is expressed by $ \langle \tilde{J}_1 \bar{J}_2 \rangle_{\R}=0$.

\paragraph{Energy and temperature dependence.}
The only reservoir correlation functions that can appear in the Wick expansion \eq{wick} for the causal fields have a simpler energy ($\omega$)
and temperature ($T$) dependence than in the representations~\eq{keld_j}-\eq{keld_g} and \eq{semi-naive_g}-\eq{semi-naive_j}.
This correlates with the physical information \HL{incorporated in these functions, i.e., in} the retarded function
\footnote{Formulating the theory in real-time~\cite{Reckermann13} one can directly identify the retarded and Keldysh 
Green functions with their usual definitions~\cite{Rammer} \HL{when setting $\Gamma=\pi$ (cf. \Eq{banticomm}),}
$\HL{-i} \langle \mathcal{T} \bar{J}_1(t) \tilde{J}_2\rangle_{\R} = -i \theta(t) \langle [ b_1(t), b_2]_{+} \rangle_{\R}$
and
$\HL{-i} \langle \mathcal{T} \bar{J}_1(t) \bar{J}_2\rangle_{\R} = -i \langle [ b_1(t), b_2]_{-} \rangle_{\R}$ \HL{where $\mathcal{T}$ is} the time-ordering superoperator.
In Laplace space, their time-evolution frequencies $X$ are convoluted together with the dot Liouvillian into ${1}/{(z-X-L)}$ in the diagram rules~\eq{sig_irr}.
The retarded function $\langle \mathcal{T} \bar{J}_1(t) \tilde{J}_2\rangle_{\R}$ thus contributes only spectral information through $X$ since \Eq{ret-contr} contains only
$  \langle \bar{J}_1 \tilde{J}_2 \rangle_{\R}
  \HL{\propto} \langle [ {b}_1 ,  {b}_2]_{+} \rangle_{\R} = \HL{\delta_{1,2}}
$, whereas the Keldysh function $\HL{-i}\langle \mathcal{T} \bar{J}_1(t) \bar{J}_2\rangle_{\R}$ additionally includes statistical information through \Eq{keld-contr} containing
$  \langle \bar{J}_1 \bar{J}_2 \rangle_{\R}
  \HL{\propto} \langle [ {b}_1 ,  {b}_2]_{-} \rangle_{\R}$.}
\begin{align}
  \label{ret-contr}
  \tilde{\gamma}_{1,2} (\eta_2\omega_2)
  :=
  \langle \bar{J}_1 \tilde{J}_2 \rangle_{\R}
  &
  = \frac{\Gamma_2}{2\pi} \delta_{1,\bar{2}}
  ,
\end{align}
and \HL{in the} Keldysh function
\begin{align}
  \label{keld-contr}
  \bar{\gamma}_{1,2} (\eta_2\omega_2)
  :=
  \langle \bar{J}_1 \bar{J}_2 \rangle_{\R}
  &
  =\frac{\Gamma_2}{2\pi}
  \tanh(\eta_2\omega_2/2T)\delta_{1,\bar{2}}
  .
\end{align}
In the causal representation, the superoperator ordering explicitly shows
that the retarded contraction $\tilde{\gamma}$ contains no information about the distribution function of the reservoirs
 and is therefore temperature independent.
Equation \eq{ret-contr} follows directly from the anticommutation relation \eq{commut} and the property \eq{trJ},
and is indeed independent of the reservoir density operator $\rho^{\R}$.
In contrast, for the Keldysh contraction \HL{$\bar{\gamma}$}, one first needs to use the property \eq{FDT} specific to the equilibrium state of the  
non-interacting reservoirs, before \Eq{commut} and \Eq{trJ} can be applied.
The above shows that the representation \eq{kr_el} reflects most explicitly the causal structure of the perturbation theory,
motivating its denotation.
It thereby automatically achieves the decomposition of the reservoir Fermi distribution function into its symmetric (trivial) and antisymmetric 
(non-trivial) parts with respect to the energy $\omega$ that was introduced in \Cite{Schoeller09a}, where it \HL{was} crucial \HL{for} setting up the RT-RG.
\key{In the following, we show that in the causal representation for the QD fields the perturbation theory also drastically simplifies and that in this  formulation the RT-RG appears quite naturally.}

\subsubsection{Perturbation series and diagram rules\label{sec:perturb}}

\begin{figure}[tbp]
  \includegraphics[width=1.0\linewidth]{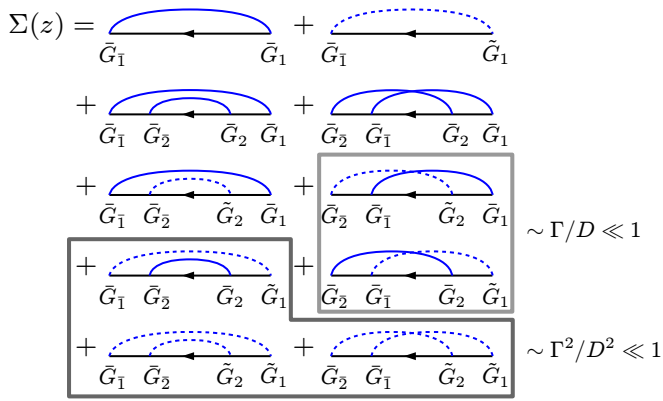}
   \caption{
     One- and two-loop contributions to the self-energy kernel $\Sigma$.
     The full black line denotes the free dot propagation, the arrow
indicates the ordering of superoperators (``late goes left'') in Laplace space
that is ``inherited'' from the time ordering.
     The curves denote the contraction of dot superoperators.
     Here, the line connects the dot superoperators whose corresponding
reservoir superoperators are contracted:
     $\langle GJ\ldots G J \rangle^{\R}
     =
     \langle J J \rangle^{\R} G\ldots G
     \longrightarrow
     \contraction{}{G}{\ldots}{G}
     G\ldots G
     $
     .
     This is in contrast to the standard technique, where a contraction line
connects contracted operators \emph{themselves}.
     The full blue line denotes a Keldysh contraction $\bar{\gamma}$ of
two $\bar{G}$ type vertices (cf. \Eq{keld-contr})
     whereas the dashed blue contraction corresponds to the retarded
contraction $\tilde{\gamma}$ of a $\bar{G}$ with an earlier $\tilde{G}$ vertex to the right
     (cf.~\Eq{ret-contr}).
     Diagrams with vanishing contractions, i.e., where $\tilde{G}$ is contracted to the
right (cf. \Eq{trJ}), are not drawn.
     Furthermore, the 2 loop diagrams in the light and dark gray boxes
     contain a $\tilde{\gamma}$ contraction enclosing $k=1$ and
respectively $k=2$ other vertices
     and therefore scale with $\Gamma^k/D^k$ and can be neglected in the wide band limit.
     In the causal representation only 5 diagrams remain, in contrast to
formulations using ``naive'' superoperators \cite{Leijnse08a} where 20
terms remain, which partially cancel out.
   }
   \label{fig:sigma-diagrams}
\end{figure}

\paragraph{Diagram rules}
When defining the causal representations \eq{kr_d} and \eq{kr_el}, we performed opposite Keldysh rotations for the dot and the reservoirs.
This is motivated by the form of the tunneling Liouvillian \Eq{lvp}, which in the causal representation can be written compactly as
\begin{align}
  \label{q-li}
  L^V=
  \bar{G}_1 \bar{J}_{\bar{1}} + \tilde{G}_1\tilde{J}_{\bar{1}}
  =\sum_{q=\pm} G^{q}_1 J^{q}_{\bar{1}}
  .
\end{align}
\hl{We note the convenient absence of a minus sign on the right-hand side:
we can treat the dot superoperators $G_i^{q_i}$ \emph{as if} their commute (rather than anticommute) 
with the reservoir superoperators $J_k^{q_k}$ as mentioned in the remark after \Eq{Vshort}.}
We can now integrate out the reservoir degrees of freedom in each term of the expansion \eq{exp} in the way discussed in~\Cite{Schoeller09a} by commuting all reservoir superoperators to the right side using the relation
\begin{align}
  J^{q}_1 L^{\R} = \left ( L^{\R}-\eta_1 (\omega_1 +\mu_1 ) \right) J^{q}_1
  .
\end{align}
The $m$th-order contributions to $\Sigma(z)$ then read as
\begin{align}
  &\frac{1}{z-L} G^{q_1}_1 \frac{1}{z-X_1-L} G^{q_2}_2\frac{1}{z-X_2-L} ... \\ \nonumber 
  &\frac{1}{z-X_{m-1}}G^{q_{m}}_{m} \frac{1}{z-L} \rho (t_0)
  \Tr{\R} \left (J^{q_1}_{\bar{1}} J^{q_2}_{\bar{2}} ... J^{q_{m}}_{\bar{m}} \rho^{\R} \right)
  .
\end{align}
Here,
$X_i=\sum_{k\leq i} x_k$,
where the summation runs over the reservoir frequencies $x_k= \eta_k (\omega_k+\mu_k)$
of the $G^{q_k}_k (J_{\bar{k}}^{q_k})$ originally standing to the left of resolvent $i$.
Applying the Wick theorem~\eq{wick}, we can represent the terms diagrammatically
by propagator\HL{-lines} connecting vertices that are contracted in pairs by lines with frequencies $x_k$.
The irreducible parts of these diagrams, i.e., those parts which cannot be cut without hitting at least one reservoir contraction,
 are collected into the irreducible kernel or \emph{self-energy superoperator} $\Sigma (z)$.
In \Fig{fig:sigma-diagrams}, we show the diagrams for $\Sigma(z)$ to one- and two-loop order.
Leaving \HL{implicit the sum over all indices, all possible configurations of pair contractions and all orders $m$,} we can write:
\begin{align}
  \label{sig_irr}
  &
  \Sigma (z) = (-1)^{P} \left(\prod \HL{\gamma_{ij}} \right)_{\mathrm{irr}}
  \times
  \\
  &
  \HL{G^{-}_{1}\frac{1}{z-X_1-L}{G}^{q_2}_{2} ...
    ...{G}^{q_{m-1}}_{m-1}\frac{1}{z-X_{m-1}-L} G^{q_{m}}_{m}
  }
  \nonumber
\end{align}
Here, $\gamma$ denotes that the function $\bar{\gamma}$ ($\tilde{\gamma}$), given by  \Eq{keld-contr} (\Eq{ret-contr}), should be written for pair contraction connecting a \HL{$\bar{G}_i$ vertex on the left with a $\bar{G}_j$  ($\tilde{G}_j$)} vertex on the right.
By \Eq{q-li} it is the \emph{earliest} vertex (rightmost) that decides the type of contraction,
i.e.,  its indices appear as the argument of the contraction function.
Here, $X_i$ is now the sum over the frequencies of all reservoir contractions which go over $i$-th resolvent 
(since contractions that start and end to the left \HL{or} right cancel out).

Importantly, since on the left we always have $\bar{G}$ (i.e., $G^{q_1}$ with $q_1=-$),
the property \eq{trG} of the causal field of type $\bar{G}$ is seen
 to guarantee the conservation of probability:
\begin{align}
  \label{trSigma} 
  \Tr{D} \Sigma (z) = 0
  .
\end{align}
Another general property restricts the frequency dependence:
\begin{align}
   K \Sigma (z) K= -\Sigma(-z^{*})
   .
  \label{hermcond}
\end{align}
Here, $K=K^{-1}$ is the antilinear superoperator that effects the Hermitian conjugation of an \emph{operator}, see \App{sec:K-pr}.
\HL{(Such a transformation of an arbitrary superoperator \HL{$S \rightarrow KSK :=S^c$ was} referred to as ``$c$-conjugation'' in \Cite{Schoeller09a}.)}
This guarantees through \Eq{kineq}-\eq{Leff} that the reduced density operator remains Hermitian during the time-evolution, i.e., $\rho(z)=\rho(-z^{*})$.
The property \HL{\eq{hermcond} derives from} the conjugation properties \eq{GbarK} \HL{below} and \HL{from} $KLK^{-1}=-L$.
\key{
The causal representation \eq{sig_irr} of the diagrammatic perturbation theory is very useful in general and
\HL{has been extended} to time-dependent problems \HL{in \Cite{Saptsov13a,Reckermann13}.}
We now discuss the main advantages
\HL{that} will be important for setting up the RG and the construction of a convenient supervector basis  in \Sec{sec:sym_Liv}.
}

\paragraph{Wide-band limit.}
First of all, the number of \emph{contributing terms} \HL{in \Eq{sig_irr}} is strongly reduced since the trivial ($\tilde{\gamma}$) and non-trivial ($\bar{\gamma}$) 
energy dependence of the contractions is automatically separated,
which in other representations \HL{has} to be done separately~\cite{Leijnse08a}.
\HL{Here, terms that} do not contribute in the \HL{wide} bandwidth limit can be identified \HL{using the} diagram topology as illustrated in \Fig{fig:sigma-diagrams}.
\HL{All} diagrams where one or more vertices are enclosed between contracted vertices $\bar{G}_1$ and $\tilde{G}_{\bar{1}}$  give contributions of order $\Gamma/D \ll 1$ or smaller and can be neglected.
This can be proved by careful examination of the poles appearing when closing all integrals in the complex upper half plane (see \Cite{Reckermann13}), or argued in the time-representation.
\footnote{In the wide-band limit the energy-independent contractions $\tilde{\gamma}$ correspond to the $\delta$-functions in time, leaving no 
phase space for the $\omega$-integrals when enclosing vertices.}
As a result $\tilde{\gamma}$ contractions can only occur inside $\bar{\gamma}$ contractions in diagrams with more than one loop,
see \Fig{fig:sigma-diagrams}.
This results in an exponential reduction in the number of contributing terms.

\key{
This feature also naturally suggests a starting point for a two-stage RG approach.
First, all terms \HL{that} contain retarded contractions $\tilde{\gamma}$ can be integrated out explicitly by a one-step diagram resummation, leading to a renormalization of $L$ discussed in \Sec{sec:discrete}.
The remaining diagrams will then contain only $\bar{G}$ superoperators with non-trivial contractions $\bar{\gamma}$, which \HL{can be eliminated by} a second, continuous RG.
This approach will be worked out in detail in \Sec{sec:rg}.
}

\paragraph{Infinite-temperature limit}
\label{sec:perturb12}
The choice of the supervector basis is simplified very much by noting that the superoperator structure of $\Sigma(z)$ is strongly restricted in the causal representation:
in the wide-band limit (see above), there is only one  diagram that starts with $\tilde{G}$ on the right and ends with $\bar{G}$ on the left.
We denote this special one-loop diagram by $\tilde{\Sigma}$.
Importantly, all other diagrams start and end with a $\bar{G}$ vertex.
The $\omega$ integral for the $\tilde{\Sigma}$ diagram can be performed by closing the integration contour in the upper/lower half-plane 
of the complex plane for $\eta=\mp$ and neglecting small corrections of order $\Gamma/D \ll 1$:
\begin{align}
  \tilde{\Sigma}(z)
  = \int d{\omega}
   \bar{G}_1 \frac{\tilde{\gamma} (\eta \omega)}{z-L-\eta\omega-\eta\mu_1}\tilde{G}_{\bar{1}}
  = 
  -i \frac{\Gamma_{1}}{2} \bar{G}_1 \tilde{G}_{\bar{1}}
  \label{tildeSigma}
  .
\end{align}
where $\Gamma_{1}=\Gamma_r$ for multiindex $1=\eta,\sigma,r,\omega$.
The self-energy $\tilde{\Sigma}$ has a clear physical meaning: it is the self-energy one obtains in the infinite temperature limit.
This follows from its definition since all contractions $\bar{\gamma}=0$ for $T=\infty$, and $\tilde{\Sigma}$ is 
the only diagram in wide-band limit without this contraction function.
As discussed in \Sec{sec:causal}, the retarded correlation function of the reservoirs $\tilde{\gamma}$ contains only spectral information (see the discussion of \Eq{ret-contr})
and is therefore independent of $T$. Therefore, $\Sigma=\tilde{\Sigma}$ at $T=\infty$.
Since at $T=\infty$ no energy scale matters any more,
$\tilde{\Sigma}$ is independent of the QD frequency $z$ or any QD energy scale in the problem
as well as the cutoff $D$.
The dependence on $\Gamma$ remains, however: in the high-temperature limit all quantum dot states are equally accessible by tunneling processes. This is described by the self-energy $\tilde{\Sigma}$.

The action of $\tilde{\Sigma}$ is very different from that of the Liouvillian $L$ of the isolated QD: it is not super-Hermitian, as $L$ is, but rather 
\HL{anti-super-Hermitian,}
\begin{align}
  \tilde{\Sigma}^\sdagger =  - \tilde{\Sigma}
  ,
  \label{tildeSigma-hc}
\end{align}
which simply follows from the Hermitian conjugation property of the causal {field superoperators} [\Eq{Gconj}].
For $T=\infty$, the effective Liouvillian thus reduces to
$L(z) = L+\tilde{\Sigma}(z)$
with stationary state 
\begin{align}
\label{ZL}
\rho
=\tfrac{1}{4}\unit,
\end{align}
which is the maximal entropy state.
This follows directly from the causal representation of $\tilde{\Sigma}$ [\Eq{tildeSigma}], which shows that $\tilde{\Sigma}$ and $\tilde{G}$ share the same right eigenvectors,
combined with the probability conservation \Eq{Gtildeunit}.
By the same argument it is clear that in this limit the current vanishes:
anticipating the result \eq{current_a} \HL{of} \Sec{sec:current}, we find for 
the self-energy required for the current equals
\begin{align}
  \tilde{\Sigma}^r(z)=-i \tfrac{\Gamma_1}{2} \bar{G}_{1}\tilde{G}_{\bar{1}}|_{r_1=r}
  \label{tildeSigmar}
  ,
\end{align}
where in the sum over $1$ we exclude the reservoir index $r$.
This also vanishes by \Eq{Gtildeunit} and therefore $\langle I^r \rangle = 0$.

\paragraph{Cutoff dependence and complete basis}

Another advantage of the causal representation is that the cutoff dependence of integrals in the  \emph{individual} diagrams that do contribute in the wide-band limit can be analyzed on the level of superoperators.
These self-energy contributions seem to depend on the energy integral cutoff $D$. However, using the causal structure of the perturbation theory, one can \emph{explicitly} see that such dependence cancels out
due to the superoperator structure of the vertices (i.e., the matrices multiplying the $D$-dependent contributions \HL{to the} integrals vanish).
The condition for this is that one keeps the complete basis of many-body eigenstates of the dot Hamiltonian.
To see this, however, one needs to consider the entire Liouville space, including all off-diagonal density operator elements and not restrict the analysis to only diagonal density matrix elements based on symmetry properties
as is often done.
The idea is best illustrated by considering the one-loop contributions in \Fig{fig:sigma-diagrams}.
By \Eq{tildeSigma}, the one-loop diagram with a $\tilde{\gamma}$ contraction is explicitly independent of $D$.
For the $\bar{\gamma}$ contraction of two $\bar{G}$-type vertices,
\begin{align}
\HL{
  \int d{\omega_1}
  \bar{G}_1 \frac{ \bar{\gamma}_{\bar{1}1} (\eta_1 \omega_1)}
  {z-L-\eta_1\omega_1-\eta_1\mu_1}\bar{G}_{\bar{1}}
 }
   ,
\end{align}
the $D$ dependence enters through the  most divergent part of the integral, obtained by neglecting $L_D$ in the denominator. This part is thus proportional \HL{to a superoperator
that is identically zero due to anticommutation relations \eq{commutG}:}
\begin{align}
  \bar{G}_{1} \bar{G}_{\bar{1}}
  =  -\bar{G}_{\bar{1}} \bar{G}_{1}
  =   \bar{G}_{2} \bar{G}_{\bar{2}} =  0
  ,
  \label{cutoff-indep}
\end{align}
\HL{Here, we} renamed the dummy summation indices \HL{$1=\eta_1,\sigma_1,r_1,\omega_1$}; in fact, only the \HL{implicit} summation over $\eta_1$ is relevant here.
Importantly, this argument breaks down as soon as many-body states have been excluded from the Hilbert space basis, e.g., based on their large energy.
For example, in the limit $U\rightarrow \infty$, one can exclude the doubly occupied QD state $|2\rangle$ and thereby eliminate Liouville-space elements with eigenvalues exceeding $D$.
This simplifies the calculations, but for vertex operators projected onto this subspace the relation \eq{cutoff-indep} does \emph{not} hold anymore (since it is a non-linear relation).
As a result, the explicit $D$ dependence remains and does not cancel out from such expressions,
and the cutoff should be set to $D \sim U$ if one makes this approximation.

The above analysis can be extended to higher-order diagrams, and one finds that also there the $D$ dependence drops out.
Here one uses that diagrams containing $\tilde{G}$ vertices are always expressible in terms of the cutoff-independent 
$\tilde{\Sigma}$ skeleton (see also \Sec{sec:discrete}).
This analysis confirms \HL{the observation}, made previously \emph{after} explicit calculations of the kernel \HL{$\Sigma(i0)$}, e.g., in \Cite{Koenig96b}\HL{, that the cutoff $D$ drops out.
Here, however, the cutoff dependence can be completely assessed on a very general level,}  based on \HL{the fermionic} superoperator algebra, without \HL{any} need for \HL{the} explicit calculation of matrix-elements.

\subsubsection{\emph{Finite} temperature perturbation theory:
 Elimination of infinite-temperature contractions
\label{sec:discrete}}
\begin{figure}[tbp]
  \includegraphics[width=1.0\linewidth]{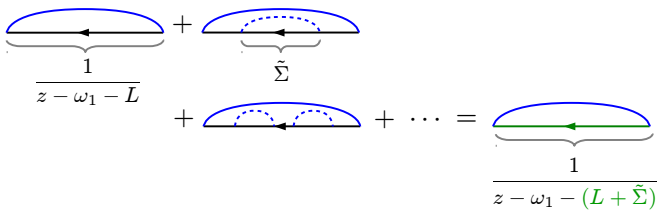}
  \caption{
    Example resummation of diagrams with a single $\bar{\gamma}$ loop (solid blue curve)
    and $n=0,1,2,\ldots$ skeleton diagrams $\tilde{\Sigma}$,
    resulting in one-loop diagram with a renormalized propagator.
    Since in the wide-band limit $\tilde{\gamma}$ contractions (dashed blue curve) can not contain other vertices,
    this can be extended to any irreducible diagrams with any number of $\bar{\gamma}$ loops.
  }
  \label{fig:zero-diagrams}
\end{figure}

As already mentioned at the end of \Sec{sec:perturb}, 
the causal structure of the perturbation theory naturally suggests to proceed in two stages.
We first eliminate all diagrams containing the retarded contraction $\tilde{\gamma}$, i.e., the skeleton diagram \eq{tildeSigma}.
\HL{As discussed in \Sec{sec:causal}, the function $\tilde{\gamma}$, given by \Eq{ret-contr},
only contains spectral information about the reservoirs.}
Since this contraction always occurs isolated and inside other Keldysh contractions (cf. \Sec{sec:perturb}), we can 
use \eq{tildeSigma} as a skeleton diagram and on each propagator line resum the series,
\begin{align}
  \frac{1}{z-L-X}\sum_{n=0}^\infty \left( \tilde{\Sigma}\frac{1}{z-L-X} \right)^n
  =  \frac{1}{z-\bar{L}-X}
  ,
  \label{tildeSigma2}
\end{align}
thereby renormalizing the dot Liouvillian to
\begin{align}
  \bar{L} = L +\tilde{\Sigma}
  .
  \label{barL}
\end{align}
This is illustrated in \Fig{fig:zero-diagrams}.
As discussed in \Sec{sec:perturb12}, $\tilde{\Sigma}$ equals the QD self-energy in the limit of infinite temperature in all reservoirs.
The lack of energy dependence of $\tilde{\Sigma}$ reflects that in the high-temperature limit, all QD states are equally accessible via transitions induced by the electrodes.
Physically, one expects that the Keldysh contractions $\bar{\gamma}$, describing the non-trivial, temperature dependent part of the distribution function~\cite{Rammer}, \HL{will} drop out in this limit.

In a second stage, we calculate the finite temperature effects, which are all incorporated through the self-energy $\bar{\Sigma}(z)$.
Its diagrammatic perturbation series has the same structure as for $\Sigma$, but is expressed entirely  in terms of the contraction $\bar{\gamma}$, 
the vertex $\bar{G}$ and the Liouvillian $\bar{L}$:
\begin{align}
  \label{barSigma}
  &
  \bar{\Sigma}(z)
  =\hl{\HL{(-1)^P\left(\prod \bar{\gamma}_{ij}\right)_{\mathrm{irr}}\times}}
  \\
  &
  \bar{G}_1 \frac{1}{z-X_1-\bar{L}}\bar{G}_2 ... \frac{1}{z-X_{n-1}-\bar{L}}\bar{G}_n
  \nonumber
  .
\end{align}
Summing this renormalized perturbation theory, one obtains, of course, the same exact result \eq{Leff} for the effective Liouvillian, which, however, is now decomposed in a different way:
\begin{align}
  \label{Leff2}
  L(z) = \bar{L}+\bar{\Sigma} (z)
  .
\end{align}
\HL{All} contractions $\tilde{\gamma}$ have been eliminated simultaneously in the above transformation of the perturbation series\HL{. This was} previously referred to as the discrete step of \HL{the real-time} renormalization group procedure.~\cite{Schoeller09a}
In the Anderson model in the wide-band limit,
only one skeleton diagram contributes to $\bar{L}$ through $\tilde{\Sigma}$
and no renormalization of vertex $\bar{G}$ is required to eliminate $\tilde{\gamma}$.
\footnote{For models not of the Anderson-type (bilinear coupling), higher skeleton diagrams and vertex renormalizations arise, requiring a more 
general discrete  renormalization.~\cite{Schoeller09a}}
This discrete first step is a necessary preparation for the second-step RG in which  we will integrate out the $\bar{\gamma}$ contractions as well in a continuous RG flow.
In contrast to the $\bar{G}$, the vertices $\tilde{G}$ do not share the left zero eigen-supervector $(Z_{L}|$ with the effective Liouvillian [cf. \Eq{Gx1}].
If they were still present during a process of continuous renormalization,
this would lead to divergences whenever the zero eigenvalue of $L$ would appear in the resolvent $(z-X-L)^{-1}$ (see also \Cite{Korb07}).
\key{
The vertices $\tilde{G}$ \emph{must} therefore be integrated out before any continuous RG can be formulated.
This is a characteristic feature of an RG for dissipative systems that exhibits \HL{a stationary state (a} zero \HL{eigen-supervector}).
}
The causal structure of the renormalized perturbation theory \eq{barSigma} makes clear that this elimination has been achieved:
the zero eigenvalue term of $\bar{L}$ always drops out due to the presence of $\bar{G}$'s adjacent to the propagator $(z-X-\bar{L})^{-1}$.

At this point, we can already make an observation that is important for the construction of an explicit expansion for the effective Liouvillian $L(z)$ \HL{later on} in \Sec{sec:expL}:
no continuous RG scheme, which reorganizes the perturbation series \Eq{barSigma} involving only $\bar{G}$ vertices, can ever generate terms of the form $\bar{G}...\tilde{G}$, i.e., with a $\tilde{G}$ superoperator standing on the far right.
In the notation of \Sec{sec:expL} this implies that the coefficient of the term $|Z_{R})(Z_{R}|$ in the expansion \Eq{obliv} is not changed during 
a RG flow [or any non-perturbative approximation to the series \eq{barSigma}].

What was done so far can be understood as a formal expansion around the infinite-temperature limit as a reference point. We start from the \HL{exact solution for the infinite-temperature, wide-band limit} and then reformulate \HL{the} perturbation theory for finite temperature (including the zero-temperature limit \HL{of interest}.)
This is \emph{not} to be confused with the expansion in \HL{inverse} powers of $T$: we formally expand in the Keldysh distribution functions
(contractions)  $\bar{\gamma}$. Actually, one meets this expansion already in most standard applications of the generalized master equation \HL{approach}
where one also calculates the kernel \HL{$\Sigma(i0)$, obtaining ``rates'' that include} Fermi's golden rule:
\HL{this corresponds to the correction to $\tilde{\Sigma}$ of first order in $\bar{\gamma}$.}
Here, we reformulated the exact perturbation theory around this reference point in order to perturbatively calculate
the important higher-order corrections to that result. To this end, we performed a \HL{first} natural step by exactly incorporating the
``trivial'' infinite-temperature fluctuations into a redefinition of the Liouvillian. 
This \HL{first} step is the crucial starting point for \HL{the RT-RG approach that we will develop to incorporate the ``non-trivial'' 
finite-temperature fluctuations also non-perturbatively: it} prevents a serious technical problem related to the  zero eigenvalue of 
\HL{the stationary non-equilibrium state} from appearing
(see Sec.\ref{sec:flow-Keldysh} and Sec.\ref{sec:expL}). 

\subsection{Basis of the Anderson-model \\
in Liouville Fock-space\label{sec:sym_Liv}}

\key{
To obtain explicit RG equations that can be \HL{solved numerically in an efficient way}
we need to construct a basis that exploits the advantages of
(i) selection rules for the self-energy superoperator induced by the global symmetries
(ii) the causal structure of $\bar{\Sigma}(z)$ and
(iii) \HL{the} fermion-parity superselection rule.
}
Again, these can be useful in \HL{the} perturbation theory as well\HL{, so we discuss them here.}

\subsubsection{Liouville space bra-ket formalism\label{sec:braket}}
\key{By \Eq{sig_irr} and \eq{barSigma} the QD self-energy $\Sigma$ ($\bar{\Sigma}$) is a functional of the QD Liouvillian $L$ ($\bar{L}$) and the vertex 
superoperators $G^q$ ($\bar{G}$),
which linearly act on the QD Hilbert space of many-body states.
}
To explicitly calculate these \HL{superoperators, it} is convenient to introduce a ``bra-ket'' formalism
analogous to that of standard quantum mechanics:
we write a 16-component supervector (representing an operator acting on the $4$-dimensional Hilbert space) as a rounded superket $A= |A)$
and introduce its dual supervector $(A|\bullet =\tr  \left(  {A}^\dagger \bullet\right)$ as a linear functional acting on operators.
An operator $ {A}$ is orthogonal to $ {B}$ if $(A|B)=\tr(  {A}^\dagger  {B} )=0$ (see, \HL{e.g.,} \Cite{Mukamel82}).
The dual vector may be written as the Hermitian conjugate of a superket:
\begin{align}
  (A| = |A)^{\sdagger}, ~~~~   |A) = (A|^{\sdagger},
  \label{hc}
\end{align}
where $\sdagger$ is not to be confused with the supervector corresponding to the Hermitian conjugate of the operator $A$, $A^\dagger = |A^\dagger)$.
An operator basis $|A_i)$, $i=1,\ldots,16$  of mutually orthonormal operators $\tr ( {A}^\dagger_i  {A}_j) =\delta_{i,j}$ is complete if any operator $ {B}$ can be expanded as
$
 {B}=\sum_i (A_i|B) {A}_i
$ with coefficients $(A_i|B)= \tr (A_i^\dagger B)$.
Any superoperator $S$ acting on such operators can then be expressed
in general as a sum of 256 terms:
\begin{align}
  \label{exp_liv}
  S=\sum_{i,j=1}^{16} (A_i|S|A_j)\, |A_i)(A_j|
  .
\end{align}

\subsubsection{Liouville Fock-space basis\label{sec:fock-liouville}}
\key{
To maximally reduce the number of terms in the expansion\HL{~\eq{exp_liv} of physical} superoperators,
we now exploit the close analogy \HL{between our} Liouville field superoperators in the causal representation
\HL{and} the usual fermionic field operators.
In this section we \HL{will} first construct a suitable orthonormal basis of operators (supervectors) in which any QD operator can be expanded.
In \Sec{sec:basis} we analyze their transformation properties under the symmetry transformations of the Anderson model.
In this basis we can then \HL{in \Sec{sec:expL}} easily construct superoperator expansions compatible with these symmetries.
}

We start from the key property \eq{trG} of the vertex operators in the causal representation.
In bra-ket notation the trace operation \HL{corresponds to the action}
\begin{align}
  \tr_{D} \bullet = \tr_{D} \unit \bullet = 2 ( Z_{L}|
  \label{trace}
\end{align}
 of the dual \HL{of} the normalized supervector
\begin{align}
  |Z_{L})= \tfrac{1}{2} \unit
  \label{state}
  .
\end{align}
Therefore,  by \Eq{trG}, $(Z_{L}|$ is a left zero eigen-supervector of $\bar{G}$ and by \eq{Gconj}
it follows that $|Z_{L})$ must be a right eigen-supervector of $\tilde{G}$:
\begin{align}
  (Z_{L}|\bar{G}_1 = 0
  ,
  & &
  \tilde{G}_1|Z_{L}) = 0
  \label{x0G}
  .
\end{align}
We can formally consider the state annihilated by operators $\tilde{G}$ as a ``supervacuum''.
(Note that the ``supervacuum'' state is the most symmetric dot operator 
similar to standard field theories where the vacuum state usually is the most symmetric one).
This vacuum supervector is proportional to the physical infinite temperature density
operator [cf. \Eq{ZL}] with  maximal von Neumann entropy \HL{$S=-\mathrm{Tr}_{D} \left\{ \rho\ln(\rho) \right\}$}.
It is then also natural to construct the corresponding right zero eigen-supervector of $\bar{G}_1$ whose dual is the left zero eigen-supervector 
of $\tilde{G}_1$:
\begin{align}
  \bar{G}_1|Z_{R}) = 0
  ,
  &&
  (Z_{R}|\tilde{G}_1 = 0
  .
  \label{Gx1}
\end{align}
Since $\bar{G}_1$ is a non-Hermitian superoperator
these eigen-supervectors are not simply related by Hermitian conjugation in Liouville space, $|Z_{R}) \neq (Z_{L}|^\sdagger$ [cf. \Eq{hc}].

The operator $Z_R$ will be found to play a key role throughout this work.
In \App{sec:grassmann} we discuss its many interesting properties, 
\HL{most prominently, its relation to the fermion-parity, its similarity to Grassmann numbers}, and its relation to the spin- and charge-rotations of \Sec{sec:basis}.
To construct $|Z_{R})$ we now exploit the close analogy to \HL{the usual} field operators:
$\bar{G}_{1}$ is a creation operator in \emph{Liouville Fock-space}, since its Hermitian superconjugate $\tilde{G}_{\bar{1}}$ annihilates the supervacuum state $|Z_{L})$ by \Eq{x0G}.
The state annihilated by $\bar{G}_{1}$ is therefore simply the maximally occupied state in the QD Liouville Fock-space, starting from the 
vacuum:\footnote{This relies on the anticommutation relations \eq{commutG}
of the bare vertices, which break down under renormalization\HL{:
$|Z_R)$ given by \Eq{ZR-constr} is thus \emph{not} an eigen-supervector of renormalized vertices.}}
\begin{align}
  |Z_{R})
  =
  \prod_\sigma \left( \prod_{\eta} \bar{G}_{\eta\sigma} \right)
  |Z_{L})
  \label{ZR-constr}
  .
\end{align}
As for the \HL{usual} fermionic field operators, \HL{the} action of creation operators preserves the normalization: this follows directly \HL{from} \Eq{hc} and the 
anticommutation relations \eq{commutG}:
\begin{align}
  (Z_{R}|Z_{R})
  &=
  (Z_{L}| \prod_\sigma \left( \prod_{\eta} \tilde{G}_{\bar{\eta}\sigma} \bar{G}_{\eta\sigma} \right)
  |Z_{L})
  \nonumber
  \\
  &
  =
  (Z_{L}|Z_{L}) = 1
  \label{ZR-normalization}
  .
\end{align}
Note that by \Eq{commutG}, reordering the $\bar{G}$'s in the definition of $Z_{R}$
only amounts to an unimportant redefinition of the overall sign.
Using \Eq{kr_d}, we can write for the explicit action on any operator (denoted by $\bullet$) with even fermion-parity
\begin{align}
  \bar{G}_{+\sigma} \bar{G}_{-\sigma} \bullet
  =
 \tfrac{1}{2} \left(
[n_\sigma,\bullet]_{+} -1
  +  d_\sigma^\dagger \bullet d_\sigma
  -  d_\sigma         \bullet d_\sigma^\dagger
  \right)
  .
\label{chi-creation}
\end{align}
Inserting this into \Eq{ZR-constr}, we obtain
\begin{align}
  |Z_{R}) = \frac{1}{2} \prod_\sigma (2n_{\sigma}- \unit)
  .
  \label{ZR-product}
\end{align}
Since $(Z_{L}|$ is a left eigenvector of $\bar{G}$ [cf. \Eq{x0G}], we immediately see that the two zero eigen-supervectors are orthonormal, $(Z_{L}|Z_{R})=0$.
Therefore,  it is natural to include $|Z_{R})$ into the orthonormal Liouville space basis.
By successively acting on the supervacuum state $|Z_{L})$, we can generate more normalized, orthogonal operators.
There are in total eight bosonic operators:
\begin{align}
 \label{basis-boson}
 |Z_{L})       & = \tfrac{1}{2} \unit, \\
 |\chi_\sigma) & = \bar{G}_{+\sigma}\bar{G}_{-\sigma}|Z_{L}),\\
 |T_\eta)     & = \eta\bar{G}_{\eta\uparrow}\bar{G}_{\eta\downarrow}|Z_{L}),\\
 |S_\sigma)    & = \bar{G}_{+\sigma}\bar{G}_{-{\bar{\sigma}}}|Z_{L}), \\ \label{zr}
 |Z_{R})      & = \bar{G}_{{+}\uparrow}\bar{G}_{{-}\uparrow}
                 \bar{G}_{{+}\downarrow}\bar{G}_{{-}\downarrow}|Z_{L})
 ,
\end{align}
and there are eight fermionic operators:
\begin{equation}
  \begin{aligned}
    |\alpha^+_{+,\sigma}) &=  \bar{G}_{+\sigma}|Z_{L}), 
    \\
    |\alpha^+_{-,\sigma}) &= \sigma\bar{G}_{-\bar{\sigma}}|Z_{L}),
    \\
    |\alpha^-_{+,\sigma}) & = \bar{G}_{+\sigma}\bar{G}_{+\bar{\sigma}}\bar{G}_{-\bar{\sigma}}|Z_{L}),
    \\
    |\alpha^-_{-,\sigma}) &= \sigma\bar{G}_{-\bar{\sigma}}\bar{G}_{-\sigma}\bar{G}_{+\sigma}|Z_{L})
    .
  \end{aligned}
  \label{basis-fermion}
\end{equation}
The labeling of these basis supervectors  is motivated by their explicit expressions in terms of the field operators $d_\sigma,d_\sigma^\dagger$ (see \Eq{X0def}-\eq{fermion-basis}) and their behavior under symmetry transformations,
which will be discussed in the next section.
We see that the index $\nu$ of the operators $|\alpha^\nu_{\eta\sigma})$ only has a meaning in Liouville space:
it distinguishes $\nu=+$ states with one excess excitation relative to the supervacuum $|Z_{L})$ (superparticles),
from $\nu=-$ states with one deficit particle with respect to the maximally occupied superstate $|Z_{R})$ (superholes). 
By construction, these 16 operators form a complete orthonormal basis of the QD Liouville space.
This basis includes only one operator that has non-zero trace, namely, $|Z_{L})= \tfrac{1}{2}\unit$.
Since all other supervectors are orthogonal to $|Z_{L})$, their corresponding operators are traceless according to \Eq{trace}.

We emphasize that the choice of basis supervectors $|Z_{L})$ and $|Z_{R})$ relies on \HL{two} general physical properties of the problem, the probability conservation (cf. \Eq{trSigma} and \Eq{x0G}) \HL{and the fermion-parity (cf. \App{sec:grassmann}).}
The choice of the signs of the remaining operators, however, is motivated by considering the symmetry transformations of the Anderson model. These will be discussed in the next section.
\HL{The related approach of Prosen was also introduced by an explicit construction of the Liouville-Fock space~\cite{Prosen08} (see \App{sec:Pros}).}
%We note that similar Liouville-Fock-space and superoperators were introduced by %Prosen~\cite{Prosen08}, see \App{sec:Pros}.

Before we proceed, we emphasize the necessity of working with a complete basis for the QD Liouville space, which includes operators that are \emph{non-diagonal} in both spin and / or charge quantum numbers.
In perturbation theory, one can disregard all  matrix elements of the self-energy $\Sigma(z)$ involving the latter operators (cf. \Sec{sec:perturb}),
due to the conservation laws \eq{selrule}.
\HL{However, the states propagating in the inner part of a diagram contributing to $\Sigma(z)$ (i.e., a virtual intermediate state) are less restricted by the 
conservation laws, requiring the matrix elements of the vertices and the QD Liouvillian $L$ between \emph{all} the off-diagonal operators.}
\footnote{We emphasize that charge off-diagonal dot operators \emph{must} be taken into account since in a virtual intermediate state of an open system, where, e.g., $G$ has acted on a even fermion parity QD density operator,
there is a corresponding fermion operator $J$ acting on the reservoir. Although the latter is no longer explicitly present in the reduced density 
operator description (\HL{the} reservoirs are integrated out), there is no change of the fermion parity of the total system described by an elementary process 
in $\Sigma$, in full agreement with the discussion of \Eq{lvp}.}
In the RT-RG that we set up below such matrix elements involving odd-fermion-parity operators cannot be avoided for $\Sigma(z)$ as well since one needs to describe the \emph{renormalization of all virtual intermediate states}, both fermionic as well as bosonic ones.

\subsubsection{Irreducible transformation\\
 under symmetry operations\label{sec:basis}}
For the total system of QD and reservoirs  both the charge and spin components along the magnetic field are conserved [cf. Eqs. \eq{chgcons} and \eq{spincons}].
The particle number conservation on the dot is conveniently expressed using the charge-polarization operator 
\begin{align}
  T_z = \tfrac{1}{2}(n-1)
  ,
\end{align}
which measures the difference of the occupation probabilities of the empty and \HL{doubly} occupied QD states $|0\rangle$ and $|2\rangle$.
It follows from \Eq{chgcons} and \eq{spincons} that the dot superoperators for charge polarization (cf. \Eq{Ln}) and spin,
\begin{align}
  L^{T_z} &= [T_z, \bullet]_{-} = \tfrac{1}{2}L^{n}
  ,
  \\
  L^{S_z} &= [S_z, \bullet]_{-}
  ,
  \label{LSz}
\end{align}
respectively, commute with each other as well as with the Liouvillian $L$ and the self-energy $\Sigma(z)$.
Therefore, the effective Liouvillian $L(z)=L+\Sigma(z)$,
which determines the time-evolution of the reduced density operator, conserves these superobservables (see \App{sec:selrule} for a derivation):
\begin{align}
  \left[ L(z) ,  L^{T_z} \right]_{-} = 0
  ,
  \\
  \left[ L(z),  L^{S_z} \right]_{-} = 0
  .
  \label{selrule}
\end{align}
Thus $L(z)$ can be simultaneously block diagonalized with the superoperators for charge- and spin polarization and they have common \HL{eigen-supervectors}.
In the basis of \Sec{sec:basis}, the initial dot Hamiltonian operator \Eq{dot_ham} has the form
\begin{align}
  H = &
  (2\epsilon + \frac{U}{2}) |Z_{L})
  + \frac{U}{2} |Z_{R})
  \nonumber
  \\
  &
  + 2 \left( \epsilon +\frac{U}{2}\right) |T_z)
  + B |S_z)
  \label{Hdot}
  .
\end{align}
In two \HL{special} cases the total system has a higher symmetry:
for  $\epsilon = -U/2$ and $\mu_L=\mu_R$  it exhibits \emph{full} charge-rotation symmetry,
\HL{whereas for} $B=0$ it has \HL{\emph{full}} spin-rotation symmetry.
These full symmetry groups are obtained by adding \HL{the operators} $S_x$ and $S_y$ to $S_z$  (cf. \Eq{SU2spin}), \HL{giving} the SU(2) spin-algebra,
 and by adding $T_x$ and $T_y$ to $T_z$ (cf. \Eq{SU2p-hs}), \HL{giving} another SU(2) algebra generating ``charge-rotations'' (see below).
The construction of \emph{superoperators} that transform in the simplest possible way under these symmetry operations is greatly simplified by first constructing basis operators that can be classified with respect to the corresponding irreducible representations.
Importantly, the Liouville Fock basis \eq{basis-boson}-\eq{basis-fermion} that we constructed using the causal field superoperators is already very close to a symmetry-adapted basis and we merely \HL{need to} complete the classification. 
We first group these operators according to their even or odd fermion parity, and refer to these \HL{as} \emph{bosonic}, \HL{respectively,} \emph{fermionic operators}.
We subsequently classify them according to the transformation behavior under the two SU(2) rotation groups
as an irreducible tensor \HL{operator} (ITO).
This allows us to identify supervectors \HL{corresponding to Hilbert-space operators that are diagonal with respect to the QD charge and / or the QD spin.}
In perturbation theory, one only needs the self-energies connecting such diagonal components of the density matrix. In the RG \HL{that} we set up below this is no longer true\HL{. Still,} it is important to single out this block of matrix elements 
of $\Sigma(z)$, $\bar{\Sigma}(z)$ and $L(z)$.

\HL{One half of the Liouville space is then spanned by eight \emph{bosonic}} operators with integer charge- and spin-ITO ranks:
\begin{itemize}
\item 
  The zero-eigenvectors of the vertex superoperators are scalars
  (rank-0 spin and charge ITOs) with respect to both spin- and charge-rotations
  since the zero-eigenvalue equations \Eq{x0G}-\eq{Gx1} are invariant under these transformations.
  They are therefore charge- and spin-diagonal operators.
  This is also clear from their explicit form
  in terms of the Casimir operators of the spin- ($S^2=\sum_i S_i^2$) and charge- rotation ($T^2=\sum_i T_i^2$) SU(2) Lie algebras:
  \begin{align}
    ~~~~~
    |Z_{L})&=\tfrac{1}{2} \unit
          & = \tfrac{4}{3} (T^2+S^2)
    \label{X0def}
    ,
    \\
    |Z_{R})& =2 n_{\uparrow} n_{\downarrow}- n + \tfrac{1}{2}\unit
          & =  \tfrac{2}{3}(T^2-S^2)
    \label{X1def}
    .
  \end{align}
\item
  The generators of rotations in spin space \HL{are}
  \begin{align}
    |S_0)&
    =\tfrac{1}{\sqrt{2}}\sum_{\sigma} \sigma |\chi_\sigma)
    =\tfrac{1}{\sqrt{2}}\sum_{\sigma} \sigma n_\sigma
    ,
    \\
    |S_\sigma)&= {d}^{\dagger}_{\sigma} {d}_{\hl{\bar{\sigma}}},
    ~~~
    \sigma=\uparrow,\downarrow\hl{=\pm}
    .
    \label{SU2}
  \end{align}
  \HL{The operators} $-|S_+), |S_0), |S_-)$ transform as the components of a rank-1 spin-ITO (vector),
  i.e., simpler than the more familiar Cartesian components of the spin-operator,
  \begin{align}
    ~~~~~
    |S_{x,y})= \tfrac{1}{2(i)} \left(|S_+)\pm |S_-)\right)
    , ~~
    |S_z)=\tfrac 1{\sqrt{2}} |S_0)
    \label{SU2spin}
    ,
  \end{align}
  which satisfy $[|S_i),|S_j)]_{-}=i\epsilon_{ijk} |S_k)$.
  They transform as a rank-0 ITO (scalar) with respect to charge-rotations
  and are therefore charge-diagonal.
\item
  The generators of rotations in charge space \HL{are}
  \begin{align}
    |T_0) &
    = \tfrac{1}{\sqrt{2}}\sum_\sigma |\chi_\sigma)
    =\tfrac{1}{\sqrt{2}}( n - 1)
    \label{T0def}
    ,
    \\
    |T_+)&=  {d}^{\dagger}_{\uparrow}  {d}^{\dagger}_{\downarrow}
    ,
    \\
    |T_-)&=  {d}_{\downarrow}  {d}_{\uparrow}
    \label{SU2p-h}
    .
  \end{align}
  \HL{The operators} $-|T_+), |T_0), |T_-)$ transform as a rank-1 ITO (vector) under rotations in charge space,
  and as  a rank 0  ITO (scalar) with respect to spin-rotations.
  Therefore, they are all spin-diagonal.
  They are  more convenient than the Cartesian components
  \begin{align}
    ~~~~~~
    |T_{x,y})= \tfrac{1}{2(i)} \left(|T_+)\pm |T_-)\right)
    , ~~~
    |T_{z}) =\tfrac{1}{\sqrt{2}}|T_0)
  \label{SU2p-hs}
  .
  \end{align}
  satisfying the SU(2) algebra $[|T_i),|T_j)]_{-}=i\epsilon_{ijk} |T_k)$.
  In contrast to the spin-operators,
  here the indices $x,y,z$ are not related to the axes in the real space,
  but merely label the components of the SU(2) generators.
\end{itemize}

In the bosonic sector, we can thus use either
\begin{align}
  |\chi_{\sigma})= n_{\sigma}-\tfrac{1}{2}\unit = \tfrac{1}{\sqrt{2}} \left[ |T_0) + \sigma |S_0) \right]
  \label{chi-basis}
  ,
\end{align}
or $|S_0)$ and $|T_0)$ as basis vectors.
Only the latter two are adapted to charge- and spin- rotation symmetry,
but the former two allow for greater notational simplicity.
Both will be used.

The other half of the QD Liouville space is spanned by \HL{eight more, \emph{fermionic}} operators with half-integer charge- and spin-ITO ranks:
\begin{itemize}
\item
  The fermionic operators, denoted by $\alpha_{\eta,\sigma}^\nu$,
  act in three subspaces of dimension 2 labeled by  $\sigma=\pm$ (spin), $\eta=\pm$ (particle-hole)
  and, additionally, $\nu=\pm$.
  These basis operators are ITOs of rank $\tfrac{1}{2}$ with respect to both charge ($\eta$) and spin-rotations ($\sigma$).
  For $\nu=+$ these are the rank $\tfrac{1}{2}$ spin-ITOs constructed from creation and annihilation operators:
  \begin{align}
    ~~~~
    |\alpha_{+,\sigma}^{+}) & =\tfrac{1}{\sqrt{2}} {d}^{\dagger}_\sigma,
    &
    |\alpha_{-,\sigma}^{+})  &= \tfrac{1}{\sqrt{2}}  \sigma{d}_{\bar{\sigma}}
    ,
    \\
    |\alpha_{+,\sigma}^{-})  &= \hl{{2}}|Z_{R} \alpha_{+\sigma}^{+}),
    &
    |\alpha_{-,\sigma}^{-})  &= \hl{{2}}|Z_{R} \alpha_{-\sigma}^{+})
    .
    \label{fermion-basis}
  \end{align}
  All these operators are both charge- and spin off-diagonal.
 \HL{The above explicit form emphasizes} that there is an additional set of fermionic operators in the charge off-diagonal subspace \HL{that} is 
linearly independent of the standard field operators $d_\sigma$ and $d^\dagger_\sigma$,
  see the discussion of the index $\nu$ \HL{after} \Eq{basis-fermion}.
  Noting the property $\left(2Z_{R}\right)^2=\unit$ we find the following explicit relation between these two sets: \HL{with $\bar{\nu}=-\nu$,}
  \begin{align}
  |\alpha_{\eta,\sigma}^\nu)=2 Z_{R}|\alpha_{\eta,\sigma}^{\bar{\nu}})
  .
  \end{align}
\end{itemize}

The above simple transformation behavior of the basis supervectors motivates all the relative signs \HL{that} we anticipated in writing \Eq{basis-boson}-\eq{basis-fermion}. The basis is therefore completely fixed up to irrelevant phases by general physical properties  (probability conservation, symmetries) together with the supervector normalization.
\Tab{tab:itos} summarizes the transformation properties.

\begin{table}
  \begin{tabular}{ | c || c | c |c  c |}
    \hline
    \textbf{Operator}         & \textbf{fermion} & \textbf{$S$-ITO}  & \textbf{$T$-ITO} &
    \\
                              & \textbf{-parity} & (rank,index)      & (rank, index)    &
    \\ \hline \hline
    $|Z_{i})$                  &     +           & (0,0)             & (0,0)      & $i=L,R$
    \\  \hline
    $|S_{m})$                  &     +           & (1,$m$)           & (0,0)      & $m=0,\pm1$
    \\ \hline
    $|T_{m})$                  &     +           & (0,0)             & (1,$m$)    & $m=0,\pm1$
    \\ \hline
    $|\alpha^\nu_{\eta,\sigma})$ &     -   & ($\tfrac{1}{2}$,$\tfrac{\sigma}{2}$)
                                         & ($\tfrac{1}{2}$,$\tfrac{\eta}{2}$)
                                         & $\eta, \sigma = \pm1$
    \\
    \hline
  \end{tabular}
  \caption{
    Fermion-parity and irreducible transformation behavior of basis operators under spin- and charge-rotations.
        Schematically denoting these operators by $|s,m_s; t,m_t)$,
    these transform as (i) spin-irreducible tensor operators ($S$-ITOs) with rank $s$ and index $m_s$,
    $L^{S_z} |s,m_s; t,m_t) = m_s |s,m_s;t,m_t)$
    and 
    $L^{S_\pm} |s,m_s; t,m_t) =  \sqrt{s(s+1)-m_s(m_s\pm1)} |s,m_s\pm 1;t,m_t)$.
    and (ii) as charge-ITOs ($T$-ITOs) with rank $t$ and \hl{index} $m_t$,
    $L^{T_z} |s,m_s; t,m_t) = m_s |s,m_s;t,m_t)$
    and 
    $L^{T_\pm} |s,m_s; t,m_t) =  \sqrt{t(t+1)-m_t(m_t\pm1)} |s,m_s;t,m_t\pm 1)$.
    All zero-index $S$-ITOs ($T$-ITOs) correspond to {operators} acting on Hilbert space that are diagonal in spin (charge).
      }
  \label{tab:itos}
\end{table}

Finally, we note the transformation behavior under \HL{Hilbert-space} Hermitian conjugation\HL{, corresponding to an antilinear superoperator }$K$
(see \Eq{hermcond}). For all diagonal basis operators $D=Z_{L},Z_{R}, \chi_\sigma,S_0,T_0$
\begin{align}
  K |D)=|D)
  \label{basisKb}
  ,
\end{align}
whereas for the off-diagonal operators
\begin{align}
  \begin{aligned}
    K|S_\pm) & =|S_\mp),
    \\
    K|T_\pm) & =|T_\mp)
    ,
    \\
    K|\alpha^\nu_{\eta,\sigma}) & =-\nu\eta\sigma|\alpha^\nu_{\bar{\eta},\bar{\sigma}})
    .
  \end{aligned}
  \label{basisKoff}
\end{align}

\subsubsection{Expansion of the vertices\label{sec:expG}}

The calculation of the bra-ket representation of the vertex \HL{superoperators} in the basis \Eq{basis-boson}-\eq{basis-fermion}
reduces entirely to the calculation of
\begin{align}
  \label{g-b+}
  \bar{G}_{+ \sigma} =&
  |\alpha^+_{+,\sigma})(Z_{L}|+\sigma|Z_{R})(\alpha^-_{-,\bar{\sigma}}|\\ \nonumber
  &+\sigma|T_+)(\alpha^+_{+,\bar{\sigma}}| +|\alpha^-_{-,\sigma})(T_-|+|\alpha^-_{+,\sigma})(\chi_{\bar{\sigma}}|\\ \nonumber
  & -\sigma|\chi_{\sigma})(\alpha^+_{-,\bar{\sigma}}| +\sigma|S_{\sigma})(\alpha^+_{-,\sigma}| -|\alpha^-_{+,\bar{\sigma}})(S_{\bar{\sigma}}|
  .
\end{align}
This is easily performed using the second quantization technique that we introduced in Liouville Fock space, i.e., 
using algebra rather than the explicit matrix representations of \Sec{sec:basis}.
All other vertices follow from general relations:
first, using the transformation under Hermitian conjugation [see \Eq{hermcond}],
\begin{align}
  \bar{G}_{\eta \sigma}=(-1)^{L^n+1}K\bar{G}_{\bar{\eta} {\sigma}} K
  \label{GbarK}
  ,
\end{align}
the result for opposite charge index $\eta=-$ follows
\HL{from} Eqs. \eq{basisKb} and \eq{basisKoff}:
\begin{align}
  \bar{G}_{- \sigma} = &-\sigma|\alpha^+_{-,\bar{\sigma}})(Z_{L}| -|Z_{R})(\alpha^-_{+,\sigma}|
  \label{g-b-} \\ 
  &-\sigma|\alpha^-_{+,\bar{\sigma}})(T_+|
  -|T_-)(\alpha^+_{-,\sigma}|+\sigma|\alpha^-_{-,\bar{\sigma}})(\chi_{\bar{\sigma}}|
  \nonumber \\ 
  &-|\chi_{\sigma})(\alpha^+_{+,\sigma}|-|S_{\bar{\sigma}})(\alpha^+_{+,\bar{\sigma}}| 
  +\sigma|\alpha^-_{-,\sigma})(S_{\sigma}|
  \nonumber
  .
\end{align}
The vertex superoperators $\tilde{G}_{\eta\sigma}$ \HL{are} obtained from $\bar{G}_{\bar{\eta}\sigma}$ by \HL{the} Hermitian conjugation \HL{relation \eq{Gconj} between the field superoperators.}
In the Liouville bra-ket formalism, this simply means that we can interchange bra and ket vectors in the expansions \eq{g-b+} and \eq{g-b-} to obtain $\tilde{G}_{\mp\sigma}$.

We point out that under the two-loop RG flow to be discussed in \Sec{sec:rg},
the structure of the vertex operators is changed
and the above relations cease to hold
[$\bar{G}$ will be modified whereas $\tilde{G}$ is not changed, implying that \Eq{Gconj} breaks down].
However, in \Sec{sec:}, we show how such vertex corrections can be incorporated effectively into the flow of the effective Liouvillian \emph{only},
 allowing us to work with the ``bare'' vertices \HL{having} the nice properties discussed above.
\HL{Two properties} of the bare vertices that \HL{remain} valid under the RG-flow \HL{are}
\begin{align}
  (\alpha_{\eta\sigma}^{+}| \bar{G}_{1} & \propto (Z_L| \text{ or 0} 
  \label{alphaG0}
  ,
  \\
  \HL{\bar{G}_1} | \alpha_{\eta\sigma}^{-}) & \propto |Z_R) \text{ or 0}
  \label{Galpha0}
  ,
\end{align}
Equation \eq{alphaG0} follows since $|\alpha^{+}_{\eta\sigma})$ is obtained from the vacuum $|Z_L)$ by the action of a single creation superoperator (cf. \Eq{basis-fermion}). Using \Eq{Gconj}
\begin{align}
  \label{alpha_zl}
  (\alpha_{\eta,\sigma}^+| = \sigma^{(1-\eta)/2} (Z_L|\tilde{G}_{\bar{\eta},(\eta\sigma)}
  .
\end{align}
When inserted in the left-hand side of \Eq{alphaG0},  this $\tilde{G}$ can be anticommuted past the $\bar{G}$  using \Eq{commutG} and the supervacuum property \eq{x0G}, $(Z_L|\bar{G}=0$.
Analogously, \Eq{Galpha0} follows by noting that \Eq{basis-fermion} for the $|\alpha_{\eta,\sigma}^-)$ can be rewritten as a single destruction 
superoperator acting on the maximally occupied state in Liouville-Fock space:
\begin{align}
  |\alpha_{\eta,\sigma}^-)=\bar{\sigma}^{(1-\eta)/2}  \tilde{G}_{\eta,(\eta\sigma)}|Z_R)
  .
\end{align}
Commuting the \HL{$\bar{G}$ to the right} and using $\bar{G}|Z_R)=0$ [\Eq{Gx1}] on the 
right of each term we obtain \Eq{Galpha0}.

\subsubsection{Expansion of the effective Liouvillian\label{sec:expL}}

\paragraph{Causal structure\label{sec:causalstruct}}
By the general properties \eq{x0G} and \eq{Gx1}, the vertices must have an expansion of the form
(confirmed by the explicit results \Eq{g-b+}-\eq{g-b-}):
\begin{align}
  \bar{G}=..|Z_{R})(\bullet|+..|\bullet)(Z_{L}|+...
  \label{barGexp}
  ,
  \\
  \tilde{G}=..|\bullet)(Z_{R}|+ ..|Z_{L})(\bullet| +...
  \label{tildeGexp}
  ,
\end{align}
where the remaining terms involve neither $Z_{L}$ \HL{nor} $Z_{R}$.
Therefore, the terms in the expansion of the effective Liouvillian involving these vectors
are strongly restricted.
Combined with the general causal structure of the perturbative series, i.e., the way $\bar{G}$ and $\tilde{G}$ can appear, this imposes further 
constraints (cf. \Sec{sec:perturb}):
\begin{itemize}
\item
  Terms of the form $|Z_{L})(\bullet|$ are prohibited by probability conservation,
  since otherwise the trace condition \eq{trSigma} would be violated:
  \begin{alignat}{3}
    (Z_{L}|\bar{L}
    & =  (Z_{L}|L(z)            &=  (Z_{L}| \Sigma(z)       &   
    \\
    & =  (Z_{L}| \tilde{\Sigma} & =  (Z_{L}|\bar{\Sigma}(z) & =  0
    \label{ZLeigen}
    .
  \end{alignat}
  This is guaranteed by the causal structure, which requires
  that the leftmost vertex is always of the type $\bar{G}$,
  with expansion \Eq{barGexp}.

\item
  Terms of form $|\bullet)(Z_{R}|$ can only appear due to the diagrams collected in
  $\tilde{\Sigma}=-i\tfrac{1}{2}\Gamma_1 \bar{G}_{1} \tilde{G}_{\bar{1}}$.
  Expanding \Eq{tildeSigma} in the basis \eq{basis-boson}-\eq{basis-fermion}, one finds that
  only the term $|Z_{R})(Z_{R}|$ with coefficient $-i 4 \Gamma$
  can appear [see also \Eq{sin}].
  Importantly, this implies that $|Z_R)$ is a right eigenvector of both $\bar{L}$
  as well as the \emph{exact} effective Liouvillian $L(z)$ and the kernel $\Sigma(z)$   
  [see \HL{the} discussion of \Eq{obliv}]:
  \begin{alignat}{2}
    \bar{L} |Z_{R})
    & =L(z) |Z_{R})           &=\Sigma(z)|Z_{R})
    ,
    \\
    & = \tilde{\Sigma}|Z_{R}) &= -4i\Gamma |Z_{R})
    \label{ZReigen}
    .
  \end{alignat}
  In contrast, 
  \begin{align}
    \bar{\Sigma}(z)|Z_{R}) =0
    \label{ZReigenbarSigma}
    .
  \end{align}
  Note that this eigenvector and eigenvalue are independent of the QD frequency $z$.
\item 
  The term $|Z_{R})(Z_{L}|$ is not forbidden by general considerations.
  However, such terms always drop out
  in the calculation of 
  the transport current, which interest us here (see \Sec{sec:current}).
  This happens because in all required expressions, the renormalized Liouvillian $\bar{L}$,
  parametrized as \eq{obliv}, is evaluated between two $\bar{G}$ vertices cf. \Eq{barSigma}.
  Therefore,  by \Eq{barGexp} the term $|Z_{R})(Z_{L}|$ with coefficient $\zeta$ always drops out.
  We emphasize, however, that $\zeta$ does enter into the stationary state (cf. \Eq{rho}) and other physical quantities, such as the 
average dot energy, and may therefore be important for, e.g., thermal transport problems.
\item
  Terms of the form $|\alpha_{\eta,\sigma}^+)(\bullet|$ and $|\bullet)(\alpha_{\eta,\sigma}^-|$
  can appear \emph{only} in the bare Liouvillian $L=[H,\bullet]_{-}$
  or the infinite-temperature kernel 
  $\tilde{\Sigma}=-i\tfrac{1}{2}\Gamma_1 \bar{G}_{1} \tilde{G}_{\bar{1}}$,
  but \emph{not} in the non-trivial kernel $\bar{\Sigma}(z)$:
  \begin{align}
    (\alpha_{\eta,\sigma}^+|\bar{\Sigma}(z) &= 0
    \label{alpha+0}
    ,
    \\
    \bar{\Sigma}(z)|\alpha_{\eta,\sigma}^-) &= 0
    \label{alpha-0}
    .
  \end{align}
  Both relations follow from the fact that $\bar{\Sigma}(z)$ contains \emph{only} vertices of the type $\bar{G}$ in \HL{the} expansion \eq{barSigma}.
  \Eq{alpha+0} follows from \Eq{alphaG0} applied to \Eq{barSigma} and \HL{then} using $(Z_L|\bar{L}=0$ and $(Z_L|\bar{G}=0$, the vacuum property \eq{x0G}.
  Analogously, \Eq{alpha-0} follows from \Eq{Galpha0} using our general result~\eq{ZReigen}, $\bar{L}|Z_R)=-4i\Gamma |Z_R)$, and $\bar{G}|Z_R)=0$, \Eq{Gx1}.
  \Eq{alpha+0}-\eq{alpha-0} allow us to make general predictions about the excitation spectrum of the \emph{exact} dot Liouvillian, $L(z)$
 [see \Sec{sec:fermion-pt} and \sec{sec:fermion}].
\end{itemize}

\paragraph{Spin- and charge-rotation symmetry
\label{sec:spin}}

We now first expand the infinite-temperature self-energy $\tilde{\Sigma}$ and the renormalized Liouvillian $\bar{L}$ in the basis~\eq{basis-boson}-\eq{basis-fermion}.
Substituting the above bra-ket expansions of the superoperators $\bar{G}_1$  and $\tilde{G}_{\bar{1}}$ 
into \Eq{tildeSigma} we get:
\begin{align} 
  \tilde{\Sigma}
   =
   &  -i\Gamma \Big{[}
 4|Z_{R})(Z_{R}|+2\sum\limits_{\sigma=\pm}|\chi_\sigma)(\chi_\sigma| \nonumber\\
  &+2\sum\limits_{\sigma=\pm}|T_\sigma)(T_\sigma|+2\sum_{\sigma=\pm} |S_{\sigma})(S_{\sigma}| \nonumber\\
  &+\sum_{\sigma=\pm}\sum_{\eta=\pm}\sum_{\nu=\pm}\left(2-\nu\right) |\alpha_{\eta,\sigma}^{\nu})(\alpha_{\eta,\sigma}^{\nu}|
  \Big{]}
  \label{sin}
\end{align}
Clearly, $\tilde{\Sigma}$ is explicitly anti-Hermitian in the superoperator sense (cf. \Eq{tildeSigma-hc}).
Combining this with the bare dot Liouvillian, obtained by expanding the commutator $L=[H,\bullet]_{-}$ of \eq{Hdot} we obtain
\begin{align}
  & 
  \bar{L}_{\Lambda} |_{\Lambda=\infty} 
  := \bar{L}
  = L + \tilde{\Sigma}
  =
  -i 4\Gamma |Z_{R})(Z_{R}| -2i\Gamma {\chi}_0
  \nonumber
  \\
  &
  + \sum_\sigma \Big[( \sigma B-2i\Gamma)|S_\sigma)(S_\sigma|
  +( \sigma(U+2\epsilon) -2i\Gamma )|T_\sigma)(T_\sigma|
  \nonumber \\
  &
  +\sum\limits_{\eta,\nu}\
  \left(
     \eta\left(\epsilon+\frac{U}{2}\right)+\sigma\frac{B}{2} 
    -i(2-\nu)\Gamma
  \right)
  |\alpha_{\eta,\sigma}^\nu)(\alpha_{\eta,\sigma}^\nu|
  \nonumber\\
  & +\frac{U}{2} \sum\limits_{\eta,\nu}  |\alpha_{\eta,\sigma}^\nu)(\alpha_{\eta,\sigma}^{\bar{\nu}}|
\Big]
,
  \label{initial}
\end{align}
where for \HL{later} reference we introduced the notation
$ \bar{L}_{\Lambda}|_{\Lambda=\infty} := \bar{L}
$
of \Sec{sec:rg}.
All non-zero eigenvalues of $\bar{L}$ have negative imaginary parts, thereby automatically regularizing all resolvents that can 
appear $\bar{G}({z-\bar{L}-X})^{-1}\bar{G}$ in the perturbation theory for $\bar{\Sigma}(z)$ for $z \rightarrow i0$.
This can be seen explicitly
since all terms are already in diagonal superoperator form, with the exception of the
odd-fermion terms (the last two lines in \Eq{initial})
whose eigenvalues are given below (set $\Delta F_{\eta,\sigma}^{-,+}=0$ in \Eq{ferm-spec}).
Note that the right zero eigenvector
 of $\bar{L}_{\Lambda=\infty}$ is  $\tfrac{1}{2}|Z_{L})$
-- it is the only basis vector missing in \Eq{initial} -- 
in agreement with the result \eq{ZL} in \Sec{sec:discrete}.
We further note that the infinite temperature kernel $\tilde{\Sigma}$ \eq{sin} contributes terms to \Eq{initial} that are diagonal in the index $\nu$,
whereas contributions off-diagonal in $\nu$ are produced in \Eq{initial} by the Coulomb interaction included in the bare dot Liouvillian $L$.
We will show in \Sec{sec:fermion} that the continuous RT-RG produces only  contributions \HL{to} the effective Liouvillian $L(z)$
\HL{that are off-diagonal in $\nu$,
which has} important consequences.

We can now write down the exact \emph{form} of the QD effective Liouvillian $L(z)$
taking into account all general restrictions that we have derived above.
In the most general case that we consider only spin- and charge-rotation symmetry
 about the $z$-axis: this implies that the effective Liouvillian must be a sum of superoperators that
(i) transform as an irreducible tensor of any rank but with index zero with respect to both the charge- and spin-rotation group
(i.e., by pairing only bras and kets of basis supervectors with the same charge and spin indices)
and
(ii) preserve the fermion-parity (i.e., by avoiding combinations of fermion and boson kets and bras).
Using \Tab{tab:itos}, which lists the transformation properties of the basis supervectors \eq{basis-boson}-\eq{basis-fermion},
we can readily construct the  most general form of superoperators of this kind, \HL{which} are furthermore compatible with the causal structure of the perturbative series \eq{sig_irr}.
In \Tab{tab:itso} we have classified all these superoperators according to their irreducible transformation properties under the \emph{full} symmetry group of both spin- and charge-rotations.
This makes it easy to impose further restrictions on the expansion coefficients in \HL{the special} cases of higher symmetry ($B=0$ and / or $\epsilon=-U/2,\mu_L=\mu_R$), see \Sec{sec:high-sym}.
The most general form of the \emph{exact} QD effective Liouvillian $L (z) = L + \Sigma (z)=\bar{L}+\bar{\Sigma}(z)$
then reads:
\begin{align}
  \label{obliv}
  &i L(z)
  =
  \\
  &{4\Gamma}|Z_{R})(Z_{R}| +\zeta |Z_{R})(Z_{L}|
  +|Z_{R})(\overrightarrow{\phi}|
  +|\overrightarrow{\psi})(Z_{L}|
  \nonumber
  \\
  &
  + \xi
  +\sum_\sigma
  \left[
    M_\sigma |T_\sigma)(T_\sigma|
    +
    E_\sigma |S_\sigma)(S_\sigma|
    + \sum\limits_\eta F_{\eta,\sigma} \right]
  \nonumber
  ,
\end{align}
where $\Gamma = \frac{1}{2} \sum_r \Gamma_r$ and $F_{\eta,\sigma}$ are superoperators acting in the two-dimensional $\nu$ space spanned by
$|\alpha^\nu_{\eta,\sigma})$ (see below).
This is a central result of the paper,
and before discussing the \HL{occurring} coefficients in detail, we point out its importance.

\begin{table}
  \begin{tabular}{ | c || c |c | }
    \hline
    \textbf{Superoperator}                         & \textbf{$S$-ITSO} & \textbf{$T$-ITSO} \\
                                                   & (rank,index)      & (rank, index) 
    \\
    \hline
    $|Z_R)(Z_R|,~|Z_R)(Z_L|$                       & (0,0)             & (0,0)
    \\
    \hline
    $\sum\limits_{m=0,\pm1}^{} |S_m)(S_m|$          & (0,0)             & (0,0) 
    \\
    \hline
    $\sum\limits_{m=0,\pm1}^{} |T_m)(T_m|$          & (0,0)             & (0,0) 
    \\
    \hline
    \hline
    $|S_0)(Z_L|, ~|Z_R)(S_0|$                     & (1,0)              & (0,0) 
    \\
    \hline
    $\sum\limits_{m=\pm1}^{} m|S_m)(S_m|$          & (1,0)              & (0,0) 
    \\
    \hline
    $|T_0)(Z_L|, ~|Z_R)(T_0|$                     & (0,0)              & (1,0) 
    \\
    \hline
    $\sum\limits_{m=\pm1}^{} m |T_m)(T_m|$         & (0,0)              & (1,0) 
    \\
    \hline
    $|S_0)(T_0|, ~|T_0)(S_0|$                     & (1,0)              & (1,0) 
    \\
    \hline
    $
    \mathop{
    \sum\limits_{\sigma,\eta=\pm}^{\phantom{1}}}
    \sigma^{\tau_\sigma} \eta^{\tau_\eta}
    |\alpha^\nu_{\eta,\sigma})(\alpha^{\nu'}_{\eta,\sigma}|
    $
                                                 & ($\tau_\sigma$,0)   & ($\tau_\eta$,0)   
    \\
                                                 & $\tau_{\sigma}=0,1$ & $\tau_{\eta}=0,1$
    \\
    \hline
    \hline
    $\sum\limits_{m=0,\pm1}^{} (3m^2-2)|S_m)(S_m|$ & (2,0)              & (0,0) 
    \\
    \hline
    $\sum\limits_{m=0,\pm1}^{} (3m^2-2)|T_m)(T_m|$ & (0,0)              & (2,0) 
    \\
    \hline
  \end{tabular}
  \caption{
    Irreducible tensor \emph{superoperators} (ITSOs) of different rank
    but with (i) \emph{zero index} with respect to both spin and charge rotations
    and (ii) satisfying the causal structure constraints (cf. \Sec{sec:causalstruct}).
    The general effective Anderson Liouvillian \eq{obliv} is a linear combination of all of these,
    where the coefficient of $|Z_R)(Z_R|$ is always fixed to $-4i\Gamma$
    in the wide-band limit (see \Sec{sec:causalstruct}).
    In the special limits of higher symmetry
    only $(0,0)$, $S$-ITSO resp. $T$-ITSOs can appear in this expansion (see \Sec{sec:high-sym}).
    The ITSOs are constructed by standard angular momentum coupling.
    For this, one takes the supervectors in \Tab{tab:itos},
    denoted schematically by $|s,m_s; t,m_t)$  where $s,t$ and $m_s,m_t$
    are the rank and index with respect to spin and charge rotations in Liouville space.
    Then, one constructs conjugate bra supervectors \HL{that} transform with the \emph{same} rank and index:
    these are  $(-1)^{s-m_s+t-m_t}(s,-m_s; t,-m_t|$.
    Coupling these kets and bras with Clebsch-Gordan coefficients gives the \emph{superoperators}
    transforming with definite rank and index with respect to spin and charge rotations.
  }
  \label{tab:itso}
\end{table}

\key{By exploiting only \HL{its} general properties
we have reduced the number of terms \HL{contributing to $L(z)$} from 256 [cf. \Eq{exp_liv}] down to just 30.
The key simplification \HL{came by} using the causal field superoperators to construct the Liouville Fock space.
\HL{The resulting} general Liouvillian can be easily diagonalized
as we show in the next section.
Furthermore, because of its general nature, the parametrization \eq{obliv} is useful in \HL{applications other} than those considered here and may be extended to more complex Anderson-type models (see \App{sec:grassmann}).}
For example, we note that $|Z_{R})$ is always a right eigen-supervector of the effective Liouvillian decaying with rate $2m\Gamma$ where $m$ is the
 number of electrodes attached to the dot (for our case $m=2$).
This mode was recently also found in a study investigating the time relaxation of the density matrix of the Anderson model.~\cite{Contreras12}
It was observed that this mode, appearing in one-loop perturbation theory, is not affected by two-loop corrections.
Our work generalizes this result: the eigenvalue of the right eigen-supervector $|Z_{R})$ is not affected by \emph{any} higher-order corrections.
We also see how this relies on assuming the wide-band limit.
Both insights directly rely on the causal representation of the field superoperators.
\HL{Further implications for the} time dependence will be discussed elsewhere~\cite{Saptsov13a,Reckermann13}.

We now \HL{list} how the expansion coefficients for $L(z)$ are incorporated in \Eq{obliv}
\HL{through the following terms:}
\begin{itemize}
\item
  \HL{The zero eigen-supervectors of the vertices, $|Z_{L})$ and $|Z_{R})$:
  the choice of their coefficients} is based on the general properties of the perturbative series (cf. \Sec{sec:perturb}).
\item
  A supervector in the two dimensional $\chi$-subspace spanned \HL{by} $|\chi_\sigma)$
  \begin{align}
    |\overrightarrow{\psi}) & = \sum_\sigma \psi_\sigma |\chi_\sigma)
  \end{align}
\item
  An independent vector in the corresponding dual space
  \begin{align}
    (\overrightarrow{\phi}| & = \sum_\sigma \phi_\sigma (\chi_\sigma|
    .
  \end{align}
\item
  A superoperator acting on the $\chi$-subspace
  \begin{align}
    \xi  & = \sum_{\sigma,\sigma} \xi_{\sigma,\sigma'} |\chi_\sigma)(\chi_{\sigma'}|
    = \sum_{i=0,1,2,3} \xi_i \chi_i
    \label{xi-operator}
    .
  \end{align}
  Here, \HL{after the second equality}, the matrix  $\xi_{\sigma,\sigma'}$ \HL{is} decomposed
  in the standard basis  $(\tau_i)_{\sigma,\sigma'}$
  of the unit ($i=0$) and three Pauli matrices ($i=1,2,3$),
  giving another $\chi$-subspace \HL{superoperator-basis}:
  \begin{align}
    \chi_i= \sum_{\sigma,\sigma'} (\tau_i)_{\sigma,\sigma'} |\chi_\sigma)(\chi_{\sigma'}|
    .
  \end{align}
\item
  Four superoperators acting on the 2 dimensional $\alpha_{\eta,\sigma}$-subspaces spanned by $|\alpha_{\eta,\sigma}^{\pm})$
  \begin{align}
    F_{\eta,\sigma}
    & =
    \sum_{\nu,\nu'}     F_{\eta, \sigma}^{\nu,\nu'}
    |\alpha_{\eta, \sigma}^\nu)( \alpha_{\eta, \sigma}^{\nu'}|
    \\
    &= \sum_{i=0,1,2,3} F_{\eta,\sigma}^{i} \alpha_{\eta,\sigma}^i
    \label{F-operators}
    .
  \end{align}
  with unit and Pauli-vector superoperators
  \begin{align}
    \alpha_{\eta,\sigma}^i =
    \sum_{\nu,\nu'} (\tau_i)_{\nu,\nu'} |\alpha_{\eta,\sigma}^\nu)(\alpha_{\eta,\sigma}^{\nu'}|
    \label{pauli-aplha}
    .
  \end{align}
  for each fixed $\sigma = \pm$ and $\eta=\pm$. 
\end{itemize}
It is convenient to use the four-vector as well as the $2\times2$ matrix representations for the superoperators $\xi$ and $F_{\eta,\sigma}$.

All the above expansion coefficients depend on the QD frequency $z$ (not written)
and satisfy the following conjugation relations, \HL{which derives} from the Hermicity condition \eq{hermcond}: $K\left[iL(z)\right]K = iL(-z^{*})$.
For the charge and spin diagonal operators these are
\begin{align}
  \overrightarrow{\phi}(z) &=  \overrightarrow{\phi}^{*}(-z^{*}),
  & 
  \overrightarrow{\psi}(z) &=  \overrightarrow{\psi}^{*}(-z^{*}),
  \nonumber
  \\
  \xi(z) &=  \xi^{*}(-z^{*}),
  & 
  \label{hc-relation-diag}
\end{align}
implying that these \HL{coefficients} are real only for zero frequency $z=i0$.
For the charge or spin nondiagonal operators we have
\begin{align}
  F_{\eta,\sigma}^{\nu,\nu'}(z) & = 
  \nu \nu' {F_{\bar{\eta},{\bar{\sigma}} }^{{\nu},{\nu}'}}^{*}(-z^{*})
  ,
  &
  \nu,\nu'=\pm
  ,
  \\
  M_\sigma(z) & = M_{\bar{\sigma}}^{*}(-z^{*})
  ,
  &
  \\
  E_\sigma(z) & = E_{\bar{\sigma}}^{*}(-z^{*}).
  &
  \label{hc-relation-nondiag}
\end{align}
Therefore, at finite dot frequency $z$ all parameters are in general complex and all $2\times2$ coefficient matrices are non-Hermitian.

Note that in \Eq{obliv} we have parametrized $iL(z)$, rather than $L(z)$, i.e., including the imaginary factor $i$.
This anticipates the application to the RT-RG, where a renormalized version of the Liouvillian $\bar{L}$
appears in the final RG equations only as the combination $i\bar{L}$ (cf. \Sec{sec:freq-dep1} and \sec{sec:freq-dep2}).
Finally, we emphasize that the simplifications that led up to the parametrization \Eq{obliv} remain valid for the RT-RG:
\Eq{trSigma}, as well as  \Eq{x0G} and \Eq{Gx1}  do not change under the continuous renormalization,
 as will be shown in \Sec{sec:rg}.

\subsection{Effective Liouvillian
\label{sec:proj}} 

\subsubsection{Spectral decomposition of $L(z)$ and $\bar{L}$}

Above we have reduced $L(z)$ to block-diagonal form as far as possible by using symmetry and general properties.
In \Sec{sec:rg}, we will see that $L(z)$ is closely related to a \emph{renormalized} version of the QD Liouvillian  $\bar{L}$ that we will denote by $\bar{L}_\Lambda$.
To make this clear, we have to anticipate a result:
 $\bar{L}_\Lambda$ interpolates between $\bar{L}$ and $L(z)$ as the flow parameter $\Lambda$ varies from $\infty$ to $0$:
$L(z)=\bar{L} + \int_\infty^0 d\Lambda \frac{d \bar{L}_\Lambda}{d\Lambda}$.
This is done by redistributing diagrams of $\bar{\Sigma}$ in \Eq{Leff2} and including a fraction of them  into a redefinition of the Liouvillian $\bar{L}$.
By construction $\bar{\Sigma}(z)$ is thus decomposed into ``pieces'' $\frac{d \bar{L}_\Lambda}{d\Lambda}$ with the same matrix structure
that are accumulated during the flow.
At the end of the flow,  $\bar{L}_\Lambda$  equals $L(z)$.
Therefore, $\bar{L}_\Lambda$ has an expansion of the same form [\Eq{obliv}] as $L(z)$.
For notational simplicity, we denote the expansion coefficients of $\bar{L}_\Lambda$ by the same variables as for $L(z)$. In cases where this leads to 
confusion, 
the coefficients of $\bar{L}_\Lambda$ are distinguished from those of $L(z)$ by indicating their $\Lambda$ dependence, e.g., $F^{\nu,\nu'}_{\eta\sigma\Lambda}$ vs. $F^{\nu,\nu'}_{\eta\sigma}$, 
which can, however, often be omitted.
All results obtained in this section thus apply to both the exact $L(z)$ as well as the renormalized $\bar{L}_\Lambda$.
\HL{This} explicit form of $\bar{L}_\Lambda$ \HL{is required since later on it} needs to be inserted into resolvent superoperators 
\HL{that appear in the RG equations.}

We therefore need to completely diagonalize $\bar{L}_\Lambda$
such that it can be expanded into its eigen projectors
$
P^{k}=|\bar{\lambda}^k)(\lambda^k| = (P^k)^2
$:
\begin{align}
  \bar{L}_\Lambda = \sum_{k} \lambda^k P^k
  \label{eigenproj}
  ,\end{align}
where the sum runs over the labels \HL{$k$} of the eigenvalues.
Here, \HL{ $({\lambda}^k|$ and $|\bar{\lambda}^k)$} are the left and right eigen-supervectors of $\bar{L}_\Lambda$ for the same eigenvalue $\lambda^k$:
$
\bar{L}_\Lambda P^{k} = P^{k} \bar{L}_\Lambda = \lambda^k P^{k}
$.
Using this complete and orthogonal set of projectors, one can then evaluate resolvent superoperators in \Eq{barSigma} explicitly:
\begin{align}
  ...\bar{G} \frac{1}{z-X-\bar{L}_\Lambda}\bar{G}...
  =
  ...\sum_i \frac{1}{z-X - \lambda^i} \bar{G}P^i\bar{G}...
  \label{someterm}
\end{align}
\HL{We note that the} diagonalization of \Eq{obliv} \HL{can also be} useful for higher orders of (renormalized) perturbation theory \Eq{sig_irr} (\Eq{barSigma}):
when expanding the QD $L$ and $\bar{L}_\Lambda$ in the form \eq{obliv} it is directly adapted to all symmetries of the problem
and one can efficiently construct explicit matrix representation of the self-energies $\Sigma(z)$ and $\bar{\Sigma}(z)$, respectively.

\HL{The application of} the above spectral decomposition to the continuous RG \HL{in} \Sec{sec:rg} involves two assumptions that should be pointed out here.
First, we always assume that the zero eigenvalue of $\bar{L}_\Lambda$ is non-degenerate,
corresponding to the unique stationary state for the density operator.
This is always found to be the case for the numerically calculated RG flows discussed in \Sec{sec:rg}.
However, in principle, it may also happen that two (or more) \emph{non-zero} eigenvalues of $\bar{L}_\Lambda$ become degenerate during this flow. If this is 
the case, and additionally the supermatrix $\bar{L}_\Lambda$ has nonzero elements on its diagonal in the normal Jordan form in the degenerate subspace, 
then no complete eigenprojector basis exists.
For the Anderson model, this presents no crucial complication: the eigenbasis can in principle be circumvented for the calculation of the two dimensional
superoperators. However,  numerically we never meet \HL{this} problem in the application of the RT-RG theory presented below.

\subsubsection{Eigenvalues, eigen-supervectors and the stationary state\label{sec:eigenval}}
We now \HL{explicitly} diagonalize $L(z)$ or $\bar{L}_\Lambda$, parametrized as in \Eq{obliv}, in each of its block-diagonals.
We \HL{first list the eigenvalues and projectors in} the block spanned by the charge- and spin-diagonal \HL{\emph{bosonic}} operators $|Z_{L})$,$|Z_{R})$,$|S_0)$, and $|T_0)$, which contains the stationary \HL{non-equilibrium} state:
\begin{itemize}
\item
  Eigenvalue $\lambda^{Z_{L}}=0$ \HL{with projector}
  \begin{align}
    P^{Z_{L}}= 2 | \rho)(Z_{L}|
    \label{zero-proj}
    ,
  \end{align}
  with the stationary density operator
  \begin{align}
    ~~~~~
    \HL{|\rho)}
    =-\frac{1}{2\xi} |\overrightarrow{\psi}) +\frac{1}{2}|Z_{L})
    -\frac{  \zeta-(\overrightarrow{\phi}| {\xi}^{-1}|\overrightarrow{\psi})}{8\Gamma}
    |Z_{R})
    \label{rho}
    .
  \end{align}
  We note that the coefficient $\zeta$ appears only in the stationary state
  and the next projector, $P^{Z_{R}}$,
  but not in any other eigenprojector or eigenvalue.
\item
  Eigenvalue
  $\lambda^{Z_{R}}= {-i} 4\Gamma$  \HL{with projector}
  \begin{align}
    \label{p1}
    &
    ~
    P^{Z_{R}}
    =
    (\overrightarrow{\phi}|
    \frac{1}{4\Gamma(4\Gamma- \xi)}
    |\overrightarrow{\psi})
    \, 
    |Z_{R})(Z_{L}|
    \\
    &
    +|Z_{R})(Z_{R}|
    +\frac{\zeta}{4 \Gamma}|Z_{R})(Z_{L}|
    +|Z_{R})(\overrightarrow{\phi}| \frac{1}{4\Gamma- \xi}
    \nonumber
    .
  \end{align}
\item
  Eigenvalues $\lambda^{\chi \pm}$ with projectors
  \begin{align}
    \label{p34}
    ~~~~~
    &P^{3,4}
    =
    P^{\chi,\pm}
    + %\nonumber \\
     \frac{ (\overrightarrow{\phi}|P^{\chi\pm}|\overrightarrow{\psi}) }
           { \left(\lambda^{\chi\pm}-4\Gamma\right)\lambda^{\chi\pm} }
    |Z_{R})(Z_{L}|
    \\
    &+\frac{1}{\lambda^{\chi,\pm}-{4\Gamma}} |Z_{R})(\overrightarrow{\phi}|P^{\chi,\pm}
    +
    \frac{1}{\lambda^{\chi,\pm}}    P^{\chi\pm}   |\overrightarrow{\psi})(Z_{L}|
    \nonumber
    .
  \end{align}
  \HL{The} eigenvalues $\lambda^{\chi\pm}$ are determined by first diagonalizing $\xi$
  in the $\chi$-subspace, i.e., \HL{by finding eigen-projectors $P^{\chi,\sigma}$ of $\xi$}
  \begin{align}
    \xi P^{\chi,\sigma} =   P^{\chi,\sigma}  \xi  = \lambda^{\chi,\sigma} P^{\chi,\sigma}
    .
  \end{align}
  Since $\xi_{\sigma,\sigma'}$ is a $2\times2$ non-Hermitian matrix, \HL{it} can be expressed
  in the vector $\vecg{\xi}=(\xi_1,\xi_2,\xi_3)$ and coefficient $\xi_0$ (cf. \Eq{xi-operator})\HL{, all of which are complex}:
  \begin{align}
    \label{pchi}
    \lambda^{\chi,\sigma} & = -i\left(\xi_0 +\sigma \sqrt{\vecg{\xi}^2}\right)
    \\ \label{pchi-p}
      P^{\chi,\sigma}
      & =\frac{1}{2}\chi_0 +\sigma \frac{\vecg{\chi}\cdot \vecg{\xi}}{2\sqrt{\vecg{\xi}^2}}
    \end{align}
    Here, the square root of the complex argument is defined
    such that the branch cut lies in the lower-half complex plane
    since \HL{the} effective Liouvillian must be regular in the upper half-plane
    according to \Eq{Laplace-rho} \HL{and} \eq{kineq}.
\end{itemize}
The remaining block-diagonals
acting on bosonic subspaces are one-dimensional: for $\sigma =\pm$
\begin{align}
  \lambda^{T_\sigma}={-i}M_\sigma, & & P^{T_\sigma}=|T_\sigma)(T_\sigma|
  ,
  \\
  \lambda^{S_\sigma}={-i}E_\sigma, & & P^{S_\sigma}=|S_\sigma)(S_\sigma|
  .
\end{align}

Finally, in the four \HL{remaining,} two-dimensional, \HL{\emph{fermionic}} subspaces labeled by $\alpha_{\eta,\sigma}$ (for fixed $\eta$ and $\sigma$),
 the eigenvalues and projectors can be calculated \HL{in the same way as for} the bosonic $\chi$ block:
\begin{align}
  \label{palpha}
  \lambda^{\alpha_{\eta,\sigma},\pm}
  & = -i\left(F_{\eta,\sigma}^0 \pm \sqrt{\vec{F}_{\eta\sigma}^2}\right)
  ,
  \\ \label{palpha-p}
  P^{\alpha_{\eta,\sigma},\pm}
  & =\frac{1}{2} \alpha_{\eta,\sigma}^0
  \pm \frac{ \vec{F}_{\eta\sigma} \cdot \vecg{\alpha}_{\eta, \sigma}}
  {2\sqrt{ \vec{F}_{\eta\sigma}^2 } }
    ,
\end{align}
where the coefficients ${F}_{\eta\sigma}^0$ and
$\vec{F}_{\eta\sigma} = ({F}_{\eta\sigma}^1,{F}_{\eta\sigma}^2,{F}_{\eta\sigma}^3)$ 
are again complex [cf. \Eq{F-operators}].

We note that \HL{it is} in principle possible that $\vecg{\xi}^2 =0$ while $\vecg{\xi} \neq \overrightarrow{0}$.
In this case, the supermatrix representation of $\chi$ has non-zero diagonal element
in its normal Jordan form.
In this case,  \HL{$\xi$} has no eigenbasis and \HL{\Eq{pchi-p}} does not apply.
Still,  the required matrix valued functions of \HL{$\xi$} can be calculated
using the Hamilton-Cayley theorem.
The same remarks apply \HL{to $\vec{F}_{\eta\sigma}$ and} \HL{\Eq{palpha-p}}.
However, in practical applications, we never meet such a situation.
We also note that during the RG flows discussed in \Sec{sec:rg},
we never run into points where $\lambda^{\chi,\pm}={-i}4\Gamma$
and one therefore need not worry about the vanishing of the denominators in \Eq{p34}
or the existence of the inverse of $\left(4\Gamma-\xi\right)$ in \Eq{p1}.

An important simplification applies to the first four bosonic projectors $P^{Z_{L}}$, $P^{Z_{R}}$, $P^{3,4}$ that involve the vectors $|Z_{L})$ and $|Z_{R})$, analogous to the corresponding terms in \HL{the} expansion \eq{obliv} of the effective Liouvillian (cf. \Sec{sec:causalstruct}).
When inserting projectors into \Eq{someterm} between two vertices $\bar{G}$, (i)
% \begin{itemize}
% \item
  The projectors  $P^{Z_{L}}$, $P^{Z_{R}}$ give no contributions;
% \item
(ii)
  The projectors $P^{3,4}$ only contribute through the first term $P^{\chi,\pm}$ in \Eq{p34}.
% \end{itemize}
As a result, in all applications below we can replace  \Eq{eigenproj} with the simpler expansion
\begin{align}
  &
  \bar{L}_\Lambda
  \overset{\bar{G} \cdots \bar{G}}{\longrightarrow}
  \label{eigenproj2}
  \\
  & 
  \lambda^{\chi} P^{\chi}+
  \sum_{\sigma}
  \Big[
     \lambda^{S_\sigma} P^{S_\sigma}
    +\lambda^{T_\sigma} P^{T_\sigma}
    + \sum_{\eta} \lambda^{\alpha_{\eta,\sigma}} P^{\alpha_{\eta,\sigma}}
      \Big]
  .
  \nonumber
\end{align}
Here, we leave implicit the sum over the two eigenvalues
$\lambda^{\chi,\pm}$ and $\lambda^{\alpha_{\eta,\sigma},\pm}$ in the 
$\chi$ and $\alpha_{\eta,\sigma}$-subspace, respectively.
\key{
A crucial stability requirement for the RG in \Sec{sec:rg} is thereby explicitly satisfied: the zero eigenprojector \eq{zero-proj}, 
corresponding to the physical  stationary state, never appears in the resolvents.
}

\subsubsection{Fermionic excitations: Spectral decomposition of $\bar{\Sigma}$
\label{sec:fermion-pt}}

The expansion \eq{obliv} can of course also be applied to $\bar{\Sigma}(z)=L(z)-\bar{L}$.
This, however, involves additional simplifications causing certain terms appearing in the expansion of $L(z)$ (and $\bar{L}_\Lambda$) to be absent.
First, due to \Eq{ZReigenbarSigma} the term $|Z_R)(Z_R|$ is missing.
Second, \HL{since Eqs. \eq{alpha+0} and \eq{alpha-0} derive} from  the causal structure (cf. \Sec{sec:causalstruct}),
most of the coefficients of the fermionic sector of $\bar{\Sigma}$
, denoted by $-i \Delta F^{\nu,\nu'}_{\eta\sigma}$,
 vanish:
$\Delta F^{\nu,\nu'}_{\eta\sigma}= \delta_{\nu,-}\delta_{\nu',+} \Delta F^{-,+}_{\eta\sigma}$.
We can express  the coefficient matrices $F^{\nu,\nu'}_{\eta\sigma}$
 of $L(z)=\bar{L}+\bar{\Sigma}(z)$ using \Eq{initial},
in terms of \HL{those of $\bar{\Sigma}(z)$}{, i.e., in terms of the $\Delta F^{-,+}_{\eta\sigma}$}:
\begin{align}
   &-i\begin{pmatrix}
      F^{+,+}_{\eta\sigma} & F^{+,-}_{\eta\sigma} \\
      F^{-,+}_{\eta\sigma} & F^{-,-}_{\eta\sigma}
    \end{pmatrix}
     \\ \nonumber&=
   \begin{pmatrix}
       \eta\left(\epsilon+\frac{U}{2}\right)+\sigma\frac{B}{2} 
    -i\Gamma &  \frac U 2 \\
      \frac U 2 -i \Delta F^{-,+}_{\eta\sigma} &  \eta\left(\epsilon+\frac{U}{2}\right)+\sigma\frac{B}{2} 
    -i3\Gamma
    \end{pmatrix}
    .
\end{align}
Converting to spherical coefficients $F_{\eta,\sigma}^{0}$ and $\vec{F}_{\eta\sigma}$ and using \Eq{palpha}, we find
\begin{align}
  \label{ferm-spec}
  \lambda^{\alpha_{\eta,\sigma},\pm} & =
  \eta\left( \epsilon+\frac{U}{2}\right)+\sigma\frac{B}{2} -2i\Gamma \\
  \nonumber
  &\pm \eta\sqrt{\frac{U^2}{4}-\Gamma^2-i\frac{U \Delta F_{\eta,\sigma}^{-,+}}{2}}
  ,
\end{align}
\begin{align}
  \label{ferm-proj}
  P^{\alpha_{\eta,\sigma}, \pm} & =
  \frac{
    \alpha_{\eta,\sigma}^0}{2}
  \pm \eta
  \dfrac{
    \frac{U}{2} \alpha_{\eta,\sigma}^{1}
    +i\Gamma \alpha_{\eta,\sigma}^{3}
    -i \Delta F_{{\eta,\sigma}} ^{-,+}\alpha_{\eta,\sigma}^{-}
  }
  {
    2 \sqrt{ \frac{U^2}{4}-\Gamma^2-i\frac{U \Delta F_{\eta,\sigma}^{-,+}}{2} }
  }
  .
\end{align}
We thus find that the functional form of the fermionic eigenvalues \HL{of $L(z)$} is severely \HL{restricted} by the casual structure of $\bar{\Sigma}$:
all the frequency-dependent renormalization effects enter into the projectors and eigenvalues solely through the four complex numbers $\Delta F_{\eta,\sigma}^{-,+}$ in the four $\alpha_{\eta\sigma}$ blocks ($\eta,\sigma=\pm$).

For a QD decoupled from the reservoirs, $\Gamma=0$, we have $\tilde{\Sigma}=0=\bar{\Sigma}(z)=0$ and thus $\Delta F_{\eta,\sigma}^{-,+}=0$:
\begin{align}
  \lambda^{\alpha_{\eta,\sigma},+} & = \eta\left(\epsilon + U\right) +\sigma B/2
  ,
  \\
  \lambda^{\alpha_{\eta,\sigma},-} & = \eta\epsilon  +\sigma B/2
  \label{ferm-spec-free}
  .
\end{align}
This is the spectrum of \emph{many-body} energy excitations when adding a single electron, starting from
either an empty QD ($ \lambda^{\alpha_{\eta,\sigma},+}$),
or a singly occupied QD with spin $\bar{\sigma}$ ($ \lambda^{\alpha_{\eta,\sigma},-}$).
This is confirmed by the eigenprojectors \HL{in the limit} $\Gamma \rightarrow 0$ for finite $U$:
\begin{align}
  P^{\alpha_{\eta,\sigma}, \pm} & =
  \frac{1}{2}\left( \alpha_{\eta,\sigma}^{0}  \pm \alpha_{\eta,\sigma}^1 \right)
  =
  \begin{cases}
    |\sigma,0)(\sigma,0|             & +,\\
    |2,\bar{\sigma})(2,\bar{\sigma}| & -,
  \end{cases}
\end{align}
where $|\sigma,0)=|\sigma\rangle\langle 0|= (1/2-Z_R)d^\dagger_\sigma$ and 
$|2,\bar{\sigma})=|2\rangle\langle \bar{\sigma}|= \sigma(1/2+Z_R)d^\dagger_\sigma$.
These are the virtual intermediate states and energies that enter through $L$ into the perturbation expansion \eq{sig_irr} for $\Sigma(z)$.

\begin{figure}[tbp]
  \includegraphics[width=0.99\linewidth]{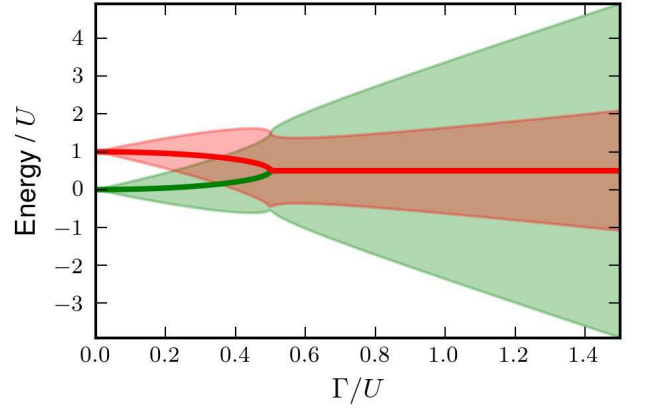}
  \caption{
    Fermionic excitation energies and widths
    of the infinite-temperature Liouvillian $\bar{L}$
    plotted as function of the tunnel coupling strength $\Gamma/U$
    for $\epsilon=B=0$.
    The energy and width are given
    by the real and imaginary parts of the fermionic eigenvalues
    $\lambda^{\alpha_{\eta,\sigma},\nu}$
   [\Eq{ferm-spec} with $\Delta F_{\eta,\sigma}^{-,+}=0$].
    The real parts for $\nu=+$ (red) and $\nu=-1$ (green), respectively, are given by the full lines,
    and the imaginary parts are indicated by the shaded width of the level
    with the corresponding color.
    For $U > 2\Gamma$, the excitations have different energies, split by $U$,
    but with the same width $2\Gamma$,
    whereas for $2\Gamma > U$ they have the same energies
    but different widths: for $\Gamma \gg U$
    $-\im \lambda^{\alpha_{\eta,\sigma},\pm} \approx \Gamma$ (red) $3\Gamma$ (green).
  }
  \label{fig:fermion-level}
\end{figure}

For any finite coupling $\Gamma$, but infinite temperature, \HL{we have} $\bar{\Sigma}=0$ and therefore again $\Delta F_{\eta,\sigma}^{-,+}=0$. \HL{In this case}, however, $\tilde{\Sigma}\propto \Gamma \neq 0$ and the eigenvalues obtained from \Eq{ferm-spec} depend qualitatively on the 
interaction strength $U$:
for $U < 2\Gamma$, the coupling to the reservoirs $\Gamma$
adds different imaginary parts to the degenerate real eigenvalues $\eta(\epsilon+U/2)+\sigma B/2$,
whereas for $U > 2\Gamma$ it
adds a uniform imaginary part $-i\Gamma$ to these eigenvalues
while differently shifting their real parts.
This dependence on $U$ is plotted in \Fig{fig:fermion-level} and shows a bifurcation at $U=2\Gamma$.
These are the energies and projectors that enter through $\bar{L}$ into the renormalized perturbation theory \eq{barSigma}.

Finally, for both finite coupling $\Gamma$ and finite temperature $T$, the complex coefficients $\Delta F_{\eta,\sigma}^{-,+}$ are non-trivial functions that need to be calculated, e.g., either perturbatively or using the RG, see \Sec{sec:fermion}. 
However, even in this case \Eq{ferm-spec} provides an exact relation:
the \emph{average} of the complex particle and hole excitation eigenvalues in each $\alpha_{\eta\sigma}$-block is independent
of $\Delta F_{\eta,\sigma}^{-,+}$ and thereby, also independent of the frequency $z$:
\begin{align}
  \label{symmetric_spec}
  \frac{1}{2}\sum_\pm \lambda^{\alpha_{\eta,\sigma}, \pm}
  =
  \eta\left(\epsilon+\frac{U}{2}\right)+\sigma\frac{B}{2}-2i\Gamma
  .
\end{align}
Physically speaking, \emph{both} the renormalized energies of single particle fermionic excitation energies (real parts) as well as their decay rates / broadenings 
(imaginary parts) lie symmetric with respect to the above average values.
For example, if the particle excitation broadens, the hole excitation \HL{must sharpen} up and \emph{vice versa}.

Finally, we can infer an important physical stability constraint on the functions $\Delta F_{\eta,\sigma}^{-,+}$: they must be such that the imaginary part of the  square root is less than $2\Gamma$ for all $\Lambda$ \HL{and $z$}.
Otherwise, inverse Laplace-transforming $L(z)$ to time space would yield terms that diverge 
for $t \rightarrow \infty$, which is unphysical.
This restricts the maximal excitation widths in the fermionic block: the negative imaginary part of \HL{the eigenvalues $\lambda^{\alpha_{\eta, \sigma},\pm}$} cannot exceed the value $4\Gamma$.

\subsubsection{High-symmetry stationary states
\label{sec:high-sym}}

As a cross-check on the results of \Sec{sec:eigenval},
 we now analyze the stationary density operator in the two special parameter regimes where the model has a higher symmetry than in general.
In both cases, the superoperator $\xi$ has no off-diagonal terms in the basis of $|S_0)$ and $|T_0)$,
\begin{align}
  {\xi} = \xi_{TT} |T_0)(T_0| +\xi_{SS} |S_0)(S_0|
  ,
\end{align}
because there are no scalars \HL{with respect to} spin- or charge-rotations \HL{that} contain such terms as components
(see \Tab{tab:itso}).
If either symmetry is broken, charge and / or spin-rotations allow for mixing terms $|T_0)(S_0|$ and $|S_0)(T_0|$ in \Eq{obliv}. 
For this reason, we work with the $|\chi_\sigma)$ basis \eq{chi-basis} in the general case.

\paragraph{Full spin-rotation symmetry, $B=0$.}
At zero magnetic field, all terms in the Liouvillian must be rank-0 ITSOs (scalars) with respect to spin rotations, i.e., the 
terms $|S_0)(S_0|$ and $|S_\pm)(S_\pm|$ must have the same coefficients, $E_+=E_-=\xi_{SS}$
and the coefficients of the rank-1 and -2 spin ITSOs must vanish, i.e., in the $\chi$ subspace
$|\phi) = \phi_T |T_0)$ and $|\psi) = \psi_T |T_0)$ with real
$\phi_T$ \hl{and $\psi_T$}  at $z=i0$ (cf. \Eq{hc-relation-diag}):
\begin{align}
  &i\bar{L}_\Lambda
  =
  ...
  +\phi_T |Z_{R})(T_0| +\psi_T |T_0)(Z_{L}| 
  \\
  &+M_+ |T_+)(T_+|+M_-|T_-)(T_-| + \xi_{TT} |T_0)(T_0|
  \nonumber \\
  &+\xi_{SS} \Big(  |S_+)(S_+|+  |S_-)(S_-|+|S_0)(S_0| \Big) + \ldots
  \nonumber
  .
\end{align}
See \Tab{tab:itso}, where the ITSOs are listed.
The expression for the stationary state $\rho$ obtained using \Eq{rho}
is then independent of $\xi_{SS}$ \HL{and is} therefore explicitly invariant under spin-rotations, as required:
\begin{align}
  \label{rhoT}
  \rho & =-\frac{1}{2} \psi_T |T_0) +\frac{1}{2}|Z_{L})
  -\frac{ \zeta- \phi_T \psi_T/\xi_{TT}}{8\Gamma}|Z_{R})
  .
\end{align}
\paragraph{Full charge-rotation symmetry, $\epsilon =- U/2$ and $\mu_L=\mu_R$.}
At the particle-hole symmetric point, a similar reduction must take place:
here $M_+=M_-=\xi_{TT}$, $|\phi) = \phi_S |S_0)$ and $|\psi) = \psi_S |S_0)$:
\begin{align}
  \label{iLambdabarph}
  &i\bar{L}_\Lambda
  =
  \ldots
  +\phi_S |Z_{R})(S_0| +\psi_S |S_0)(Z_{L}| 
  \\
  &+ \xi_{TT} \Big( |T_+)(T_+|+ |T_-)(T_-| +  |T_0)(T_0| \Big)
  \nonumber \\
  &+E_+ |S_+)(S_+|+  E_{-} |S_-)(S_-| + \xi_{SS} |S_0)(S_0|
  \nonumber
\end{align}
and the stationary state is explicitly invariant under charge rotations:
\begin{align}
  \label{rhoDph}
  \rho&=-\frac{1}{2} \psi_S |S_0) +\frac{1}{2}|Z_{L})
  -\frac{\zeta - \phi_S \psi_S / \xi_{SS}}{{8\Gamma}}|Z_{R})
  .
\end{align}

\subsection{Current superoperator and its irreducible self-energy\label{sec:current}}
\key{
Our main objective is to calculate the stationary current \HL{that} flows through the QD.
Having set up the perturbation theory formalism for the density operator, the expression for the average current can now be compactly derived.
Moreover, we give a general proof that in general at zero bias the current vanishes, as it should, independent of the way the self-energy is calculated.
}

The current flowing out of reservoir \HL{$r=L$ or $R$} is obtained using \hl{Heisenberg operators (with index $\mathrm{H}$):
$
I^r_{\mathrm{H}}   = -\frac{d}{dt}  {n}^r_{\mathrm{H}}
                = -i [ H^\tot ,  {n}^r_{\mathrm{H}}]_{-}
                = -i \left[ {V}^r_{\mathrm{H}} ,  {n}^r_{\mathrm{H}}\right]_{-}
$.} Note that there is no summation over the electrode index $r$ [cf. \Eq{Vrdef}].
\hl{The expectation value of the Schr{\"o}dinger-picture current operator
\begin{align}
  \label{current}
  I^r   = -i \left[ {V}^r ,  {n}^r\right]_{-}
  ,
\end{align}
}can be expressed in superoperators using the cyclic invariance of the total system trace:
\begin{align}
  \langle I^r \rangle (t) & = \Tr{D} \Tr{\R} \left(I^r \hl{ \rho^{\tot} (t) } \right)
  \nonumber
  \\
  &=-i   \Tr{D} \Tr{\R} \left( L^{I^r}  e^{-i L^\tot (t-t_0)} \rho (t_0)\rho^{\R} \right)
  \label{average}
  .
\end{align}
We note that observable averages involve anticommutators of the corresponding operator (see, e.g., \Cite{Mukamel03}):
\begin{align}
  \label{classic}
  L^{I^r}=\frac{i}{2} [ I^r , \bullet]_{+}
  .
\end{align}
This is in contrast to time-evolution superoperators, which involve commutators of the Hamiltonian operator.
This difference is exploited below.
If one uses \Eq{current} in \Eq{average}, the evaluation of the reservoir trace is unnecessarily complicated,
since \Eq{current} involves two operators acting on the reservoir, ${V}^r$ and ${n}^r$.
Here, we proceed differently: we first use that the tunneling through junction $r$ conserves the particle number of the dot and the reservoir $r$, \HL{i.e.,}
$
[ {n}^r + n ,  {V}^r]_{-}=0
$,
\HL{to} eliminate one electrode operator:
\begin{align}
  \label{current_d}
  I^r = i \left[ {V}^r ,  n\right]_{-} = -i \left[ n ,  {V}^r\right]_{-}
  .
\end{align}
Then, using the identity:
$
\left[\left[ {A} ,  {B}\right]_{-} , \bullet\right]_{+}
= \left[ {A} , [ {B} , \bullet]_{-} \right]_{+}
 - \left[ {B} , [ {A} , \bullet]_{+} \right]_{-}
$
the current superoperator anticommutator can be decomposed as
\begin{align}
  L^{I^r} = \frac{1}{2} \left(L^{n +} L^{V,r}  - L^{V,r} L^{n +} \right)
  \label{general_current}
  .
\end{align}
Here, we introduced the \emph{anti}commutator superoperator for the particle number (cf. \Eq{Ln}),
\begin{align}
  L^{n+}= \left[ n , \bullet \right]_{+}
  ,
\end{align}
and decomposed the tunneling interactions into the junction contributions,
$L^V = \sum_r L^{V,r}$.
Importantly, the last term of \Eq{general_current} is irrelevant when inserted into \Eq{average} since $\mathrm{Tr}_{D}\mathrm{Tr}_{\R} \, L^{V,r} \bullet =0$
due to the commutator form of $L^{V,r}$.
We obtain
\begin{align}
  \langle I^r \rangle (t) 
  = -i\frac{1}{2} \Tr{D} L^{n+}  \left( \Tr{\R} L^{V,r} \HL{\rho^{\tot}(t)} \right)
  .
\end{align}
Integrating out of the reservoirs and collecting terms into irreducible blocks [cf. \Sec{sec:perturb}], one now obtains
\begin{align}
  \label{current_aa}
  \langle I^r \rangle (z)
  &= \tfrac{1}{2} \Tr{D} L^{n+} \Sigma^r (z) \frac{1}{z-L (z)} \rho (t_0)
  \nonumber \\ 
  &= -\tfrac{1}{2} i\Tr{D}  L^{n+}  \Sigma^r (z) \rho (z)
  .
\end{align}
where $\Sigma^r$ is just that part of the irreducible self-energy  $\Sigma$ \HL{for which} the latest (leftmost) 
vertex is associated with reservoir $r$.
We can thus decompose
\begin{align}
  \Sigma(z)=\sum_r \Sigma^r(z)
  \label{Sigmadecomp}
  .
\end{align}
\key{For the stationary current
$\langle I^r \rangle
 = {\mathsf{lim}}_{t-t_0 \rightarrow \infty} \langle I^r \rangle (t)
 = {\mathsf{lim}}_{z\rightarrow i0} -i z \langle I^r \rangle (z)
$, we then obtain the central result of this subsection:
\begin{align}
  \label{current_a}
  \langle I^r \rangle
  = -i \tfrac{1}{2} \Tr{D}  L^{n+}  \Sigma^r (i0) \rho
  ,
\end{align}
where $\rho$ is the stationary density operator (cf. \Eq{rho}).
}

The first advantage of the \Eq{current_a} is that it allows one to explicitly see that the current is always zero at zero bias.
For any number of electrodes, at zero voltage bias all electrochemical potentials are equal, $\mu_r=0$, implying that all partial self-energies $\Sigma^{r}$ are proportional to the total self-energy:
\begin{align}
  \Sigma^{r} \propto \frac{\Gamma_r}{\sum_r \Gamma_r} \Sigma(i0)
  .
  \label{sigmarpropto}
\end{align}
\HL{Next, we} add $L$ to $\Sigma(i0)$
in \Eq{current_a} \HL{without changing its value}
since  by local charge conservation, \Eq{nconserv},
$
\mathrm{Tr}_{D}\, L^{n+} L \bullet = \mathrm{Tr}_{D}\, n [H,\bullet]_{-}
                          =  \mathrm{Tr}_{D}\, [n,H]_{-} \bullet = 0
$ when acting on any dot operator.
\HL{We can thus express the stationary, zero-bias current in terms of the effective Liouvillian $L(z)=L+\Sigma(z)$
and directly see that it must vanish,}
\begin{align}
  \langle I^r \rangle \propto -i\Tr{D}  L^{n+} L(i0) \rho=0
  ,
  \label{Irzero}
\end{align}
since the stationary state $\rho$ is the zero eigen-supervector of $ L(i0)$ (cf. \Eq{zeroeigenvector}).
\HL{We note, however, that the relation \eq{sigmarpropto} relies on the assumption of reservoir-frequency independent spectral densities
 $\Gamma_r(\omega) =\Gamma_r$ which we make throughout this paper.
Apart from that,} the above proof holds no matter what approximations one makes for the calculation of the self-energy $\Sigma\HL{(i0)}$ as long as all reservoirs are treated in the same way.
This applies to perturbation theory up to any finite order, as well as to the RT-RG approach that we set up in \Sec{sec:rg}.

A second advantage of \Eq{current_a} is that we can directly relate the current to \HL{just} a few supermatrix elements of the \HL{zero-frequency, effective Liouvillian $L(i0)$} in the basis introduced in \Eq{obliv}.
The dot trace combined with the action of $L^{n+}$ can be expressed in the dual supervectors of \eq{X0def} and \eq{T0def}
\begin{align}
  \frac{1}{2}\Tr{D} (L^{n+} \bullet)
  = \Tr{D} ( n \bullet)
  = \Big( \sqrt{2}(T_0|+{2} (Z_{L}| \Big) \bullet
  .
\end{align}
Equation \eq{x0G} implies probability conservation, \Eq{trSigma},
but also, more strongly, that
$(Z_{L}|\Sigma^r(i0)\bullet = 2 \mathrm{Tr}_{D}\Sigma^r(i0) \bullet = 0$
\HL{for fixed $r$}.
From this, we obtain
\begin{align}
  \langle I^r \rangle = -\sqrt{2} i (T_0|\Sigma^{r}|\rho)
  \label{relevant_current}
  .
\end{align}
Clearly, the partial self-energy $\Sigma^{r}$ has the same general form as \Eq{obliv} and we distinguish its parameters 
\HL{(except $\Gamma_r$) by} an additional reservoir superscript $r$:
\begin{align}
  \label{obliv-current}
  &i \Sigma^{r}
  =
  \\
  &{2\Gamma_r}|Z_{R})(Z_{R}| +\zeta^r |Z_{R})(Z_{L}|
  +|Z_{R})(\overrightarrow{\phi^r}|
  +|\overrightarrow{\psi^r})(Z_{L}|
   \nonumber
  \\
  & +
  \xi^r
  +\sum_\sigma
  \left[
    \, M_\sigma^r |T_\sigma)(T_\sigma|
    +
    E_\sigma^r |S_\sigma)(S_\sigma|
    +\sum\limits_\eta F_{\eta,\sigma}^ r
  \right]
  \nonumber
  .
\end{align}
 Inserting this form and the explicit expression for the stationary state \Eq{rho} into \Eq{relevant_current}, we obtain the final explicit expression for the average stationary current in terms of the self-energy expansion coefficients:
\begin{align}
  \label{currentI}
  \langle I^r \rangle = \frac{1}{\sqrt{2}} \Big{[}
  (T_0|\overrightarrow{\psi}^r)-(T_0| {\xi}^r  {\xi}^{-1}|\overrightarrow{\psi}) 
 \Big{]}
 .
\end{align}
We emphasize that this equation is exact, given that the coefficients $\psi_\sigma^r$ and $\xi^r$ of the partial self-energies $\Sigma^r$ are known,
from which $\psi_\sigma = \sum_r \psi_\sigma^r$ and $\xi=\sum_r \xi^r$ also follow.
We can thus calculate the current easily if we perform all \HL{self-energy} calculations \HL{separately for each fixed value of the} reservoir index $r$ at the latest (leftmost) vertex and sum over $r$ to obtain \Eq{Sigmadecomp}.
Finally, we note that using \Eq{currentI}, one can check explicitly that \HL{if one imposes} particle-hole symmetry on the expansion coefficients \HL{in \Eq{obliv-current},
then the current \eq{relevant_current} vanishes:}
one finds that $|\psi^r)$ and $\xi^r$ have no components involving $|T_0)$.
\HL{This is} in agreement with the result \eq{Irzero} obtained above making explicit use of $\mu_L=\mu_R$.

%%% Local Variables: 
%%% mode: latex
%%% TeX-master: "paper"
%%% End: 
\section{Real-time renormalization group\label{sec:rg}}
In this section we set up the \HL{real-time renormalization group (RT-RG)} calculation of the effective Liouvillian $L(z)$. It is based on the perturbative expansion 
\HL{of} the ``finite-temperature'' self-energy $\bar{\Sigma}$ in the causal representation introduced in the previous section, cf.~\Sec{sec:discrete}.
The procedure is to calculate $L(z)$ by introducing an RG flow of the ``infinite-temperature'' Liouvillian $\bar{L}$ and the vertex $\bar{G}$ as function 
of a decreasing energy scale cutoff $\Lambda$, with the initial \HL{conditions given by} $\bar{L}=\tilde{\Sigma}+L$ (thereby including the vertex of the type 
$\tilde{G}$) and $\bar{G}$, the vertex of \HL{the} perturbation theory.
There are a number of motivations for performing such an RG treatment of the perturbative series for $\bar{\Sigma}$.
\begin{itemize}
\item
  First of all, treating $\bar{\Sigma}$ perturbatively in $\bar{G}$, while infinite orders of $\tilde{G}$ have already been resummed into $\tilde{\Sigma}$, would
  amount to an inconsistent treatment.
  We note that even in the non-interacting case $U=0$
  we already need to do a renormalized perturbation theory \eq{barSigma} \HL{\emph{up to two-loop terms}} to recover the exact result \HL{for} all quantities 
  (i.e., not just the current).
  For strong interaction $U$, \HL{higher-order} corrections \HL{in $\bar{G}$} become \HL{increasingly important}: at finite and especially at low $T$ the strong interaction $U$ leads to different lifetime broadening for single- (SET) and two-electron \hl{inelastic cotunneling (ICT)} excitations, with a non-trivial
  voltage dependence.
  \HL{This is not described by $\tilde{\Sigma}$: it leads to a broadening  of the various   excitations of the QD  that is energy \emph{independent} and $\sim \Gamma$. The nontrivial, energy-dependent corrections due to 
quantum fluctuations contained in the high-order contributions to $\bar{\Sigma}(z)$ are required.}
\item
  Second,
\HL{although the bare perturbation theory breaks down at these resonances as $T\rightarrow0$,
in the renormalized perturbation theory \eq{barSigma}, the low 
  energy cutoff $T$ is replaced by the imaginary parts $\sim \Gamma$ in $\bar{L}$.}
 Still, the resonance due to the Kondo effect causes even the renormalized perturbation theory to break down and three-loop corrections result in the 
  enhancement of Kondo exchange processes.
  These have been studied extensively using the RT-RG \HL{based
on an effective Kondo-model obtained by a Schrieffer-Wolff transformation from the Anderson model.~\cite{Schoeller09b}}
  In the regime of large applied bias voltage \HL{and / or} magnetic field
  such three-loop corrections can be neglected due to the dephasing of the Kondo correlations.
  This is the regime of interest in this work \HL{for which we expect} the one-plus-two-loop RG approach to give a good first 
approximation that deserves to be studied.
\item
  In order to study the Kondo effect in the Anderson model
  beyond the \Cite{Schoeller09b},
  at least a three-loop treatment is necessary.
  The renormalization of the one- and two-loop terms that we study here will then couple to the 3-loop terms and still play \HL{an} important role.
  Therefore, our study of the two-loop RG provides an important starting point for such a much more involved study, which in particular can address the low bias and low 
  magnetic field regime.
\end{itemize}

\subsection{Flow of Keldysh contractions: Continuous RG\label{sec:flow-Keldysh}}
In general, RG approaches to transport aim to eliminate the \emph{effect} of reservoirs states, starting at high energies, by incorporating it into a 
redefinition of the system parameters.
Typically, one eliminates \emph{the states themselves}.
Here, in contrast, we successively suppress the \emph{occupations of the states} using an RG procedure
while keeping the states.
Before we specify the details and advantages of this cutoff scheme,
we first outline the main idea of the functional renormalization group approach
when applied to the real-time perturbation series \eq{barSigma}.
By our causal reformulation of the perturbation theory (cf. \Sec{sec:causal}), it is clear that all the information about the occupations of the reservoir 
states is contained in the Keldysh components of \HL{the} correlation functions, i.e., in the $\bar{\gamma}$ contraction.
Therefore, we introduce a cutoff-dependent contraction function $\bar{\gamma}_\Lambda$ \HL{that} monotonously flows from the initial, full contraction 
function $\bar{\gamma}_\Lambda|_{\Lambda=\infty}=\bar{\gamma}$ given by \Eq{keld-contr}
to the trivial final function $\bar{\gamma}_\Lambda|_{\Lambda=0}=0$
where all occupations are suppressed.
During this flow, we demand that the effective Liouvillian remains invariant:
for every value of the cutoff parameter $\Lambda$,
\begin{align}
  L(z)
  &= \bar{L} + \bar{\Sigma} \left( \left\{ \bar{\gamma} , \bar{L}, \bar{G} \right\} \right)
  \\
  & = \bar{L}_{\Lambda} + \bar{\Sigma} \left( \left\{ \bar{\gamma}_\Lambda , \bar{L}_{\Lambda}, \bar{G}_\Lambda \right\} \right)
  \label{invariance}
  \\ 
  & = \left. \bar{L}_{\Lambda}\right|_{\Lambda=0}
  .
  \label{flowfinal}
\end{align}
Thus, $L(z)$ has the same functional dependence on the  contractions $\gamma_\Lambda$, the Liouvillian $\bar{L}_\Lambda$, and vertices $\bar{G}_\Lambda$.
The latter two now acquire a $\Lambda$ dependence to maintain invariance.
The same diagram rules are thus valid for any value of $\Lambda$.
As a result, at the end of the flow where $\bar{\gamma}_\Lambda|_{\Lambda=0}=0$, the effective Liouvillian is given simply by  $\bar{L}_{\Lambda}|_{\Lambda=0}$.
The information about the reservoir degrees of freedom, previously incorporated into the self-energy $\bar{\Sigma}$, has now been incorporated fully 
into the dot Liouvillian [cf. \Eq{flowfinal}].

\begin{figure}[tbp]
  \includegraphics[width=1.0\linewidth]{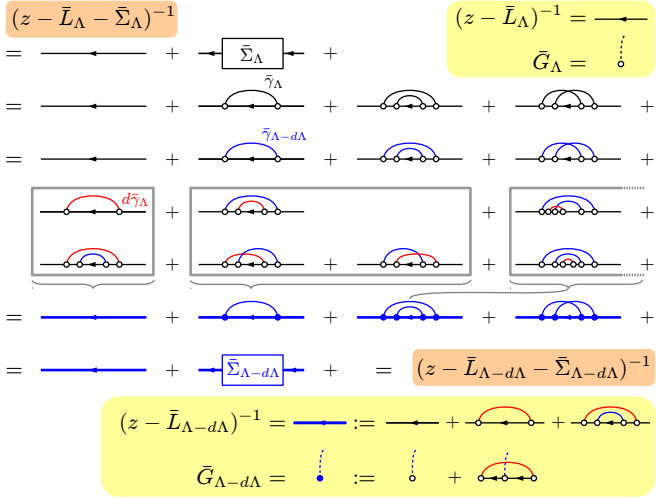}
  \caption{
    Renormalization group transformation with an infinitesimal change  $d\Lambda > 0$ of the flow parameter,
    $\Lambda \rightarrow \Lambda - d\Lambda$.
    The physical restriction is that the effective Liouvillian,
    $L(z)=\bar{L}_{\Lambda}+\bar{\Sigma}_{\Lambda}$,
    or equivalently, the density operator propagator $(z-L(z))^{-1}$ remains unchanged.
    In lines 1-2
    we start from the perturbation theory at scale $\Lambda$
    and split up the contraction function
    $\gamma_\Lambda$ (black curved lines)
    into the contraction function with reduced flow parameter ($\gamma_{\Lambda-d\Lambda}$, blue curved line),
    which should appear in the renormalized perturbation series
    and the change the contraction function ($-d\gamma_{\Lambda}$, red curve line).
    Next, in lines 3-5
    the latter terms of linear order in $d\gamma_\Lambda$ are collected into
    1- and 2-loop renormalizations $d\bar{L}_\Lambda$ of the Liouvillian in the resolvents
    and 1-loop renormalization $d\bar{G}_\Lambda$ of the vertices.
    Finally, the perturbation series is rewritten in terms of new Liouvillian $\bar{L}_{\Lambda - d\Lambda}$
    and vertices $\bar{G}_{\Lambda - d\Lambda}$ defined on the new scale $\Lambda-d\Lambda$ (indicated by blue).
    This transformation is exact if one also accounts for the generation of higher order vertices~\cite{Schoeller09a},
    which we, however, neglect here (they are not drawn).
    We do account for the renormalization of the original vertices, i.e., of \emph{single-charge fluctuations}.
  }
  \label{fig:rg-derivation}
\end{figure}

The RG flow is generated by making an infinitesimal change $d \Lambda$ of the cutoff,
resulting in a infinitesimal change
$d \bar{\gamma}_\Lambda \approx (d \bar{\gamma}_\Lambda/ d\Lambda) d \Lambda$
of the Keldysh contraction function.
In the perturbation series at scale $\Lambda$, one splits up the contraction function as
 $\bar{\gamma}_\Lambda
=\bar{\gamma}_{\Lambda-d\Lambda} +d\bar{\gamma}_\Lambda$
and collects all terms in a perturbation series of the same form but containing only $\bar{\gamma}_{\Lambda-d\Lambda}$ contractions.
In this process, the terms containing one infinitesimal contraction $d\bar{\gamma}_\Lambda$ can be identified with renormalizations $d\bar{L}$ of the Liouvillian, $d\bar{G}$ of the vertices, and newly generated higher-order vertices with more than one leg.
The process is illustrated in \Fig{fig:rg-derivation}.
The Liouvillian  $\bar{L}_{\Lambda-d\Lambda}=\bar{L}_{\Lambda}-d\bar{L}_{\Lambda}$,
 and vertices $\bar{G}_{\Lambda-d\Lambda}=\bar{G}_{\Lambda}-d\bar{G}_{\Lambda}$ of the new perturbation
series are then all defined for the new, lower cutoff scale $\Lambda-d\Lambda$.
We obtain differential equations for these quantities describing their renormalization as one continuously reduces the cutoff $\Lambda$.
These are the real-time renormalization group (RT-RG) equations.

A key requirement in setting up this continuous RG is that for any $\Lambda$ the zero-eigen vector of $\bar{L}^\Lambda$ does not appear in the resolvents $(z+X-\bar{L}^\Lambda)^{-1}$,
to avoid divergences as function of the frequencies.
The RG thus has to be formulated such that the property \eq{x0G} of the vertices in the causal representation is preserved.
This can be shown to be the case [see \Eq{GZ_L} below].

The final key point is the choice of a cutoff-dependent distribution function in the contraction $\bar{\gamma}$.
The numerical integration of the RG equations is more stable when we introduce a contraction function with a cutoff on the \emph{imaginary frequency axis},
\begin{align}
  \label{cutoff} 
  \bar{\gamma}_{12,\Lambda} (\eta\omega) =
  & \delta_{1\bar{2}}\frac{\Gamma}{\pi} T\sum_{l=0}^\infty
  \frac{ \Theta_{T} (\Lambda - |\omega^l| ) }{\eta\omega-\bar{\mu}_r -i\omega^l} 
  ,
\end{align}
through the function
\begin{align}
  \Theta_{T} (\omega)  = 
  \begin{cases}
      \Theta (\omega) &   |\omega|>\pi T \\
      \frac{1}{2}+\frac{\omega}{2\pi T} & |\omega|<\pi T
  \end{cases}
  ,
\end{align}
where $\Theta(\omega) $ is the \HL{step}-function and
\begin{align}
  \omega^l = (2l+1) \pi T
  \label{Mat_fr}
\end{align}
is the $l$-th Matsubara frequency ($l=0,1,2,\ldots$).
In the limit $\Lambda \rightarrow +\infty$ we 
recover the partial fraction expansion of the meromorphic function $(\Gamma /{2\pi})\tanh (\omega/2T)$ as required.
Imposing this cutoff in Matsubara space leads to a suppression of the tails of $\bar{\gamma}^\Lambda(\eta\omega)$ on the real frequency axis 
as $\Lambda \rightarrow 0$
 rather than sharp truncation above frequency $\Lambda$.
This implements the suppression of contributions from states above energy scale $\Lambda$.
From here on, we will consider the zero-temperature limit $T\rightarrow 0$ for which the contraction function \eq{cutoff} reduces to the simple form:
\begin{align}
\label{cutoffT0}
\bar{\gamma}_{12,\Lambda} (\eta\omega) =
\delta_{1\bar{2}} \frac{\Gamma}{\pi} \int_{-\Lambda}^{\Lambda} d\omega' \frac{1}{\eta\omega-\bar{\mu}_r-i\omega'}
.
\end{align}
See \Cite{Schoeller09a} for a detailed discussion.\footnote{Our definition of the contraction $\bar{\gamma}_\Lambda$ differs from that in 
\Cite{Schoeller09a} by \HL{a} factor $-2\Gamma$ due to our
\HL{(i)} our sign convention for the $\eta$ index$^{\byhand{67}}$, cf.~\Eq{etadef} \HL{(factor $-1$)},
\HL{(ii)} our normalization of the \emph{field} superoperators$^{\byhand{76}}$ such that $[\bar{G}_{\bar{1}},\tilde{G}_1]_{+}=1$ \HL{(factor $2$)}, and
\HL{(iii)} our inclusion of the coupling into the reservoir fields \HL{(factor $\Gamma$)}.}

\begin{figure}[tbp]
  \includegraphics[width=1.0\linewidth]{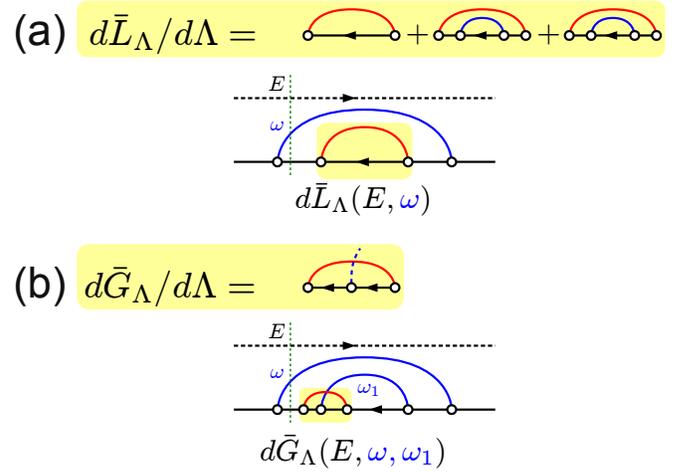}
  \caption{
    RG-equations and frequency dependence
    generated by the renormalization transformation of the diagrammatic perturbation theory in \Fig{fig:rg-derivation},
    (using the same red and blue colors).
    (a) Liouvillian renormalization by one- and two-loop corrections.
    The example diagram illustrates that the correction depends
    on the external frequency $E$ of the diagram (Laplace variable)
    and on the sum of the reservoir frequencies $\omega$ running over the $d\bar{L}/d\Lambda$ block (marked yellow).
    These frequencies are read off at the vertical cut (green dashed line) to the left of this block.
    (b) Vertex  renormalization by one-loop corrections.
    The example diagram illustrates that the vertex correction depends additionally on frequency $\omega_1$
    of the vertex leg.
  }
  \label{fig:rg-freqdep}
\end{figure}

\subsection{RG in frequency space\label{sec:rgfreq}}

\subsubsection{Non-equilibrium Matsubara representation
and frequency dependence\label{sec:Non_Matsu}}

To formulate the RG equations, we need a more compact notation for the various frequencies.
Since the contraction functions $\bar{\gamma}^\Lambda$ depend on $\eta \omega=x-\eta\mu$, we re-express the reservoir energies in the resolvents in 
$\bar{\Sigma}$ [\Eq{barSigma}]
in explicit calculations as
\begin{align}
  E-X_i =  E-x_{1...n} = E_{1...n}-\bar{\omega}_{1...n}
  \label{E-X}
  .
\end{align}
Here, $1,..,n$ are the multiindices of the contractions going over diagram segment $i$.
The frequencies are now taken relative to the electrochemical potentials and we write their sums as repeated multiindices:
for \HL{$k=\eta_k,\sigma_k,r_k,\omega_k$,}
\begin{align}
  x_k  & = \bar{\omega}_k+\bar{\mu}_{k},
  &
  x_{1...n} &= x_1+\ldots+x_n,
  \\ 
  \bar{\omega}_k & = \eta_k\omega_k,
  &
  \bar{\omega}_{1...n} &= \bar{\omega}_1 + \ldots+\bar{\omega}_n,
  \\
  \hl{ \bar{\mu}_{{k}} }  &= \eta_k {\mu}_{_k}
  &
  \bar{\mu}_{1...n} & = \bar{\mu}_1+\ldots  +\bar{\mu}_n.
\end{align}
Similarly, we express the dot energies relative to these chemical potentials as
\begin{align}
  E_{1...n}=E-\sum\limits_{i=1...n}\bar{\mu}_i
  .
\end{align}

A key advantage of the  cutoff parametrization \eq{cutoff} is that it allows us to
analytically perform all the integrations over the reservoir frequencies $\bar{\omega}$  in \Eq{barSigma} by closing each integration contour in the complex lower half-plane
and \HL{by} using the residual theorem.
As a result, we can replace all integrations over real frequencies in the resolvents by summations over Matsubara frequencies lying in the lower half-plane: \Eq{E-X} becomes
\begin{align}
  E_{1...n}+i\omega_{1...n}
  \label{dot_fr}
  .
\end{align}
For finite $T$ the $\omega_k$-sum runs over the positive discrete frequency values $(2l_k+1)\pi T$, $l_k=0,1,2,..$, which turns into an integral over positive $\omega_k$ for $T=0$.\footnote{The positive imaginary part (integration contour in the lower half-plane) contrasts with \Cite{Schoeller09a}, but the final notation \Eq{dot_fr} agrees again with \Cite{Schoeller09a} \HL{due to compensating differences in the notation.}}
As always, we do not explicitly indicate this integral.

\subsubsection{RG equations\label{sec:rgeq}}

During the RG flow, the Liouvillian develops a non-trivial dependence on both the real energy $E$ of the QD
and on \HL{$i\omega$, the sum of the imaginary frequencies of all the reservoir-contractions} that pass over the propagator in a diagram.
This is illustrated in \Fig{fig:rg-freqdep}(a).
We separately keep track of these frequency dependencies by writing
\begin{align}
  \bar{L}_\Lambda (E,\omega) &:= \bar{L}_\Lambda (E+i\omega),
  \\
  \Pi_\Lambda (E,\omega) &:= \frac{1}{E+i\omega - \bar{L}_\Lambda (E,\omega)}
  .
  \label{resolvent}
\end{align}
The renormalization of the vertex $\bar{G}_1$ introduces similar dependencies and an additional dependence on the reservoir frequency $\omega_1$ of the vertex ``leg'', i.e., the contraction connecting it to another vertex, see \Fig{fig:rg-freqdep}(b):
\begin{align}
  \bar{G}_{1,\Lambda} (E,\omega,\omega_1)
  :=
  \bar{G}_{1,\Lambda} (E+i\omega,i\omega_1)
  .
\end{align}

The formally exact, infinite hierarchy \HL{of} RT-RG equations \HL{was} derived in general form in \Cite{Schoeller09a}.
Here, we restrict ourselves to  the one- and two-loop order approximation for the Liouvillian and the limit of $T\rightarrow 0$:
\begin{widetext}
\begin{align}
  \label{RG-L}
  &\frac{d \bar{L} (E,\omega)}{d \Lambda} =
  i\frac{\Gamma }{\pi}
  \bar{G}_1 (E,\omega,\Lambda)
  \Pi (E_1,\omega+\Lambda)
  \bar{G}_{\bar{1}} (E_1,\omega+\Lambda,-\Lambda)
  +\frac{\Gamma^2  }{\pi^2} \times
  \\ \nonumber
  & \bar{G}_1 (E,\omega,\Lambda)
  \Pi (E_1,\omega+\Lambda)
  \bar{G}_2 (E_1,\omega+\Lambda,\omega_2)
  \Pi(E_{12},\omega+\Lambda+\omega_2)
  \bar{G}_{\bar{2}} (E_{12},\omega+\Lambda+\omega_2,-\omega_2)
  \Pi (E_1,\omega+\Lambda) \bar{G}_{\bar{1}} (E_1,\omega+\Lambda,-\Lambda)
  .
\end{align}
This approximation  requires that one also accounts for the renormalization of vertices to one-loop order:
\begin{align}
  \label{RG-G}
  &\frac{d\bar{G}_1 (E,\omega,\omega_1)}{d\Lambda}
  =
  -i \frac{\Gamma}{\pi}
  \bar{G}_2 (E,\omega,\Lambda)
  \Pi (E_2,\omega+\Lambda)
  \bar{G}_1 (E_2,\omega+\Lambda,\omega_1 )
  \Pi ( E_{12},\omega+\omega_1 +\Lambda)
  \bar{G}_{\bar{2}} ( E_{12},\omega+\omega_1+\Lambda,-\Lambda)
  .
\end{align}
\end{widetext}
\HL{Here} and in the following, we leave implicit \HL{both} the $\Lambda$ dependence of the renormalized  $\bar{L}$ and $\bar{G}$ \HL{as well as} the summation / integration over all internal indices.
\footnote{\Eq{RG-L} and \eq{RG-G} differ from those in \Cite{Schoeller09a} by a factor 2 for each loop due to our different normalization$^{\byhand{74}}$
of $\bar{G}$ (reduced by factor $\sqrt{2}$)
\HL{The} sign difference in \HL{the} definition of the cut-off function$^{\byhand{74}}$ is compensated by \HL{our} closing \HL{of the integration-contour}$^{\byhand{75}}$ in \HL{the} lower (instead of the upper) half-plane.}

By construction, the self-energy $\bar{\Sigma}(z)$ is obtained by directly integrating \Eq{RG-L}:
\begin{align}
  \bar{\Sigma}(E+i\omega)
  = L(E+i\omega)-\bar{L}
  = \int_\infty^0 d \Lambda \frac{d \bar{L}_\Lambda}{d\Lambda}(E,\omega)
  \label{barSigmaintegral}
  .
\end{align}
For the transport current we need \HL{the} reservoir-resolved parts $\bar{\Sigma}^r(z)$ of this self-energy [cf. \Eq{Sigmadecomp}], which \HL{are} obtained in the same way:
for $r=L,R$
\begin{align}
  \bar{\Sigma}^r(E+i\omega)
  = L^r(E+i\omega)-\bar{L}^r
  = \int_\infty^0 d \Lambda \frac{d \bar{L}_\Lambda^r}{d\Lambda}(E,\omega)
  \label{barSigmaintegralr}
  ,
\end{align}
where
$\bar{L}^r_\Lambda|_{\Lambda=\infty} = \bar{L}^r :=\tilde{\Sigma}^r$
is given by \Eq{tildeSigma}.
The RG equations determining $\bar{L}_\Lambda^r$ and its associated vertex $G^r$
have the same structure as the RG equations \Eq{RG-L}-\eq{RG-G}
and are simply obtained from the latter~\cite{Schoeller09a}
by providing the left-most vertex \HL{with a superscript $r$}
on the right-hand sides of \Eq{RG-L} and \eq{RG-G} and \HL{on} the left-hand side of \Eq{RG-G}.
We emphasize that 
$
\sum_r \bar{L}^r =\tilde{\Sigma} = \bar{L}-L
,
$
\HL{whereas}
$
\sum_r \bar{L}^r_\Lambda|_{\Lambda=0} = \Sigma(z) = L(z)-L
$.
\footnote{\HL{For the density operator time-evolution, we need the Liouvillian $L$,}
$
L+\Sigma(z)
 = \bar{L} + \bar{\Sigma}(z)
 = \bar{L}_\Lambda + \bar{\Sigma}_\Lambda(z)
$,
but for the current \HL{we do not}:
$
\Sigma^r(z)
 = \tilde{\Sigma}^r + \bar{\Sigma}^r(z)
 = \bar{L}_\Lambda^r + \bar{\Sigma}_\Lambda^r(z)
$}

The frequency dependence in the RG equations \Eq{RG-L}-\eq{RG-G} in the one and two-loop approximation needs to be carefully discussed.
Before we turn to this in \Sec{sec:1loop}-\sec{sec:2loop},
we discuss \HL{some} important general properties of the RG-equations
 in \Sec{sec:eigenvectors}
and their implications for the fermionic eigenvalues in \Sec{sec:fermion}.
\key{
Using these results, we can decouple some of the RG equations (see \Sec{sec:1loop})
and show that in the one-loop approximation we already obtain the exact solution for the current in the limit $U=0$,
even though in this limit two-loop corrections are \emph{non-zero} (see \Sec{sec:nonint}).
}

\subsubsection{Exact eigenvectors\label{sec:eigenvectors}}

\hl{The exact properties derived in \Sec{sec:causalstruct} follow from the causal structure of the perturbation series \eq{barSigma}.
Since the RG equations \eq{RG-L} and \eq{RG-G} represent nothing but a reorganization of the terms in that expansion
(as explicitly shown in \Cite{Schoeller09a}),
we expect that they preserve these causal structure properties.
We now show that this is indeed the case.}
Indeed, an exact left and right eigen-supervector of the \emph{renormalized} $\bar{L}$ is given by
\begin{align}
  (Z_L| \bar{L}(E,\omega) &= 0,
  \label{LZ_L}
  \\
  \bar{L}(E,\omega) |Z_R) &= -i 4\Gamma |Z_R)
  \label{LZ_R}
  ,
\end{align}
respectively.
Since at $\Lambda=\infty$ we have the initial condition
$
(Z_L| \bar{L}
=0$
and
$ \bar{L}|Z_R)=-i4\Gamma|Z_R)$,
we only need to show that
$
(Z_L| \tfrac{d}{d\Lambda} \bar{L}
=
\tfrac{d}{d\Lambda} \bar{L} |Z_R) =0
$,
respectively,
to prove Eqs. \eq{LZ_L} and \eq{LZ_R}.
We note that since \Eq{barSigmaintegral} is an exact relation,
Eqs.~\eq{LZ_L} and \eq{LZ_R} must hold for the exact, infinite hierarchy of RT-RG equations
(i.e., including all higher vertices generated during the RG flow that we neglect here).
In our two-loop approximation this relation directly follows
by letting \Eq{RG-L} act on these vectors and using 
\begin{align}
  (Z_L|\bar{G}(E,\omega,\omega_1) & =0
  \label{GZ_L}
  ,
  \\
  \bar{G}(E,\omega,\omega_1) |Z_R) & = 0
  \label{GZ_R}
  .
\end{align}
Equations \eq{GZ_L} and \eq{GZ_R} follow by assuming that they hold for a given $\Lambda$,
and by
$
(Z_L| \tfrac{d}{d\Lambda} \bar{G}
=
\tfrac{d}{d\Lambda} \bar{G} |Z_R) =0$,
obtained  acting with \Eq{RG-G} on these vectors.
Since Eqs. \eq{GZ_L} and \eq{GZ_R} hold initially for $\Lambda=\infty$ the result follows
for any $\Lambda$, $E$, $\omega$, and $\omega_1$.

Similarly, we now show that for any $\Lambda$, $E$, $\omega$, and $\omega_1$:
\begin{align}
  \tfrac{d }{d\Lambda}\bar{L}(E,\omega) |\alpha_{\eta \sigma}^-) &= 0,
  \label{alpha-0L}
  \\
  (\alpha_{\eta \sigma}^+| \tfrac{d }{d\Lambda} \bar{L}(E,\omega) &= 0,
  \label{alpha+0L}
  .
\end{align}
In our two-loop approximation for $d\bar{L}/d\Lambda$, this follows from the property of the renormalized 1-leg vertices
\begin{align}
  \bar{G}(E,\omega,\omega_1) |\alpha_{\eta \sigma}^-) & \propto |Z_R) \text{ or 0}
  \label{alpha-0G}
  ,
  \\
  (\alpha_{\eta \sigma}^+|\bar{G}(E,\omega,\omega_1) & \propto (Z_L| \text{ or 0}
  \label{alpha+0G}
  .
\end{align}
The proof of \Eq{Galpha0}-\eq{alphaG0} can be extended to the \emph{renormalized} vertices as follows.
We start by observing that the right-hand sides of the RG equations \eq{RG-L}-\eq{RG-G} have the same structure as $\bar{\Sigma}(z)$ (cf. \Eq{barSigma}).
Assuming that \Eq{alpha-0G} holds for a given scale $\Lambda$,
the RG equation \eq{RG-G} implies that it is preserved under the flow:
$
(\alpha_{\eta \sigma}^+| \frac{d}{d\Lambda}\bar{G}
=
\frac{d}{d\Lambda}\bar{G}  |\alpha_{\eta \sigma}^-)
= 0
$.
Here, we used that by \Eq{LZ_L}-\eq{LZ_R} both $|Z_R)$ and $(Z_L|$ are eigen-supervectors of the \emph{renormalized} $\bar{L}$ for all $\Lambda,E,\omega,\omega_1$.
Since \Eq{alpha-0G}-\eq{alpha+0G} hold initially for $\Lambda=\infty$, this then implies it holds for all $\Lambda$.
From this, \Eq{alpha-0L}-\eq{alpha+0L} follow directly.
The above proofs are readily extended to the infinite hierarchy of \emph{exact} RT-RG equations for vertices with multiple legs, confirming that \Eq{LZ_L}, \eq{LZ_R}, \eq{alpha+0L} and \eq{alpha-0L} hold exactly (and not just in our two-loop approximation).

\subsubsection{Fermionic excitations\label{sec:fermion}}

\paragraph{Fermionic eigenvalues}
We can now pick up the discussion of \Sec{sec:proj}.
Since the supermatrix structure of $d\bar{L}_\Lambda/d\Lambda$ in the fermionic sector is preserved under the RG flow and is the same as that of $\bar{\Sigma}(z)$,
 we can now directly relate the coefficient $\Delta F^{-+}_{\eta\sigma}$ introduced in \Sec{sec:fermion-pt} 
using \Eq{barSigmaintegral}:
\begin{align}
  \Delta F^{-+}_{\eta\sigma}(E+i\omega)
  = \int_\infty^0 d \Lambda \frac{d F^{-+}_{\eta\sigma,\Lambda}}{d\Lambda}(E+i\omega)
  \label{DeltaFrg}
\end{align}
This coefficient determines the fermionic excitations at arbitrary complex frequency as given by \Eq{ferm-spec} for the \emph{exact} $L(z)$
when the infinite hierarchy of RG-equations is used to compute the right-hand side.
We see that the $\Lambda$-dependent coefficient $F^{-+}_{\eta\sigma,\Lambda}$ of $\bar{L}_\Lambda$
interpolates between the infinite temperature limit, where $\Delta F^{-+}_{\eta\sigma}=0$,
and the exact value $\Delta F^{-+}_{\eta\sigma}$ of $\Sigma(z)$ through \Eq{DeltaFrg} as was anticipated in \Sec{sec:proj}.
All renormalization effects enter into the fermionic excitations through the renormalization of the four complex coefficient $F_{\eta,\sigma}^{-+}$ of $\bar{L}_\Lambda$.
During this flow, the qualitative features of these excitations,
discussed in \Fig{fig:fermion-level}, 
may change.
During the continuous RG, the complex parameters $\Delta F_{\eta,\sigma}^{-,+}$ will grow from zero
and modify both real and imaginary parts in \Eq{ferm-spec}.
This happens only for the interacting Anderson model, $U \neq 0$ since $U$ multiplies these coefficients in \Eq{ferm-spec}.
This flow  may include bifurcations as function of the flow parameter $\Lambda$,
but for large enough $U \gg 2\Gamma$ the excitations energies (real parts) remain non-degenerate.
However, the general result \eq{symmetric_spec} shows that during this non-trivial flow the \emph{average} of the complex eigenvalues stays fixed for all frequencies.
We conclude generally that \emph{the fermionic excitation energies and decay rates are renormalized symmetrically} with respect to the average values
$\epsilon+\frac{U}{2}+\sigma\frac{B}{2}$ and $2\Gamma$, respectively,
for any complex frequency $E+i\omega$ with $\omega > 0$.

Finally, we note that the stability constraint discussed in \Sec{sec:fermion-pt} imposes a constraint on the RG flow:
since at \emph{any} stage of the RG flow the effective Liouvillian $L(z)$ \eq{invariance} can be calculated from the perturbative expansion \eq{barSigma},
the imaginary parts of all non-zero eigenvalues of the renormalized $\bar{L}_\Lambda$ must be  negative to avoid unphysical divergence of the time-dependent density operator.
Such behavior would not go unnoticed in the RG since zero denominators would appear in the resolvents in \Eq{RG-decomposition},
\HL{leading to an instability in the RG-flow.}
This provides a simple criterion for the stability of the RG flow for the Anderson model that can be checked easily
in numerical approximations. Although in previous applications of the RT-RG no instabilities have been reported, and in the present study none were
encountered either, the general conditions for stability are currently not known.

\paragraph{Fermionic supermatrix elements}
The properties \Eq{alpha-0G}-\eq{alpha+0G} strongly restrict the fermionic matrix elements of \HL{the} resolvents $\Pi$ that can appear in the RG equations.
We will see that implies that quite generally the RG equations decouple into smaller sets of equations (see \Sec{sec:1loop}) and that important simplifications arise in the $U=0$ limit (see \Sec{sec:nonint}).
These simplifications arise since in general on the right-hand side of RG equations such as \Eq{RG-L} and \eq{RG-G} the resolvent $\Pi$ always appears sandwiched between pairs of $\bar{G}$ vertices (all are renormalized quantities but $\Lambda$ is not written).
\HL{We list the different cases:}

\begin{itemize}
\item
  In matrix elements of terms with only one resolvent,
  $(X|\bar{G}\Pi\bar{G}|Y)$,
  there are no restrictions only if
  $X=Z_R, Y=Z_L$.
  Indeed, upon inserting the completeness relation
  $
  1 = \sum_{\eta,\sigma,\nu}  |\alpha_{\eta \sigma}^\nu)(\alpha_{\eta \sigma}^\nu|+
  \text{(bosonic terms)}
  $
  left and right of $\Pi$,
  we see that according to \Eq{alpha-0G}-\eq{alpha+0G},
  all intermediate fermionic supermatrix elements contribute.
  However, when the basis supervectors $X$, $Y$ involve one \HL{of the supervectors $Z_L, Z_R$, then}
  only certain matrix elements contribute: for $\nu=\pm$ \HL{these are}
  \begin{align}
    (\alpha_{\eta \sigma}^\nu|\Pi^\alpha|\alpha_{\eta \sigma}^-),
    &  &X=Z_R,    &Y\neq Z_L
    \label{alpha-factor1}
    ,
    \\
    (\alpha_{\eta \sigma}^+|\Pi^\alpha|\alpha_{\eta \sigma}^\nu),
    &  &X\neq Z_R,& Y=Z_L
    \label{alpha-factor2}
    ,
  \end{align}
  whereas, if \HL{the supervectors $Z_L, Z_R$ are not involved,} only one factor is possible:
  \begin{align}
    (\alpha_{\eta \sigma}^+|\Pi^\alpha|\alpha_{\eta \sigma}^-),&  &X\neq Z_R,&Y\neq Z_L
    .
    \label{alpha-factor3}
  \end{align}
\item
  In terms with $n \geq 3$ resolvents
  $(X|\bar{G}\Pi\bar{G} \ldots \bar{G}\Pi\bar{G}|Y)$,
  \Eq{alpha-factor1} and \eq{alpha-factor3} apply
  to the leftmost ``boundary'' resolvent.
  Otherwise, the expression vanishes for any $\Lambda$
  since by \Eq{alpha+0G}
  $
  (X|\bar{G} \Pi |\alpha^{+})(\alpha^{+}| \bar{G}\Pi\bar{G} \ldots
  \propto
  (X|\bar{G} \Pi |\alpha^{+}) (Z_L|\Pi\bar{G}  \ldots
  =0
  $.
  Here, we used that $(Z_L|$ an exact eigenvector of $\bar{L}$,
  and thereby of $\Pi$ by \Eq{LZ_L}, 
  and a zero eigenvector of $\bar{G}$ by \Eq{GZ_L}.
%\item
  \newline
  Similarly,
  \Eq{alpha-factor2} and \eq{alpha-factor3} \HL{also} apply to
  the rightmost ``boundary'' resolvents
  since $\ldots \bar{G} \Pi \bar{G}\Pi |\alpha^{-})(\alpha^{-}| \bar{G}|Y) =0$
  by \Eq{alpha-0G}, \eq{LZ_R}, and \eq{GZ_R}.
\item
  Finally, in terms
  with $n \geq 3$ resolvents,
  the resolvents \HL{that} are not at the boundary
  can \emph{only} contribute with fermionic matrix element
  $ (\alpha_{\eta \sigma}^+|\Pi^\alpha|\alpha_{\eta \sigma}^-)$, irrespective of $X$ and $Y$:
  this factor \emph{must always occur} at least $n-2\geq 1$ times.
\end{itemize}

\subsection{1 loop RG equations\label{sec:1loop}}
\subsubsection{Frequency dependence\label{sec:freq-dep1}}
Since our goal is to calculate the stationary state from the effective Liouvillian
$L(z) =\left. \bar{L}(E,\omega)\right|_{\Lambda=0}$ at $z=E+i\omega =i0$,
we first consider  the RG equation for this quantity in the one-loop approximation and at frequency $E=0$
\begin{align}
  \label{1loop}
  \frac{d \bar{L}^0 (0)}{d \Lambda} =
  i\frac{\Gamma }{\pi}
  \bar{G}_1^{0}
  \Pi^0(\bar{\mu}_1,\Lambda)
  \bar{G}_{\bar{1}}^0
  .
\end{align}
 Here, the superscript 0 indicates that we also evaluate the Liouvillian at zero reservoir frequency, $\omega=0$: $\bar{L}^0 (E) := \bar{L} (E,0)$.
Similarly, in \Eq{1loop} we approximate the vertices by their initial values, as given by Eqs. \eq{g-b+} and \eq{g-b-}\HL{, or by} \Eq{kr_d},
\begin{align}
  \bar{G}_1 (0,0,\Lambda)
  \approx
  \bar{G}_{\bar{1}} (\bar{\mu}_1,\Lambda,-\Lambda)
  \approx
  \bar{G}_1^0
  \label{freq-neglect}
\end{align}
neglecting their dependence on the dot frequency ($E_1=\bar{\mu}_1$), the reservoir frequency ($\omega=\Lambda$), and the vertex-leg frequency ($\omega_1=\pm \Lambda$).
Such frequency dependencies arises only when accounting for the renormalization of the vertices: for small frequencies we can approximate in \Eq{RG-G}
\begin{align}
\label{H-approximation}
  \frac{d\bar{G}}{d\Lambda} \sim \frac{\Gamma}{\Lambda^2} \bar{G}^3
\end{align}
on the right-hand side $\bar{G} \sim \bar{G}^0 \sim 1$, giving
 $\bar{G}=\bar{G}^0+O(\Gamma/\Lambda)$.
In 1-loop order for $\bar{L}$ one must therefore consistently neglect the renormalization of $\bar{G}$, \Eq{freq-neglect},
with respect to the log-corrections to the Liouvillian that arise from \Eq{1loop}.
This will be checked later on.
The resolvent is likewise evaluated at $\omega=\Lambda$. Note that \HL{the} $\omega$ dependence of the resolvent in first approximation,
\begin{align}
  \Pi^0 (E,\omega)  = \frac 1{E+i\omega-\bar{L}^0 (E)}
  ,
\end{align}
does not originate from the Liouvillian.

In contrast, \Eq{1loop} depends on QD frequency $E$ in an important way: Due to the finite bias voltage the renormalization of the zero $E$-frequency 
Liouvillian couples to the  finite frequency Liouvillian $E \rightarrow E_1=\bar{\mu}_1=\eta_1 r_1 V/2$
appearing in the \eq{resolvent} on the right-hand side.
We therefore need to consider instead the \HL{following} RG equations
\HL{on a discrete grid of \emph{finite} QD frequencies
$E=k_L \bar{\mu}_L+ k_R \bar{\mu}_R=\eta(k_L-k_R)V/2$:}
\begin{align}
  \label{1loopE}
  \frac{d \bar{L}^0 (E)}{d \Lambda} =
  i\frac{\Gamma }{\pi}
  \bar{G}_1^0
  \Pi^0 (E_1,\Lambda)
  \bar{G}_{\bar{1}}^0,
\end{align}
\HL{where $k_L,k_R=0,1,2,\ldots$.}
In the numerical calculations, we keep as many equations as required to make the solution converge with respect to the $k_r$.
This coupling of the RG flow of the Liouvillian at energies differing by multiples of the voltage arises
because the Matsubara frequencies of the different reservoirs are shifted by different electrochemical potentials:
this is typical feature of renormalization in a non-equilibrium system~\cite{Schoeller09a}.
We discuss the effect of neglecting the QD energy $E$-dependence in the RG equations in detail when we analyze the numerical results in \Sec{sec:set}.

\subsubsection{Explicit 1-loop RG-equations for the Liouvillian}
Inserting the spectral decomposition for $\bar{L}(E)$, cf. \Eq{eigenproj2},
into \Eq{1loopE}
\begin{align}
  \frac{d\bar{L}^0 (E)}{d\Lambda}
  =
  \frac{\Gamma}{\pi}
  \frac{1}{\Lambda-i\Theta^k_1}
  \bar{G}^0_1 P^k_1\bar{G}^0_{\bar{1}}
  ,
\label{RG-decomposition}
\end{align}
we obtain, abbreviating $\Theta^k_1 = E_1 - \lambda^k(E_1)$,
\begin{align}
  \frac{d
  }{d\Lambda}
  (\kappa_3| \bar{L}^0 (E) |\kappa_0)
  = \frac{\Gamma}{\pi}
  (\kappa_3| \mathcal{M}_{} |\kappa_0)
  ,
  \label{1loopexplicit}
\end{align}
where $|\kappa_i)$ are elements of the basis \eq{basis-boson}-\eq{basis-fermion} and the super matrix elements are sums ($\kappa_{1,2}$ sums implicit) 
of factored contributions:
\begin{align}
  &(\kappa_3| \mathcal{M}_{} |\kappa_0)
  =
  &
  i
  \sum_i
  (\kappa_3| \bar{G}^0_1 |\kappa_2)
  (\kappa_1| \bar{G}^0_{\bar{1}} |\kappa_0)
  (\kappa_2| {\Pi}^i_1 |\kappa_1)
  .
\end{align}
The product of $\bar{G}^0$ matrix elements gives a simple numerical factor $0$ or $\pm 1$ (see expansions \eq{g-b+}-\eq{g-b-}),
whereas the super matrix elements
\begin{align}
  i
 (\kappa_2| {\Pi}^{i}_1 |\kappa_1)
 =
  \frac{1}{\Lambda-i\Theta^i_1}
  (\kappa_2| {P}^i_1 |\kappa_1)
\end{align}
arise from the spectral decomposition of the resolvent,
$
\Pi^0(E_1,\Lambda) = \sum_i \Pi^{i}_1
$.
\hl{To explicitly sum over $\eta$ (contained in the multiindex $1=\eta,\sigma,r$),
which enters the resolvents only through
$
E_{\eta r} := E_1 = E-\eta\mu_r
$,
we abbreviate
$\lambda^i(E_1):=\lambda_{\eta r}^{\hl{i}}$,
$P^i(E_1):=P_{\eta r}^{\hl{i}}$
and}
\begin{align}
\hl{
  \label{Pii}
  i \Pi^i_{\eta r}  := \frac { P^i_{\eta r} }{\Lambda -iE_{\eta r}+i\lambda^i_{\eta r}}
  .
  }
\end{align}
Expanding \Eq{1loopE} in \HL{the} basis \Eq{basis-boson}-\eq{basis-fermion}, we obtain:
\begin{widetext}
\begin{align}
  &\frac{d \bar{L}^0(E)}{d \Lambda}
  =
  \frac{i\Gamma}{\pi}\Big[
  - \Big(
    (\alpha_{- \bar{\sigma}}^-| \Pi_{-r}^{\alpha_{- \bar{\sigma}}} |\alpha_{- \bar{\sigma}}^+) 
    + 
    (\alpha_{+ {\sigma}}^-   | \Pi_{+r}^{\alpha_{+ {\sigma}}}   |\alpha_{+ {\sigma}}^+)
  \Big)
  |Z_R)(Z_L|
  + \Big(
     (\alpha_{- \sigma}^-     | \Pi_{-r}^{\alpha_{- \sigma}}      |\alpha_{- \sigma}^-)
    -(\alpha_{+ \bar{\sigma}}^-| \Pi_{+r}^{\alpha_{+ \bar{\sigma}}} |\alpha_{+ \bar{\sigma}}^-)
  \Big)
  |Z_R)(\chi_\sigma|
  \nonumber
  \\
  &+
  \Big(
    (\alpha_{- \bar{\sigma}}^{+}| \Pi_{-r}^{\alpha_{- \bar{\sigma}}} |\alpha_{- \bar{\sigma}}^{+})
    -
    (\alpha_{+ {\sigma}}^{+}   | \Pi_{+r}^{\alpha_{+ {\sigma}}}     |\alpha_{+ {\sigma}}^{+})
  \Big) 
  |\chi_\sigma)(Z_{L}|
  -
  \Big(
    (\alpha_{+ {\sigma}}^{+}   | \Pi_{+r}^{\alpha_{+ {\sigma}}} |\alpha_{+ {\sigma}}^{-})
    +
    (\alpha_{- \bar{\sigma}}^{+}| \Pi_{-r}^{\alpha_{+ {\sigma}}} |\alpha_{- \bar{\sigma}}^{-})
  \Big)
  |\chi_{\sigma})(\chi_{\bar{\sigma}}|
 \nonumber
  \\ 
  &+
  \Big(
    (\alpha_{+ {\sigma}}^{+}| \Pi_{+r}^{\alpha_{+ {\sigma}}} |\alpha_{+ {\sigma}}^{-})
    +
    (\alpha_{- {\sigma}}^{+}| \Pi_{-r}^{\alpha_{- {\sigma}}} |\alpha_{- {\sigma}}^{-})
  \Big)
  |S_\sigma)(S_\sigma|
  -
  (\alpha_{+ \bar{\sigma}}^{+}| \Pi_{-r}^{\alpha_{+ \bar{\sigma}}} |\alpha_{+ \bar{\sigma}}^{-})
  |T_+)(T_+|
  -
  (\alpha_{- {\sigma}}^{+}    | \Pi_{+r}^{\alpha_{- {\sigma}}}    |\alpha_{- {\sigma}}^{-})
  |T_-)(T_-|
  \nonumber \\
  &-\left(
    (\chi_{\bar{\sigma}} | \Pi_{-r}^{\chi}       |\chi_\sigma)
    +
    (T_{+}             | \Pi_{+r}^{T_{+}}       |T_{+})
    -
    (S_{\sigma}         | \Pi_{-r}^{S_{\sigma}}  |S_{\sigma})
  \right)
  |\alpha_{+ \sigma}^{-})(\alpha_{+ \sigma}^+|
  \nonumber
  \\ 
  &
  -\left(
    (\chi_{{\sigma}} | \Pi_{+r}^{\chi}       |\chi_{\bar{\sigma}})
    +
    (T_{-}          | \Pi_{-r}^{T_{-}}      | T_{-})
    -
    (S_{\sigma}      | \Pi_{+r}^{S_{\sigma}} |S_{\sigma})
  \right)
  |\alpha_{- \sigma}^-)(\alpha_{- \sigma}^+|
  ~~\Big]
 \label{1loopRG}
 .
   \end{align}
\end{widetext}
Here, we leave implicit the  summation over $\sigma$ and $r$,
as well as the summation over the two eigenvalues in the $\chi$ and $\alpha_{\eta\sigma}$ subspaces [cf. \Eq{eigenproj2}].
The first two terms in the equation do not contribute to the calculation of the remaining terms of the effective Liouvillian or the transport current,
but \HL{they} are written here for completeness.
Due to $(Z_{L}|\bar{G}=0$ [cf. discussion of \Eq{x0G}], the zero eigen projector $P^{Z_L}$ does not appear in \Eq{1loopRG}, ensuring that none of 
the resolvents can diverge during the RG-flow.

Using the relations
\HL{
$
K ( \Lambda-i(z -\bar{L}(z) ) )^{-1}K
=
(\Lambda-i(-z^* -\bar{L}(-z^*) ))^{-1}
,
$}
[cf. \Eq{hermcond}, \eq{basisKoff}], and by expanding resolvents into eigen-projectors using \Eq{Pii},
we obtain the explicit RG equations for the Liouvillian expansion coefficients.
\HL{We have} a set of equations for 10 complex coefficients
on an infinite, discrete grid of frequencies $E$
\begin{widetext}
\begin{align}
  \frac{d
    \xi_{\sigma,\bar{\sigma}}(E)
  }{d\Lambda}
  &=
  i \frac{\Gamma}{\pi}
  \left(
  \frac{
    (\alpha_{- \bar{\sigma}}^{+}|P_{-r}^{\alpha_{- \bar{\sigma}}}|\alpha_{- \bar{\sigma}}^{-})
    }
    {
     \Lambda - i(E_{-r}-\lambda_{-r}^{\alpha_{- \bar{\sigma}}})
    }
    +
  \frac{
    (\alpha_{+ \sigma}^{+}|P_{+r}^{\alpha_{+ \sigma}}|\alpha_{+ \sigma}^{-})
    }
    {
     \Lambda - i(E_{+r}-\lambda_{+r}^{\alpha_{+ \sigma}})
    }
  \right)
    \label{xi-RG}
    ,
  \\
  \frac{d
    E_{\sigma} (E)
  }{d\Lambda}
  &=i \frac{\Gamma}{\pi}\left(
  \frac{
    (\alpha_{+ \sigma}^{+}|P_{+r}^{\alpha_{+ \sigma}}|\alpha_{+ \sigma}^{-})
    }
    {
     \Lambda - i(E_{+r}-\lambda_{+r}^{\alpha_{+ \sigma}})
    }+\frac{
    (\alpha_{- {\sigma}}^{+}|P_{-r}^{\alpha_{- {\sigma}}}|\alpha_{- {\sigma}}^{-})
    }
    {
     \Lambda - i(E_{+r}-\lambda_{+r}^{\alpha_{+ \bar{\sigma}}})
    } \right)=\frac {d E^*_{\bar{\sigma}} (-E^*)}{d\Lambda}
  ,
  \\
  \frac{d
    M_{+}(E)
  }{d\Lambda}
  &=-i\frac {\Gamma}{\pi} \frac{(\alpha^+_{+ \bar{\sigma}}|P_{-r}^{\alpha_{+ \bar{\sigma}}}|\alpha^-_{+ \bar{\sigma}})}{\Lambda-i\left(E_{-r}-\lambda_{-r}^{\alpha_{+ \bar{\sigma}}}\right)}
  ,
  \\
  \frac{d
    M_{-}(E)
  }{d\Lambda}
  &
  =-i\frac {\Gamma}{\pi} 
  \frac{(\alpha^+_{-,{\sigma}}|P^{\alpha_{-,{\sigma}}}_{+r}|\alpha^-_{-,{\sigma}})}{\Lambda-i\left(E_{+r}-\lambda_{+r}^{\alpha_{-,{\sigma}}}\right)}
=
\frac {d M^*_+ (-E^*)}{d\Lambda}
\label{M-RG}
,
  \\
  \frac{d
    F_{+, \sigma}^{+-}(E)
  }{d\Lambda}
  &=-i\frac \Gamma \pi 
  \left(
    \frac{(\chi_{\bar{\sigma}}|P_{-r}^{\chi}|\chi_\sigma)}{\Lambda-i(E_{-r}-\lambda_{-r}^\chi)}
    +\frac{(T_+|P_{+r}^{T_+}|T_+)}{\Lambda-i(E_{+r}-\lambda_+^{T_+})}
    -\frac{(S_\sigma|P_{-r}^{S_\sigma}|S_\sigma)}{\Lambda-i(E_{-r}-\lambda_{-r}^{S_\sigma})}
  \right)
  \label{F-RG}
  ,
  \\
  \frac{d
    F_{-, \sigma}^{+-}(E)
  }{d\Lambda}
  &=-i\frac {\Gamma}{\pi}
  \left(
    \frac{(\chi_{{\sigma}}|P_{{+}}^{\chi}|\chi_{\bar{\sigma}})}{\Lambda-i(E_{+r}-\lambda_{{+}}^\chi)}
    +\frac{(T_-|P_{-r}^{T_-}|T_-)}{\Lambda-i(E_{-r}-\lambda_{-r}^{T_-})}
    -\frac{(S_\sigma|P_{+r}^{S_\sigma}|S_\sigma)}{\Lambda-i(E_{+r}-\lambda_{{+}}^{S_\sigma})}
  \right)
  = -\frac{d
    \left(F_{+,\bar{\sigma}}^{+-}(-E^*) \right)^*
  }{d\Lambda}
\label{F-RG-}
.
  \end{align}
\end{widetext}
Importantly, the eigen projectors of the Liouvillian $\bar{L}^0$,
$\HL{P^i_{\eta r}}=P^i(E-\eta\mu_r)$  with eigenvalues
$\HL{\lambda^i_{\eta r}}=\lambda^i(E-\eta\mu_r)$
[cf. \Eq{eigenproj}] depend on the frequency $E-\eta\mu_r$.
The explicit expressions for the projector matrix elements on the right-hand side are given in Eqs. \eq{zero-proj}, \eq{p1}, \eq{p34}, \eq{pchi} and \eq{palpha}
and involve only the 10 coefficients appearing on the left-hand side.
\Eq{xi-RG}-\eq{F-RG-} thus form a closed set of equations.
This derives from the fact that the eigenprojectors of $\bar{L}_\Lambda$ that
involve the zero eigen-supervectors $Z_L$ and $Z_R$ drop out on the right-hand side 
by \Eq{eigenproj2}.
Note that the fermionic matrix elements in the equations for the coefficients of bosonic terms that do not involve a $Z_L$ or $Z_R$ ($\xi_{\sigma \sigma'}$, $E_\sigma$ and $M_\eta$)
\HL{illustrate the simplification brought by} \Eq{eigenproj2}.

The following 5 complex coefficients do not appear in the eigenvalues and projector matrix elements on the right-hand side of \Eq{xi-RG}-\eq{F-RG-}. Their RG-equations
\begin{align}
  &
  \frac{d
    \psi_{\sigma}(E)
  }{d\Lambda}
  %&
  =
  \label{psi-RG}
  \\
  &
    ~~~~~~~
  i \frac{\Gamma}{\pi}
  \left(
  \frac{
    (\alpha_{- \bar{\sigma}}^{+}|P_{-r}^{\alpha_{- \bar{\sigma}}}|\alpha_{- \bar{\sigma}}^{+})
    }
    {
     \Lambda - i(E_{-r}-\lambda_{-r}^{\alpha_{- \bar{\sigma}}})
    }
    -
  \frac{
    (\alpha_{+ \sigma}^{+}|P_{+r}^{\alpha_{+ \sigma}}|\alpha_{+ \sigma}^{+})
    }
    {
     \Lambda - i(E_{+r}-\lambda_{+r}^{\alpha_{+ \sigma}})
    }
  \right)
  \nonumber
  ,
  \\
  &
  \frac {d\zeta(E)}{d\Lambda}
  %&
  = 
   \\
  &
  ~~~~
 - i \frac{\Gamma}{\pi}
  \left(
  \frac{
    (\alpha_{- \bar{\sigma}}^{-}|P_{-r}^{\alpha_{- \bar{\sigma}}}|\alpha_{- \bar{\sigma}}^{+})
    }
    {
     \Lambda - i(E_{-r}-\lambda_{-r}^{\alpha_{- \bar{\sigma}}})
    }
    +
  \frac{
    (\alpha_{+ \sigma}^{-}|P_{+r}^{\alpha_{+ \sigma}}|\alpha_{+ \sigma}^{+})
    }
    {
     \Lambda - i(E_{+r}-\lambda_{+r}^{\alpha_{+ \sigma}})
    }
  \right)
  \nonumber
  ,
  \\
  &
  \frac{d\phi_\sigma(E)}{d\Lambda}
  %&
  =
 \label{phi-RG}
   \\
  &
  ~~~~~~~~
   i \frac{\Gamma}{\pi}
  \left(
  \frac{
    (\alpha_{- {\sigma}}^{-}|P_{-r}^{\alpha_{- {\sigma}}}|\alpha_{- {\sigma}}^{-})
    }
    {
     \Lambda - i(E_{-r}-\lambda_{-r}^{\alpha_{- {\sigma}}})
    }
    -
  \frac{
    (\alpha_{+ \bar{\sigma}}^{-}|P_{+r}^{\alpha_{+ \bar{\sigma}}}|\alpha_{+ \bar{\sigma}}^{-})
    }
    {
     \Lambda - i(E_{+r}-\lambda_{+r}^{\alpha_{+ \bar{\sigma}}})
    }
  \right)
  \nonumber
  ,
\end{align}
are therefore not required for the solution of \Eq{xi-RG}-\eq{F-RG-},
\HL{but these coefficients \emph{do} renormalize
 and depend on this solution.}
For the calculation of the current only $\psi_\sigma$ is required.
\HL{In contrast, the coefficients $\zeta$ and $\phi_\sigma$ are only} required if one wishes to calculate, e.g., the stationary density matrix \eq{rho}.

The remaining coefficients do not flow under the one-loop RG, and remain at their initial values.
For the two diagonal matrix elements of the coefficient matrix $\xi$ in the bosonic sector we have
\begin{align}
  \frac {d\xi_{\sigma \sigma}}{d\Lambda}(E)=0
  ,
  \label{bose_null}
\end{align}
which is valid only within the present one-loop approximation.
In contrast, for the remaining 16 fermionic coefficients, we have in general (e.g., also in two-loop order)
\begin{align}
  \frac{d F_{\eta, \sigma}^{+\pm} }{d\Lambda}(E)
  &=
  \frac{d F_{\eta, \sigma}^{\pm-} }{d\Lambda}(E)=0
  ,
  \label{fermi_null}
\end{align}
due to the causal structure (cf. \Sec{sec:fermion-pt} and \Sec{sec:fermion}).
We furthermore note that \Eq{xi-RG}-\eq{phi-RG} explicitly satisfy the Hermicity conditions~\Eq{hc-relation-diag} and \eq{hc-relation-nondiag}.
Moreover, the proper transformation under charge and spin-rotations is explicitly guaranteed by our use of irreducible tensor superoperators, cf. \Sec{sec:spin}.

Finally,
for \HL{the} calculation of the current at a specific electrode $r=\pm$ \HL{(corresponding to $L,R$)} we need \HL{the} RG equations for the coefficients of \HL{the} self-energy components $\bar{L}^r$, cf. \Eq{barSigmaintegralr}.
These are simply obtained from the above equations by
(i) giving all coefficients a superscript $r$ and
(ii) suppressing the summation over $r$ contained in the multiindex $1$ on the right-hand side, i.e., by setting $1=\eta,\sigma,+,\omega$.

Before we proceed to calculate the two-loop corrections to \Eq{1loop},
\key{
we first show that already in the above 1-loop approximation we obtain the exact solution for the current in the limit $U=0$.
This is important since it demonstrates that \emph{for the current}  the two-loop corrections to $\bar{L}$, the one-loop corrections to the vertex $\bar{G}$, and the $\omega$ frequency dependence that we neglected here, are intimately connected with interaction effects.
We note, however, that for $U=0$ there are \emph{nonzero} two-loop corrections
which do not affect the current~\cite{Saptsov13a}.
}

\subsubsection{Non-interacting case $U=0$: Exact solution\label{sec:nonint}}
Without local interaction, $U=0$, the Hamiltonian \Eq{tot_Ham} is quadratic in the fermionic operators and the non-equilibrium Anderson model 
can be solved exactly in this limit. A solution using the real-time approach was reported in \Cite{Schoeller97hab,Schoeller99tut}.
We now show \HL{that}
(i) using \HL{the causal field superoperator} algebra one can obtain this solution within RT-RG framework and
(ii) within the 1-loop, frequency-independent approximation \eq{1loop} this result is recovered upon careful inspection.

\paragraph{Exact current}
In general, to calculate the current according to \Eq{currentI}, we need the elements $\xi$ and $\psi_\sigma$ of the bosonic part of the effective Liouvillian $\bar{L}_\Lambda$.
From these we can then easily find the other required coefficients $\xi^r$ and $\psi_\sigma^r$ of $\bar{L}^r_\Lambda$, which we do at the end.
The coefficients  $\xi$ and $\psi_\sigma$ in turn require the solution of 
\Eq{xi-RG}-\eq{F-RG-}, which we first discuss. \HL{Then we show} that higher order correction as well as frequency corrections \HL{that} we neglected in deriving 
\Eq{xi-RG}-\eq{F-RG-} have no influence on the stationary current.
We start by noting that on the right-hand side of
\HL{the RG-}\Eq{xi-RG}-\eq{M-RG} for the bosonic sector
\HL{only \emph{fermionic} intermediate states appear in the resolvent matrix elements.}
We therefore first calculate the eigenvalues of the fermionic projectors to which these coefficients couple. From \Eq{ferm-spec} it follows that for $U=0$
\begin{align}
  \lambda^{\alpha_{\eta \sigma},\pm}  & = \eta\epsilon +\sigma \frac{B}{2}- 2 i \Gamma \pm i\Gamma
  \label{eigen_free}
  ,
  \\
  P^{\alpha_{\eta \sigma},\pm}
  &
  = \frac{1}{2} \left(
    \alpha_{\eta \sigma}^0
    \pm \eta  \alpha_{\eta \sigma}^3
    \mp \eta\frac{\Delta F_{\eta, \sigma}^{-+}}{\Gamma} \alpha_{\eta \sigma}^{-}
  \right)
  \label{proj_free}
\end{align}
\HL{There are three important points.}
(i) Since $U=0$ the eigenvalues are independent of $\Delta F_{\eta, \sigma}^{-+}$,
i.e., they are not renormalized (cf. \Sec{sec:fermion-pt})
 and therefore do not acquire a frequency dependence.
(ii) The right-hand side of the RG equation \HL{for} any bosonic superoperator that is relevant to the current, i.e., excluding $\zeta,\phi$, but \emph{with the exception \HL{of}} $\psi_\sigma$ contains the off-diagonal super matrix elements
\begin{align}
  (\alpha_{\eta \sigma}^{+}| \Pi_1^{\alpha_{\eta \sigma}} | \alpha_{\eta \sigma}^{-})
  =0
  \label{alpha-factor-zero}
\end{align}
as a factor \HL{by the general property \eq{alpha-factor2}.}
\HL{This matrix-element} vanishes for $U=0$ by \Eq{proj_free},
\HL{implying} that these coefficients do not renormalize \emph{in any higher loop order}
since such \HL{coefficients} always contain this factor \HL{on the right-hand side of \Eq{RG-L}} at least once by \Eq{alpha-factor3}.
(iii) \HL{In contrast,} the renormalization of the quantities $\psi_\sigma$ at $E=0$ involve fermionic virtual states with the simple factors
\begin{align}
  &  i(\alpha_{+ \sigma}^{+}| \Pi_{+r}^{\alpha_{+ \sigma}} | \alpha_{+ \sigma}^{+})
  \nonumber
  \\
  & =
  \sum_r (\alpha_{+ \sigma}^{+}|
  \frac{1}{\Lambda-irV/2+i\lambda_{+r}^{\alpha_{+ \sigma}} (rV/2)}
  |\alpha_{+ \sigma}^{+})
  \\
 &=
 \sum_r  \frac{1}{\Lambda+\Gamma + i (\epsilon_\sigma -r V/2)} 
 ,
\end{align}
where \HL{we defined} $\epsilon_\sigma:= \epsilon+\sigma B/2$.
Importantly, for $U=0$ these matrix elements \HL{do not vanish, but they become} independent of $\Delta F_{\eta,  \sigma}^{-+}$.
Higher loop corrections for $\psi_\sigma$ vanish since they contain the factor \eq{alpha-factor-zero}  at least once by \Eq{alpha-factor1}. 
Therefore, \Eq{psi-RG} is the exact RG equation for $\psi_\sigma$ at \hl{$E=0$} and for $U=0$:
\begin{align}
  \frac{d \psi_\sigma (0)}{d \Lambda}
  &=\sum\limits_r\frac{2\Gamma}{\pi}
  \rm{Im}
  (\alpha_{+ \sigma}^{+}| i\Pi_{+ r}^{\alpha_{+ \sigma}} | \alpha_{+ \sigma}^{+})
  \\
  &= 
  \sum_r\frac{2 \Gamma}{\pi}\frac{\epsilon_\sigma-rV/2}{(\Lambda+\Gamma)^2+(\epsilon_\sigma-rV/2)^2}
  ,
\end{align}
where we again used the $K$-conjugation properties \Eq{hermcond}, \eq{basisKoff}.
With the initial value  $\psi_{\sigma,\Lambda=\infty}=0$ (cf. \Eq{sin}.) we obtain
\begin{align}
  \nonumber
  \psi_\sigma (0)
  = -\sum_r \frac{2 \Gamma}{\pi}\arctan\left( \frac{\epsilon_\sigma-rV/2}{\Gamma}\right)
  .
\end{align}
Leaving out the summation over the electrode $r$ in the above calculation, we obtain the coefficients of the self-energy component $\bar{L}^r_{\Lambda}|_{\Lambda=0}=\Sigma^r(0)$, required for the current (cf. \Eq{currentI}),
\begin{align}
  \label{currentIrg}
  \nonumber
  \langle I^r \rangle
  &=
  \tfrac{1}{\sqrt{2}}
  \left[
    (T_0|\overrightarrow{\psi}^r)-(T_0| {\xi}^r {\xi}^{-1}|\overrightarrow{\psi})
  \right]
  \\
  &
  = \tfrac{1}{4}\sum\limits_\sigma\left({\psi}^r_\sigma-{\psi}^{\bar{r}}_\sigma \right)
  ,
\end{align}
using $\frac{1}{\sqrt{2}}(T_0|=\frac{1}{2}\sum_\sigma (\chi_\sigma|$ and $\xi^r=\xi/2$,
\HL{giving the current}
\begin{align}
  \langle I \rangle = \sum_{r,\sigma=\pm}
  r \frac{\Gamma}{2\pi}
  \arctan{\left(\frac{\epsilon_\sigma+r V/2}{\Gamma}\right)} 
  \label{currentU0}
\end{align}
and the non-linear differential conductance:
\begin{align}
  \label{nonint-conductance}
  \frac{dI}{dV}
  =\frac{1}{4\pi}\sum_{r,\sigma=\pm} \frac{\Gamma^2}{\Gamma^2+\left( \epsilon_\sigma+rV/2\right)^2}
\end{align}
in our units $e=1, ~ \hbar=1$.
Restoring Gaussian units, the current and conductance prefactors become
$\frac{\Gamma}{2\pi} \rightarrow \Gamma \frac{e}{\hbar}$
$\frac{\Gamma}{2\pi} \rightarrow \Gamma \frac{e^2}{2\hbar}$
giving in linear response a conductance of $e^2/h$ per spin channel.
The importance of recovering this exact result for $U=0$ is that already at this level of approximation our the RT-RG approach captures correctly the weak interaction limit $U\ll\Gamma$ while treating the tunneling non-perturbatively in $\Gamma$.
Moreover, it shows that the 2-loop corrections to the Liouvillian and 1-loop corrections to the vertices \emph{\HL{that} affect the stationary current} are generated by the Coulomb interaction.
In general, however, the non-interacting limit requires a 2-loop treatment.
%%% Local Variables: 
%%% mode: latex
%%% TeX-master: "paper"
%%% End: 
\subsection{2 loop RG equations\label{sec:2loop}}
\subsubsection{Vertex frequency dependence\label{sec:freq-dep2}}
\HL{We} concluded in \Sec{sec:freq-dep1} \HL{that} in the 2-loop approximation for the Liouvillian \HL{one} should consider the vertex renormalization and $\omega$-dependence of both $\bar{G}$ and  $\bar{L}$.
Indeed, we find below that these effects are comparable and involve important cancellations.
By systematically expanding about the frequency-independent bare vertex $\bar{G}^0$, we can incorporate the vertex corrections into a single effective equation for the 2-loop Liouvillian, \Eq{2loopRG} below.
We proceed in three steps:

\emph{Step 1}
\HL{The} starting point is the 1-loop approximation defined by \Eq{1loopE} for any $E$.
This we use to calculate a first approximation for the $\omega$-dependence of both the Liouvillian and the propagator.
We expand
\begin{align}
  \bar{L} (E,\omega) & \approx  \bar{L}^0 (E) + \bar{L}^1 (E,\omega)
  \label{LDexpand}
  ,
  \\
  \Pi (E,\omega)       & \approx  \Pi^0 (E,\omega) + \Pi^1 (E,\omega)
  .
\end{align}
The 1-loop equation accounting for the leading frequency dependence is obtained by setting $G \approx G^0$ and $\Pi(E,\omega) \approx \Pi^0(E,\omega)$ \HL{in}  the 1-loop part of \Eq{RG-L}:
\begin{align}
  \label{1loopEomega}
  \frac{d \bar{L} (E,\omega)}{d \Lambda} =
  i\frac{\Gamma }{\pi}
  \bar{G}_1^0
  \Pi^0 (E_1,\Lambda+\omega)
  \bar{G}_{\bar{1}}^0
  .
\end{align}
Subtracting \Eq{1loopE}, we obtain
\begin{align}
  \frac{d \bar{L}^1 (E,\omega) }{d \Lambda}
  &\approx
  i\frac{\Gamma}{\pi}
  \bar{G}^{0}_1 
  \left( 
    \Pi^0(E_1,\omega+\Lambda)-\Pi^0(E_1,\Lambda)
  \right)
  \bar{G}^{0}_{\bar{1}}
  .
\nonumber
\end{align}
Shifting the integration variable in the $\Pi(E_1,\Lambda)$ term we obtain \HL{in wide band limit (which} we assume throughout):
\begin{align}
  &\bar{L}^1 (E,\omega) \approx
  \int \limits_{\Lambda-\omega}^\Lambda
  i\frac{\Gamma}{\pi}
  \bar{G}^{0}_1 
  \Pi^0(E_1,\omega+\Lambda)
  \bar{G}^{0}_{\bar{1}}
  \label{LD1simple}
  .
\end{align}
\Eq{LD1simple} does not need to be \HL{evaluated} further since it cancels out below.
Note that the correction vanishes at zero frequency, $\bar{L}^1 (E,0)=0$, for all $E$ as required.
Expanding the full resolvents \eq{resolvent} with the approximation \Eq{LDexpand}
to the first order in $\bar{L}^1 (E,\omega)$, we obtain:
\begin{align}
  \label{Pi1}
  \Pi^1 (E,\omega) & = \Pi^0 (E,\omega) \bar{L}^1 (E,\omega) \Pi^0 (E,\omega)
  .
\end{align}

\emph{Step 2} In a similar way, we now calculate the $\omega$ corrections in the leading 1-loop order for the vertices:
\begin{align}
  \bar{G}_1(E,\omega,\omega_1) \approx \bar{G}^0+\bar{G}^1_1 (E,\omega,\omega_1)
  .
\end{align}
Keeping only the leading term on the right-hand side of \Eq{RG-G}, we obtain with the same approximations as above
\begin{widetext}
\begin{align}
  \label{G1}
  \bar{G}_1^1 (E,\omega,\omega_1)
  =
  -i \frac{\Gamma}{\pi}
  \int \limits_{D}^\Lambda
  d{\Lambda'}
  \bar{G}_2^0
  \Pi^0 (E_2,\omega+{\Lambda'})
  \bar{G}_1^0
  \Pi^0( E_{12},\omega+\omega_1 +{\Lambda'})
  \bar{G}_{\bar{2}}^0
  .
\end{align}
\end{widetext}
We stress that the argument under the integral depends on $\Lambda'$ both through the explicit arguments
as well as through the cutoff dependence of $\bar{L}_{\Lambda'}$ ($G^0$ is the bare vertex). Restoring the latter explicitly,
\begin{align}
  \Pi^0_{\Lambda'}(E,\omega+\Lambda') =
  \frac{1}{\omega+\Lambda'-i\bar{L}_{\Lambda'}^0(E)}
  \label{implicitLambda}
  .
\end{align}
\HL{Since this is only important at this point}, we stick with the implicit notation.

\emph{Step 3}
Using the expansions for the resolvents \Eq{Pi1} and vertices \Eq{G1} we can now calculate an approximation to the right-hand side of \Eq{RG-L}, keeping the leading order frequency corrections:
\begin{align}
  \frac{d \bar{L}}{d\Lambda}  =
  \frac{d \bar{L}^\mathsf{(1\,loop)}}{d\Lambda}
  +\frac{d \bar{L}^\mathsf{(2\,loop)}}{d\Lambda}
  +\frac{d \bar{L}^\mathsf{(vertex)}}{d\Lambda}
  \label{2loop-terms}
  ,
\end{align}
\HL{where all terms are} written at frequencies $E$ and $\omega$ \HL{and}
\begin{align}
  \label{1loopterm} 
  \frac{d \bar{L}^\mathsf{(1\,loop)}}{d\Lambda}
  =
  i\frac{\Gamma}{\pi} \Big[ \bar{G}^0_1\Pi^0 (E_1,\Lambda) \bar{G}^0_{\bar{1}}
  +\bar{G}^0_1\Pi^1 (E_1,\Lambda) \bar{G}^0_{\bar{1}} \Big]
  ,
\end{align}
are the terms appearing from \HL{the} expansion of \HL{the} 1-loop diagram in $\Pi^1$,
\begin{align}
  \label{2loopterm}
  & \frac{d \bar{L}^\mathsf{(2\, loop)}}{d\Lambda}
  =
  \frac{\Gamma^2 }{\pi^2}\bar{G}^0_1 \Pi^0 (E_1,\Lambda)\bar{G}^0_2
  \\
  \nonumber
  &\times\Pi^0 (E_{12},\Lambda+\omega_2) \bar{G}^0_{\bar{2}} \Pi^0 (E_1,\Lambda) \bar{G}^0_{\bar{1}}
\end{align}
is 2-loop term with bare vertices, and
\begin{align}
  \label{vertexterm}
  \frac{d \bar{L}^\mathsf{(vertex)}}{d\Lambda}
  = i\frac{\Gamma}{\pi}  \bar{G}^1_1 (E,0,\Lambda) \Pi^0 (E_1,\Lambda) \bar{G}^0_{\bar{1}}
  \\
  +i\frac{\Gamma}{\pi}  \bar{G}^0_1  \Pi^0 (E_1,\Lambda) \bar{G}^1_{\bar{1}}  (E_1,\Lambda,-\Lambda) 
\end{align}
are the terms appearing from the expansion of $\bar{G}$ in $\bar{G}^1$ in the 1-loop diagram.
Inserting \Eq{LD1simple} into the 1-loop Liouvillian frequency correction [last term in  \Eq{1loopterm}], we see that it exactly cancels the 2-loop term \eq{2loopterm}.
Neither term therefore needs to be calculated, simplifying the approach to a great extent.

The above holds for any $E$ and $\omega$: integrating the \Eq{2loop-terms} at $\omega=0$ using the above calculated right-hand side we obtain a new approximation for $\bar{L}^0(E)$, 
improving over our initial approximation based on the one-loop equation \Eq{1loopE}. In principle, steps 1 and 2 should be repeated, resulting in corrections of higher orders, which we neglect.
We thus equate $d\bar{L}/d\Lambda \approx \bar{L}^0/d\Lambda$ on the left-hand side of \Eq{2loop-terms}.
We obtain a central result of this section: a \HL{single} effective two-loop RG equation for the Liouvillian at $\omega=0$:
\begin{align}
  \frac{d \bar{L}^0 (E)}{d \Lambda}
  & =
  i\frac{\Gamma }{\pi}
  \bar{G}_1^0
  \Pi^0 (E_1,\Lambda)
  \bar{G}_{\bar{1}}^0
  \nonumber \\
  & + i\frac{\Gamma}{\pi}  \bar{G}^1_1 (E,0,\Lambda) \Pi^0 (E_1,\Lambda) \bar{G}^0_{\bar{1}}
  \nonumber \\
  &+i\frac{\Gamma}{\pi}  \bar{G}^0_1  \Pi^0 (E_1,\Lambda) \bar{G}^1_{\bar{1}}  (E_1,\Lambda,-\Lambda) 
  \label{final}
  .
\end{align}
Notably, due to the cancellation the entire leading reservoir ($\omega$) frequency dependence comes from the vertex corrections \eq{G1}.
This single equation yields a significant simplification over the coupled \HL{integro-differential} equations \Eq{RG-L} and \Eq{RG-G}.
\HL{The equation \eq{final} for $\bar{L}$} can be converted into a differential RG equation by analytically performing the integral in \Eq{G1}, see \Sec{sec:Exp_RG}.
Furthermore, since we can work with the bare vertex superoperators, we can make use of their simple anticommutation relations \eq{commutG}, 
\HL{which} are \emph{not} preserved under the RG
(in contrast to other useful properties of the vertex, see \App{sec:K-pr}).
The QD frequency ($E$) dependence in \Eq{final} is of the same type as for the 1-loop equations \Eq{1loopRG}:
the RG equation for $L^0(0)$ depends on $L^0(\bar{\mu}_1)=L^0(\eta_1 r_1V/2)$, etc.
It therefore has to be solved in the same way by including multiple Matsubara axes and converging the energy-hierarchy of equations, see 
\Sec{sec:freq-dep1} and \Sec{sec:results}.
Finally, we also note that one can indeed neglect the frequency dependence generated by vertex renormalization since it is indeed small, 
as we assumed in our derivation of the 
1-loop equations. 
This can be seen if one substitutes the calculated correction \Eq{G1} into \Eq{RG-G}. Here, we anticipate the projector expansion of $\bar{G}^1$, \Eq{G1-proj}:
it is seen that $\bar{G}^1$ is a well behaved function of the cutoff and frequency, decaying at small $\Lambda$.\HL{It} generates only small corrections
in agreement with our approximation \Eq{H-approximation}. 
\subsubsection{Explicit 2-loop RG equations for the Liouvillian\label{sec:Exp_RG}}

To obtain a \HL{\emph{differential}} RG equation from \Eq{final}, the integration \HL{in} \Eq{G1} needs to be performed.
This \HL{is} complicated by the implicit $\Lambda'$ dependence \HL{of} the propagators \HL{that we} pointed out with \Eq{implicitLambda}.
We now make an adiabatic approximation by expanding \emph{only} this dependence about $\Lambda'=\Lambda$, i.e., we substitute
\begin{align}
  \bar{L}_{\Lambda'} \approx \bar{L}_{\Lambda}
\end{align}
in \Eq{implicitLambda} and neglect corrections $\sim d\bar{L}_{\Lambda'}/d\Lambda$, which, by the RG equation \Eq{final}, are of higher order and should therefore be neglected.
To preserve the compact form of the equations we define
\begin{align}
  \Theta^i_{1...n} &= E_{1...n} -\lambda^i (E_{1...n})
  ,
  \\
  P^i_{1...n} &= P^i (E_{1...n})
  ,
\end{align}
where $P^i (E)$ and $\lambda^i (E)$ are eigenprojectors and eigenvalues \HL{of} \Eq{eigenproj} at cutoff $\Lambda$ (not $\Lambda'$) and we will implicitly sum over all appearing eigenvalue labels $i,j,k$ below.
To perform the integral we insert the projector expansion \Eq{eigenproj} of $\bar{L}$ evaluated at $\Lambda$ under the integral
and obtain the explicit $\omega$-dependent vertex correction:
\begin{align}
\label{G1-proj}
  &\bar{G}_1^1 (E,\omega,\omega_1)
  =
  \\
  &-i \frac{\Gamma}{\pi}
  \int \limits_{D}^\Lambda
  d{\Lambda'}
  \frac{
%    \bar{G}^0_2 P^i(E_2)\bar{G}^0_1 P^j(E_{12}) \bar{G}^0_{\bar{2}}
    \bar{G}^0_2 \HL{P^i_2} \bar{G}^0_1 \HL{P^j_{12}} \bar{G}^0_{\bar{2}}
  }{\left(
      \Lambda'+\omega-i\Theta^i_2\right)\left(\Lambda'+\omega+\omega_1-i\Theta^j_{12}
    \right)}
  \nonumber \\
  &
    \nonumber \\
  = &i \frac{\Gamma}{\pi}
  \frac{
%    \bar{G}^0_2 P^i(E_2)\bar{G}^0_1 P^j(E_{12}) \bar{G}^0_{\bar{2}}
    \bar{G}^0_2 \HL{P^i_2} \bar{G}^0_1 \HL{P^j_{12}} \bar{G}^0_{\bar{2}}
  }{\omega_1-i \left(\Theta^j_{12}-\Theta^i_{2} \right)}
  \ln \left (\frac{\Lambda+\omega-i\Theta^i_2}{\Lambda+\omega+\omega_1-i\Theta^j_{12}} \right)
  \nonumber
\end{align}
Combining the rest of \Eq{1loopterm} with \Eq{vertexterm} we obtain:
\begin{align}
  \label{2loopRG}
  &\frac{d\bar{L}^0 (E)}{d\Lambda}
  =
  \frac{\Gamma}{\pi} \frac{1}{\Lambda-i\Theta^k_1}\bar{G}^0_1 P^k_1\bar{G}^0_{\bar{1}}
  \\
  &-i\frac{\Gamma^2 }{\pi^2}\frac{1}{\Lambda-i\Theta^k_1}\frac{1}{\Lambda-i( \Theta^j_{12} -\Theta^i_2)}
  \ln \left( \frac{2\Lambda-i\Theta^j_{12}}{\Lambda-i\Theta^i_2}\right)
   \nonumber
   \\
   &\times\Big[
   \bar{G}^0_1 P^k_1 \bar{G}^0_2 P^j_{12} \bar{G}^0_{\bar{1}}P^i_2\bar{G}^0_{\bar{2}}
   +
   \bar{G}^0_2 P^i_2 \bar{G}^0_1 P^j_{12} \bar{G}^0_{\bar{2}}P^k_1\bar{G}^0_{\bar{1}}
   \Big]
   \nonumber
   .
\end{align}
This is a central result of \HL{the} paper.
The explicit evaluation of \Eq{2loopRG} for the Anderson model is required for our numerical implementation, but also allows us to draw some general conclusions about the 2-loop (and higher) corrections.
The 1-loop part is given by \Eq{1loopE} and we proceed analogously for the 2-loop part:
\begin{align}
  \left. \frac{d\bar{L}^0 (E)}{d\Lambda} \right|_{\mathsf{2\,loop}} 
  = -i\frac{\Gamma^2 }{\pi^2}  (\kappa_7| \mathcal{M}_{} |\kappa_0) |\kappa_7)(\kappa_0|
  \label{2loopexplicit}
  ,
\end{align}
where the super matrix elements are factored as follows:
\begin{align}
  &(\kappa_7| \mathcal{M}_{} |\kappa_0)
  =
  \\
  &\sum_{\kappa_6..\kappa_1}
  (\kappa_7| \bar{G}^0_1 |\kappa_6)
  (\kappa_5| \bar{G}^0_2 |\kappa_4)
  (\kappa_3| \bar{G}^0_{\bar{1}} |\kappa_2)
  (\kappa_1| \bar{G}^0_{\bar{2}} |\kappa_0)
  \nonumber \\
  &
  \times
  \mathcal{N}
  \left(
    (\kappa_6| {P}^k_1 |\kappa_5),
    (\kappa_4| {P}^j_{12} |\kappa_3),
    (\kappa_2| {P}^i_2 |\kappa_1)
  \right)
  \nonumber
  .
\end{align}
The product of $\bar{G}^0$ matrix elements gives a simple numerical factor, whereas the remaining part,
\begin{align}
  \mathcal{N}
  \left(
    (\kappa_6| {P}^k_1 |\kappa_5),
    (\kappa_4| {P}^j_{12} |\kappa_3),
    (\kappa_2| {P}^i_2 |\kappa_1)
  \right)
  =
  \\
  \mathcal{S}(\Theta_1^k,\Theta_2^i,\Theta_{12}^j)
  ~
  (\kappa_6| {P}^k_1 |\kappa_5)
  (\kappa_4| {P}^j_{12} |\kappa_3)
  (\kappa_2| {P}^i_2 |\kappa_1)
  \nonumber
  ,
\end{align}
contains the product of the non-trivial projector matrix elements and the propagator factors
\begin{align}
  \mathcal{S}(\Theta_1^k,\Theta_2^i,\Theta_{12}^j)
  & =
  \frac{
  \ln \left( \dfrac{2\Lambda-i\Theta^j_{12}}{\Lambda-i\Theta^i_2} \right)    
  }{
    \left( \Lambda-i\Theta^k_1 \right)
    \left(\Lambda-i\left( \Theta^j_{12} -\Theta^i_2\right) \right)
  }
  \nonumber \\
  &+(\Theta^k_1 \leftrightarrow \Theta^i_2)
  .
\end{align}
The argument of this function is constructed by formally putting the variables containing the eigenvalues of the three projectors in the argument \HL{of} 
$\mathcal{N}$ into the corresponding arguments of the scalar function $\Theta$.
With this we can give explicit expressions for \Eq{2loopexplicit}.
\HL{It is shown in \App{sec:K-pr}  how the conservation of Hermiticity by the self-energy can be used to minimize the number of terms to be calculated.}
For the right-hand side of \Eq{2loopexplicit},
we now \HL{explicitly list} half of the terms:
\begin{align}
  &-|\chi_\sigma)(\chi_\sigma|
  \times
  \label{bosonfirst}
  \\
  &\left[
    \mathcal{N}\left( (\alpha^+_{+   \sigma}|P_{1}^{\alpha_{+    \sigma}}|\alpha^-_{+     \sigma}) ,  (T_+| P_{12}^T|T_+) ,   (\alpha^+_{+   \bar{\sigma}}|P_{2}^{\alpha_{+   \bar{\sigma}}} |\alpha^-_{+   \bar{\sigma}})\right)
  \right.
  \nonumber
  \\
  & \left. +
    \mathcal{N}\left( (\alpha^+_{+   \sigma}|P_{1}^{\alpha_{+   \sigma}}|\alpha^-_{+   \sigma}) ,  (S_\sigma|P_{1\bar{2}}^S|S_\sigma) , (\alpha^+_{-   \sigma}| P_{\bar{2}}^{\alpha_{-   \sigma}}|\alpha^-_{-   \sigma})  \right)
  \right]
  \nonumber
  \\
  &
  +|\chi_\sigma)(\chi_{\bar{\sigma}}|
  \times
  \\
  &\mathcal{N}\left( (\alpha^+_{+   \sigma}|P_{1}^{\alpha_{+   \sigma}}|\alpha^-_{+   \sigma}) , (\chi_{\bar{\sigma}} |P_{1\bar{1}}^\chi|\chi_\sigma) ,(\alpha_{-   \bar{\sigma}}^+ | P_{{\bar{1}}}^{\alpha_{-   \bar{\sigma}}}|\alpha_{-   \bar{\sigma}}^-) \right)
  \nonumber
  \\
  &
  -|\chi_\sigma)(Z_{L}|
  \times
  \\
  &\left[
    \mathcal{N} \left((\alpha_{+   \sigma}^+|P_{1}^{\alpha_{+   \sigma}}| \alpha_{+   \sigma}^-),(T_+| P_{12}^T|T_+) ,  (\alpha_{+   \bar{\sigma}}^+|P_{2}^{\alpha_{+   \bar{\sigma}}}|\alpha_{+   \bar{\sigma}}^+) \right)
  \right.
  \nonumber\\
  & -
  \mathcal{N} \left( (\alpha_{+   \sigma}^+|P_{1}^{\alpha_{+   \sigma}} | \alpha_{+   \sigma}^-),  (S_\sigma|P_{1\bar{2}}^S|S_\sigma)  ,  (\alpha_{-   \sigma}^+|P_{\bar{2}}^{\alpha_{-   \sigma}} | \alpha_{-   \sigma}^+)\right)
  \nonumber
  \\
  &\left.+
    \mathcal{N} \left( (\alpha_{+   \sigma}^+|P_{1}^{\alpha_{+   \sigma}}|\alpha_{+   \sigma}^-)  ,  (\chi_{\bar{\sigma}}|P_{1\bar{1}}^\chi|\chi_\sigma)  ,  (\alpha_{-   \bar{\sigma}}^+|P_{\bar{1}}^{\alpha_{-   \bar{\sigma}}}| \alpha_{-   \bar{\sigma}}^+) \right)
  \right]
 \nonumber
  \\
  &
  -|S_\sigma)(S_\sigma|   \times
  \\
  &\left[
    \mathcal{N}\left((\alpha_{+   \sigma}^+|P_{1}^{\alpha_{+   \sigma}}|\alpha_{+   \sigma}^-) ,(T_+| P_{12}^T|T_+) , (\alpha_{+   \sigma}^+| P_{2}^{\alpha_{+   \sigma}}|\alpha_{+   \sigma}^-) \right)
  \right. 
  \nonumber \\
  &\left.+
    \mathcal{N}\left((\alpha_{+   \sigma}^+|P_{1}^{\alpha_{+   \sigma}}|\alpha_{+   \sigma}^-) , (\chi_{\bar{\sigma}}|P_{1\bar{2}}^\chi|\chi_{\bar{\sigma}}) , ( \alpha_{-   \sigma}^+|P_{\bar{2}}^{\alpha_{-   \sigma}}|\alpha_{-   \sigma}^-) \right)
  \right]
  \nonumber\\ 
  &
  -|T_-)(T_-|
  \times
  \label{bosonlast}\\ \nonumber
  &\left[
    \mathcal{N}\left((\alpha_{-   \sigma}^+|P_{1}^{\alpha_{-   \sigma}}|\alpha_{-   \sigma}^-) , (S_\sigma|P_{12}^S|S_\sigma) ,  (\alpha_{-   \sigma}^+|P_{2}^{\alpha_{-   \sigma}}|\alpha_{-   \sigma}^-) \right)
  \right.
 \label{fermionfirst} \\
  &\left. +
    \mathcal{N}\left((\alpha_{-   \bar{\sigma}}|P_{1}^{\alpha_{-   \bar{\sigma}}}|\alpha_{-   \bar{\sigma}}^-) , (\chi_{\bar{\sigma}}|P_{1{2}}^\chi|\chi_{\bar{\sigma}}) , (\alpha_{-   \sigma}^+ |P_{{2}}^{\alpha_{-   \sigma}}|\alpha_{-   \sigma}^-) \right)
  \right]
  \nonumber \\
  &
  +|\alpha_{+   \sigma}^{-})(\alpha^{+}_{+   \sigma}|
  \times
  \\
  &\left[
    \mathcal{N}\left((T_+| P_{1}^T|T_+)     (\alpha_{+   \sigma}^+|P_{1\bar{1}}^{\alpha_{+   \sigma}}|\alpha_{+   \sigma}^-) , (S_\sigma|P_{\bar{1}}^S|S_\sigma) \right)
  \right.
  \nonumber\\
  &\left. +
    \mathcal{N}\left((T_+|P_{1}^T|T_+) , (\alpha_{+   \bar{\sigma}}^+|P_{1\bar{2}}^{\alpha_{+   \bar{\sigma}}}|\alpha_{+   \bar{\sigma}}^-) , (\chi_\sigma| P_{\bar{2}}^\chi|\chi_\sigma) \right)
  \right]
  \nonumber\\ 
  &
  +|\alpha_{-   \sigma}^{-})(\alpha_{-   \sigma}^{+}|
  \times
  \\ \label{fermionlast}
  &\left[
    \mathcal{N}\left( (\chi_\sigma|P_{1}^\chi|\chi_\sigma) ,  (\alpha_{+   \sigma}^+|P_{12}^{\alpha_{+   \sigma}}|\alpha_{+   \sigma}^-) , (S_\sigma |P_{2}^S|S_\sigma) \right)
  \right.
  \nonumber \\
  &+
  \mathcal{N}\left( (\chi_\sigma|P_{1}^\chi|\chi_\sigma) , (\alpha_{-   \bar{\sigma}}^+ |P_{1\bar{2}}^{\alpha_{-   \bar{\sigma}}}| \alpha_{-   \bar{\sigma}}^-) ,  (T_-|P_{\bar{2}}^T|T_-) \right)
  \nonumber       \\ 
  &+
  \mathcal{N}\left( (S_\sigma|P_{1}^S|S_\sigma)  , (\alpha_{-   \sigma}^+ |P_{1\bar{1}}^{\alpha_{-   \sigma}}|\alpha_{-   \sigma}^-) ,  (T_-|P_{\bar{1}}^T|T_-) \right)
  \nonumber\\
  & + \mathcal{N}\left(( S_\sigma|P_{1}^S|S_\sigma) , (\alpha_{+   \sigma}^+| P_{1{2}}^{\alpha_{+   \sigma}}|\alpha_{+   \sigma}^-)  , (\chi_{\bar{\sigma}} |P_{2}^\chi|\chi_{\bar{\sigma}}) \right)
  \nonumber\\ \nonumber
  &\left. -\mathcal{N}\left( (\chi_\sigma|P_{1}^\chi|\chi_{\bar{\sigma}})  , (\alpha_{+   \bar{\sigma}}^+| P_{1{2}}^{\alpha_{+   \bar{\sigma}}}|\alpha_{+   \bar{\sigma}}^-)  ,  (\chi_\sigma|P_{2}^\chi|\chi_{\bar{\sigma}})  \right)
  \right]
  \end{align}
Here we use the notation
\begin{align}
  P_i^k=P^k(-\eta_i\mu_i),~
  &&
  P^k_{{i} j}=P^k(-\eta_i\mu_i-\eta_j\mu_j)
  .
\end{align}
We fixed the particle-hole index $\eta$ in the multiindices as
\begin{align}
  \eta_{i}=
  \begin{cases}
    +, & i={1},{2}\\
    -, & i=\bar{1},\bar{2}
  \end{cases}
\end{align}
All other indices in Eqs. \eq{bosonfirst}-\eq{fermionlast} are \HL{implicitly} summed over.
The other half of the terms of  \Eq{2loopexplicit} can be constructed in the same way
by taking into account the opposite sign of $\eta_1$ using the recipe of \App{sec:K-pr}.
For the calculation of the current only the left-most reservoir index $r_1$ should not be summed over.

% Fig. 6 moved to here from include-results.tex

\begin{figure}[tbp]
  \includegraphics[width=0.99\linewidth]{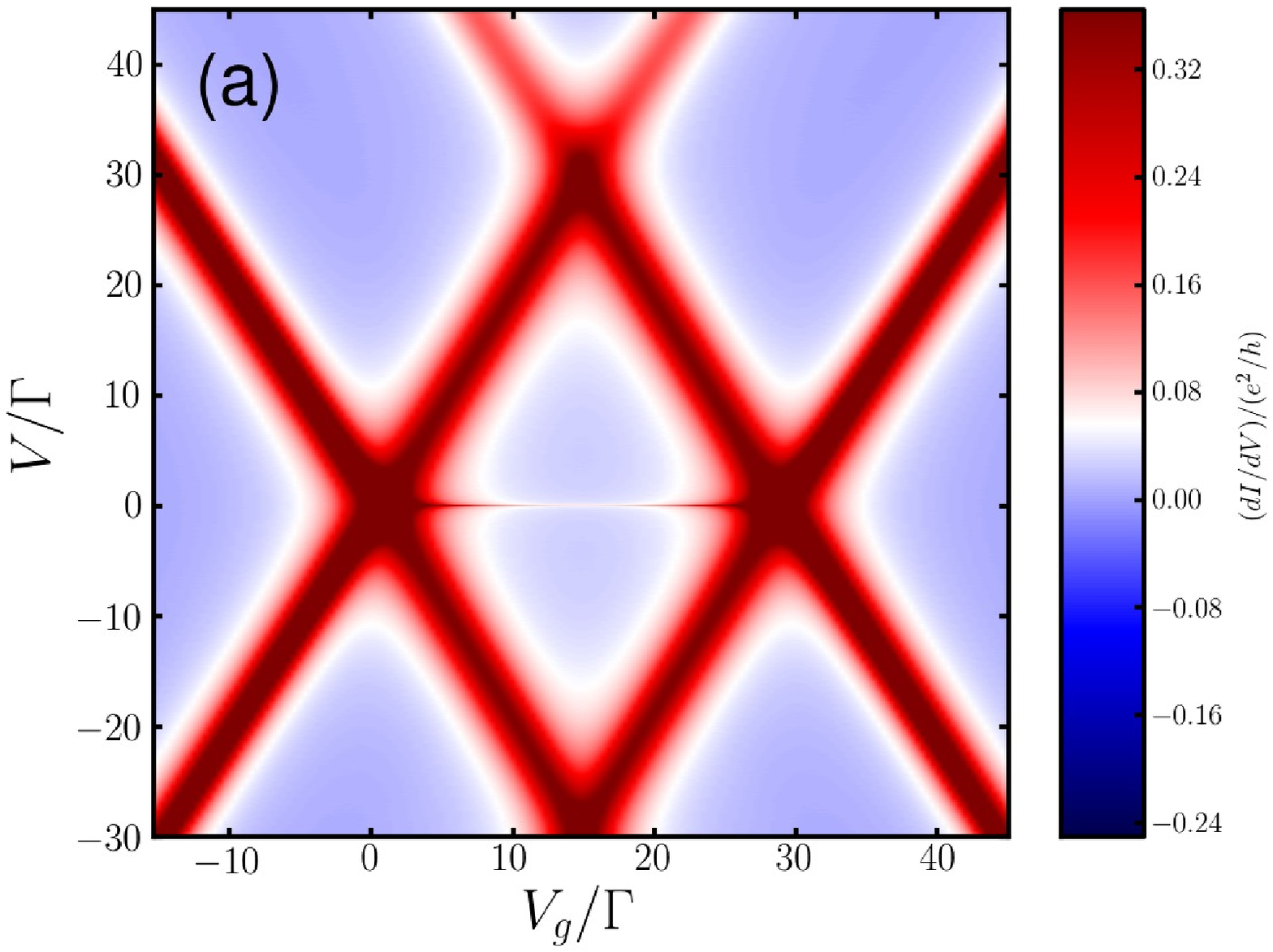}
  \includegraphics[width=0.99\linewidth]{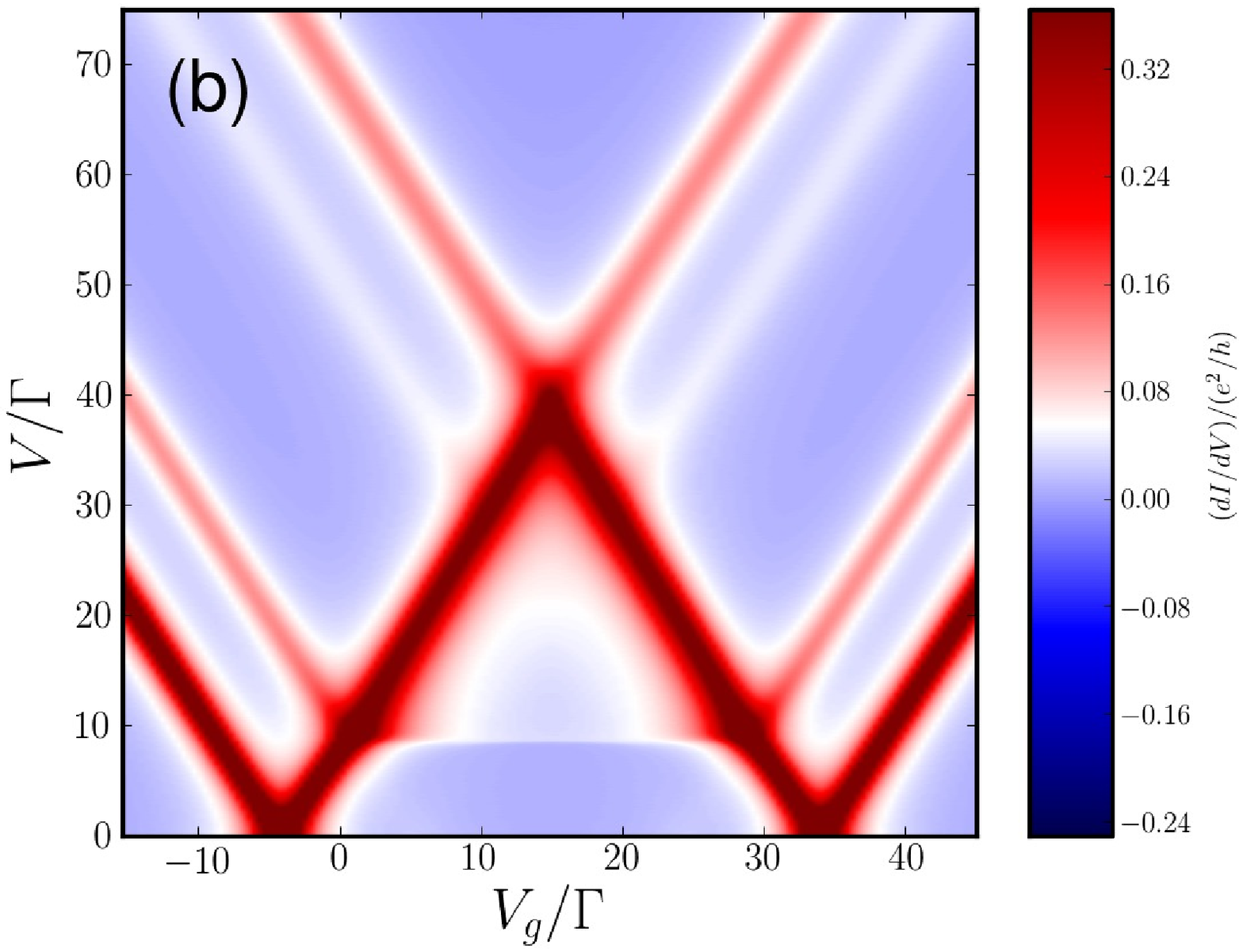}
  \caption{
    Zero-temperature non-linear conductance, $dI/dV$, vs bias $V$ and gate voltage $
    V_g = -\epsilon$ for strong interaction $U=30\Gamma$ and
    (a) zero magnetic field $B=0$ and
    (b) finite field $B=9\Gamma$.
    Both figures are calculated in 2-loop RG [Eqs. \eq{1loopE} and \eq{2loopexplicit}]
    and are converged with respect to the number of non-equibrium  Matsubara axes
    (cf. Sect.\ref{sec:Non_Matsu})
    using a sufficiently large bandwidth $D=10^3\Gamma$.
    The linear conductance in (b) shows good agreement with the Friedel sum rule
    (cf. \Fig{fig:friedel}).
  }
  \label{fig:largeU}
\end{figure}

We presented Eqs. \eq{bosonfirst}-\eq{fermionlast} in order to show a number of important general properties.
First, in the fermionic sector \eq{fermionfirst}-\eq{fermionlast}, the structure is the same as for the one-loop equation, i.e., only terms
$|\alpha_{\eta \sigma}^{-})(\alpha_{\eta \sigma}^{+}|$ appear
 as discussed (\Sec{sec:eigenvectors}).
This property holds in any order of RG and is manifestation of the general properties 
\Eq{alpha+0}-\eq{alpha-0}.

Second, \HL{the terms \eq{bosonfirst}-\eq{fermionlast} are only those on the right-hand side of} the two-loop RG equations \HL{that} are relevant to the current.
\HL{They are all} proportional to the matrix elements
$(\alpha_{\eta \sigma}^{+}|P^{\alpha_{\eta \sigma}}|\alpha_{\eta \sigma}^{-})$
as was anticipated in \Sec{sec:nonint}.
This is also a general property that holds in any loop order of the RG.
In the non-interacting limit, $U=0$, this implies that  all two-loop (and higher loops) corrections \emph{relevant to the current} vanish exactly [cf. \Eq{proj_free}]. 
We emphasize \HL{that} there are additional terms not listed in  Eqs. \eq{bosonfirst}-\eq{fermionlast}
\HL{that} are irrelevant to the current.
These describe the 2-loop renormalization of the $\zeta$ coefficient
\HL{and} involve factors $(\alpha_{\eta \sigma}^{-}|P^{\alpha_{\eta \sigma}}|\alpha_{\eta \sigma}^{+})$
and therefore do \emph{not vanish}, even for $U=0$.
However, \HL{the} corrections to this coefficient beyond the two-loop order do contain the factors \eq{alpha-factor-zero} and again vanish for $U=0$.
See for more details \Cite{Saptsov13a}.
 
%%% Local Variables: 
%%% mode: latex
%%% TeX-master: "paper"
%%% End: 

% Fig 6 --> moved to include-rg2.tex

\section{Results\label{sec:results}}
In this section we perform a detailed numerical investigation of the zero-temperature two-loop RT-RG equations
$
{d\bar{L}^0 (E)}/{d\Lambda} =
{d\bar{L}^0 (E)}/{d\Lambda} |_{\mathsf{1\,loop}}
+
{d\bar{L}^0 (E)}/{d\Lambda} |_{\mathsf{2\,loop}}
$,
where the right-hand sides are given by \Eq{1loopE} and \eq{2loopexplicit}.
\hl{We calculate the current as explained after \Eq{barSigmaintegralr}.}
We focus on the dependence on the interaction $U$ and the magnetic field $B$
as \HL{a} function of both the bias \hl{$V$} and the gate voltage \hl{$V_g=-\epsilon$}.
To clearly structure the discussion we first summarize the central features of the calculated
conductance as exemplified in \Fig{fig:largeU}(a) and (b) for zero and finite magnetic field $B$, respectively,
and assess the limits of applicability.
\begin{figure}[tbp]
  \includegraphics[width=0.99\linewidth]{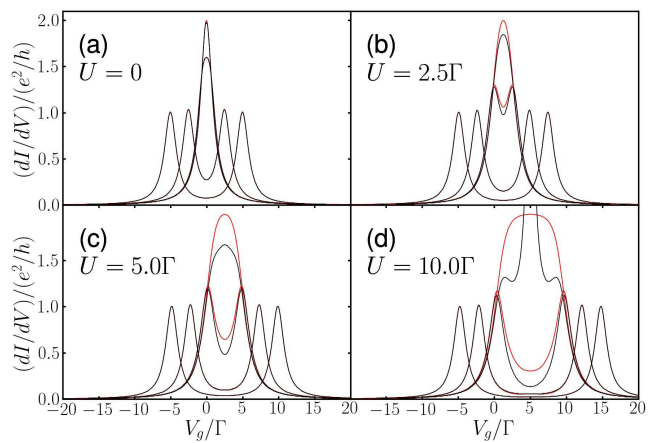}
  \caption{
    Linear conductance, $dI/dV|_{V=0}$, as function of the gate voltage $V_g$,
    obtained numerically for bias $V=0.001\Gamma$
    for which the response was checked to be linear for all $V_g$.
    Our 2-loop RT-RG results (black curves) are compared with \HL{the} Friedel sum rule
    conductance $\sum_\sigma g_\sigma$ (red curves), which are obtained from \Eq{friedel} 
    using the occupations $\langle n_\sigma\rangle$ calculated within our RG.
    Across panels (a)-(d) the interaction increases, $U/\Gamma = 0.0, 2.5, 5.0, 10.0 $.
    Within each panel the magnetic field is increased, $B/\Gamma =0.0, 1.0, 5.0, 10.0$.
    Panel (a) for $U=0.0$ serves as a reference, numerically confirming the analytic result \eq{nonint-conductance} that already in 1-loop \HL{RG} we \HL{attain} the 
exact non-interacting result \HL{for the current}.
    }
  \label{fig:friedel}
\end{figure}

\subsection{Overview and limits of applicability}
At zero magnetic field the dominant features in \Fig{fig:largeU}(a) are the Coulomb blockade diamonds defined by lines along which a 
\hl{single-electron tunneling resonance} appears.
In our calculated results, these $dI/dV$ peaks are broadened on the scale $\Gamma$ due to non-perturbative tunneling processes
and have a peak height $e^2/h$, i.e., the quantum conductance.
Going into either of the Coulomb blockade regimes, where the charge is quantized to $N=0,1$ and $2$, the current decays non-exponentially due to higher order
tunneling (cotunneling and higher order processes).
At very small bias, however, the conductance shows a pronounced anomaly,
but only in the $N=1$ regime, where the dot has an unpaired spin.
We stress from the start that this should not be naively identified with the Kondo anomaly of the Anderson model:
the correct description of the Kondo peak requires three-loop RT-RG corrections, which are beyond the scope of this work.~\cite{Schoeller09b,Pletyukhov12a}
To clarify in which regimes of voltages our results apply,
we first discuss the linear conductance through the spin channel $\sigma$: $g_\sigma=\left(d{I_\sigma}/dV\right)_{V=0}$, in particular the consistency with the Friedel sum rule,
\begin{align}
  g_\sigma=\frac { e^2} h \sin^2 (\pi\langle n_\sigma\rangle)
  \label{friedel}
  .
\end{align}
In \Fig{fig:friedel}, it is clearly seen that at zero field $B=0$ the conductance increasingly violates the Friedel sum rule between the two SET peaks 
for larger $U$. The violation \HL{becomes} maximal at the particle-hole symmetry point: our result reaches $4e^2/h$ instead of $2e^2/h$.
At best, in this regime our two-loop approach can be a starting point for further three-loop corrections containing the log-divergent Kondo corrections:
However, it should be noted that the violation is \emph{finite}, even at zero $T$:
The key observation in \Fig{fig:friedel} is that beyond a magnetic field $B \sim \Gamma$, only a \emph{renormalized} elastic cotunneling background remains and our results rapidly become consistent with the Friedel sum rule.
Clearly, at voltages above $B \sim \Gamma$,  the three-loop Kondo renormalization is expected to be negligible compared to the one- and two-loop corrections that we accounted for here.
In all further analyses, the low-bias regime $V < \Gamma$ will thus be ignored for magnetic fields $B \lesssim \Gamma$.
We do, however, show our results in this bias regime for two reasons:
(i) knowing the behavior of the two-loop scheme is of interest as it presents a starting point for future three-loop calculations,
and  (ii) the behavior of the two-loop approach can be compared with that for other methods in this regime. We note, e.g., that for $U=2.5\Gamma$, the violation of the Friedel sum rule is still rather modest, even at zero field.
Our two-loop RG thus accounts non-perturbatively for the strong tunneling effects at zero temperature, covering the complete \emph{finite-bias} 
stability diagram, where previous perturbative generalized master / kinetic equation approaches~\cite{Leijnse08a,Koller10} break down.

\begin{figure}[tbp]
  \includegraphics[width=0.99\linewidth]{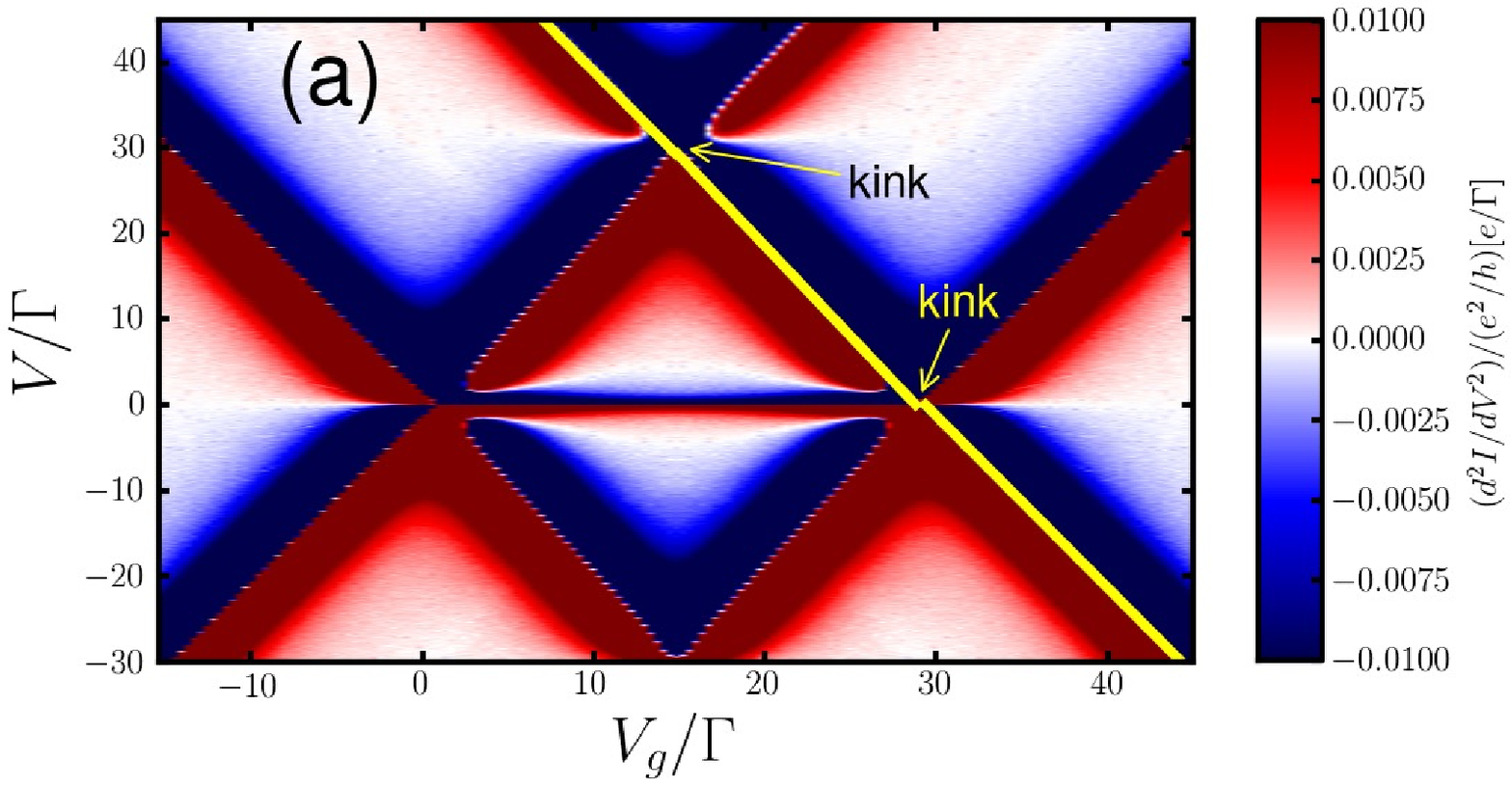}
  \includegraphics[width=0.99\linewidth]{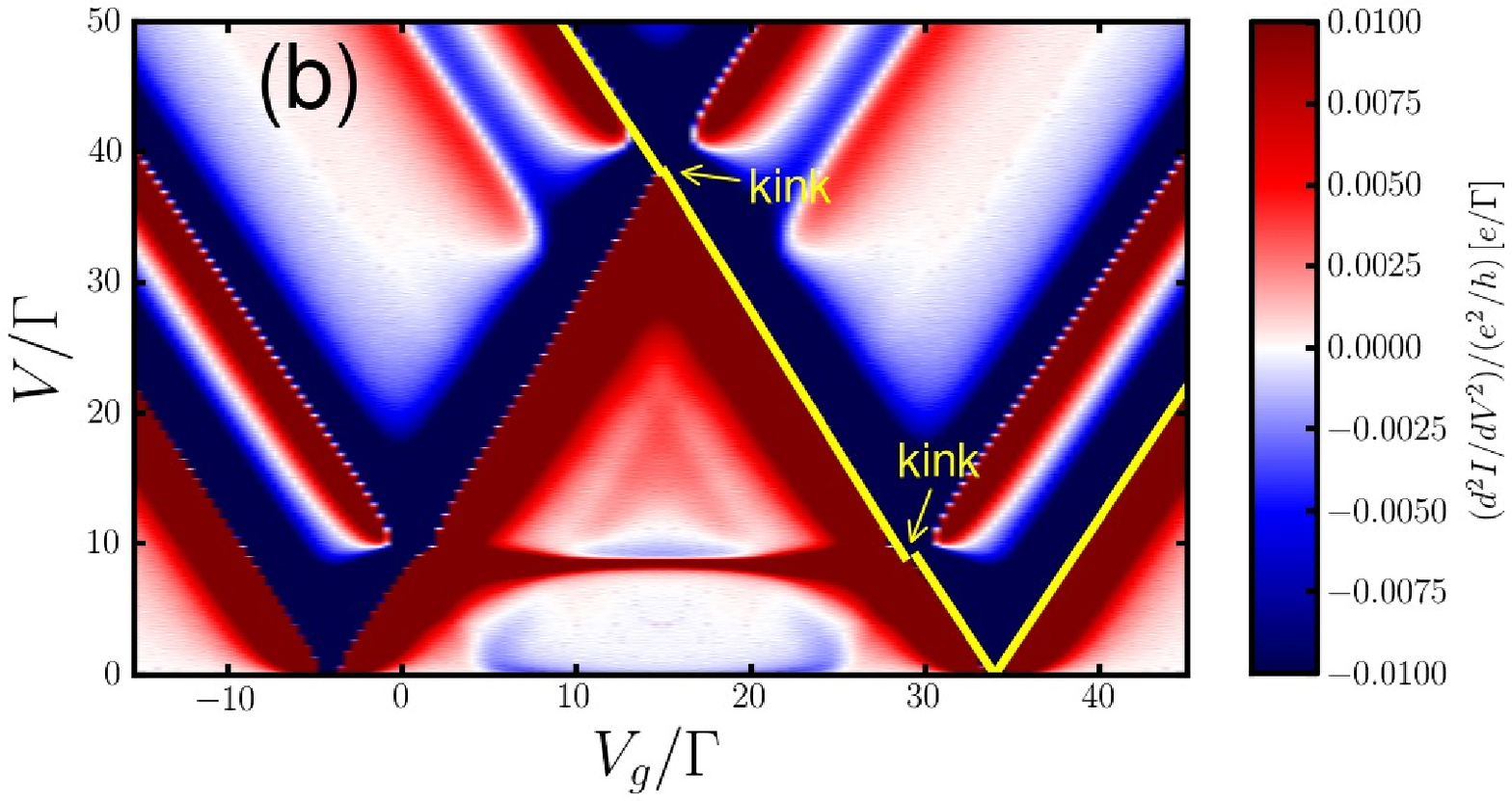}
  \caption{
    Peak positions of $dI/dV$:
      shown is $dI^2/dV^2$ in a zoom of \Fig{fig:largeU},
    making the zeros of $dI/dV$ stand out as curves
    separating red (positive) and blue (negative) regions.
    The yellow lines are guides to the eye
    obtained by accurate extrapolation of the linear parts of the resonance lines
    (including data points outside the figure).
    These emphasize the change of the slope of the linear parts of the  resonance positions,
    in addition to the non-linear renormalizations close to the kinks.
    In (a) kinks occur at $V\approx 0$ and $V\approx \pm U$,
    whereas in (b) they occur at  $V\approx \pm B$ and $V = \pm U+B$.
    Note that in (b) there is no discernible kink at $V=\pm(U-B)$:
    at this energy there is a SET resonance ``hitting'' the Coulomb blockade diamond edge
    but there is \emph{no onset of inelastic cotunneling},
    in contrast to $V=\pm B$, where there is such an excitation.
    This signals the importance of inelastic cotunneling for the appearance of such kinks.
  }
  \label{fig:largeU_kink}
\end{figure}

Based on the above, we expect that for a finite magnetic field $B \gtrsim \Gamma$, the two-loop RT-RG calculations reliably address transport features, 
illustrated in \Fig{fig:largeU}(b), at all applied voltages.
Clearly, the SET resonance peaks have been split due the Zeeman effect.
The zero-bias anomaly splits into two inelastic cotunneling resonances at finite bias $V \approx B$ (Zeeman excitations).
A  much smaller zero-bias anomaly remains, which should be ignored, as mentioned above.
The above-mentioned features are of course known from previous studies and have been observed in many experiments.
Our approach, however, includes renormalization effects of these basic transport signatures, which are non-perturbative in $\Gamma$ far from 
equilibrium. The following more detailed analysis, bearing the above restrictions in mind, indeed reveals several low-temperature renormalization 
effects
that may be of experimental relevance.

\begin{figure}[tbp]
  \includegraphics[width=0.99\linewidth]{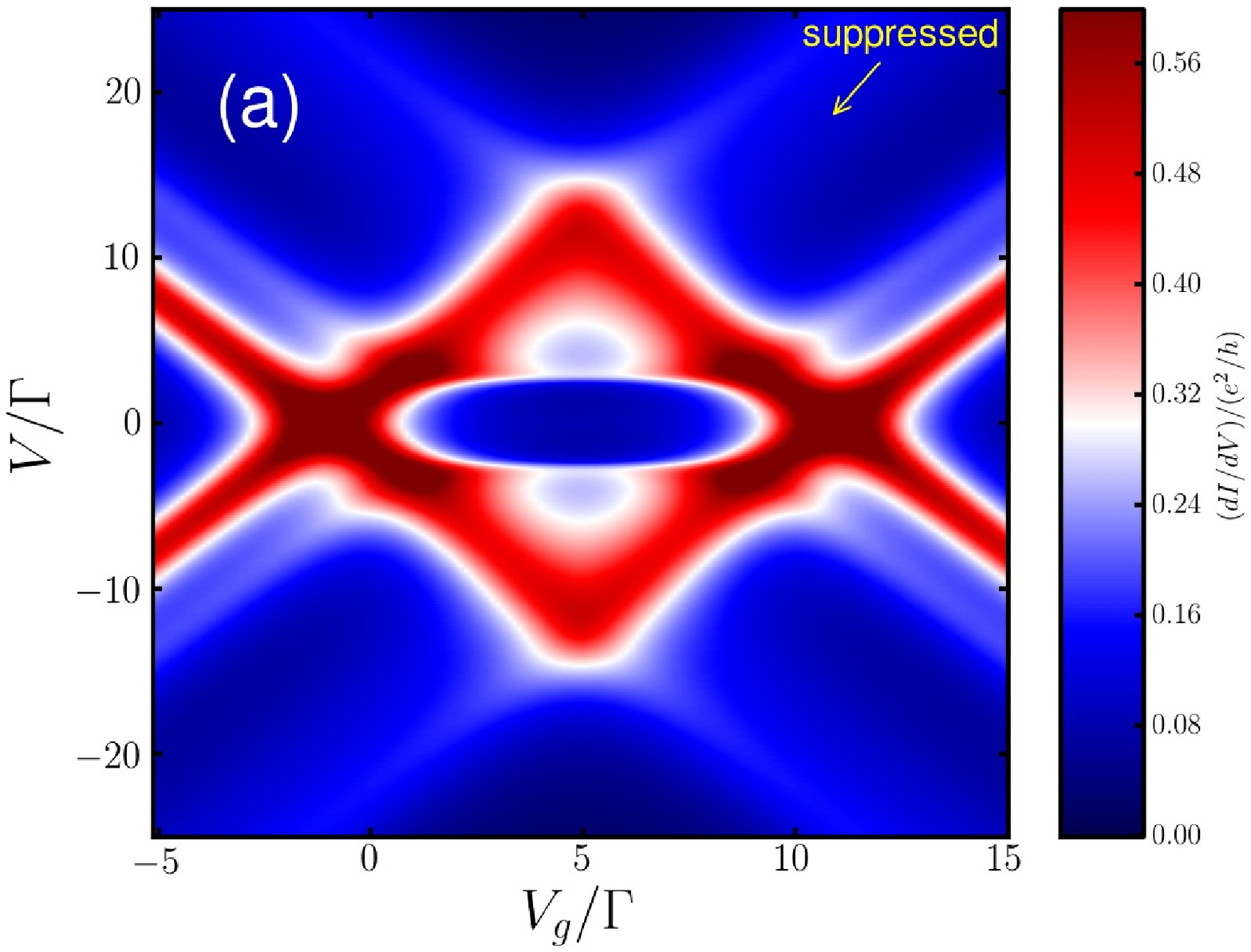}
  \includegraphics[width=0.99\linewidth]{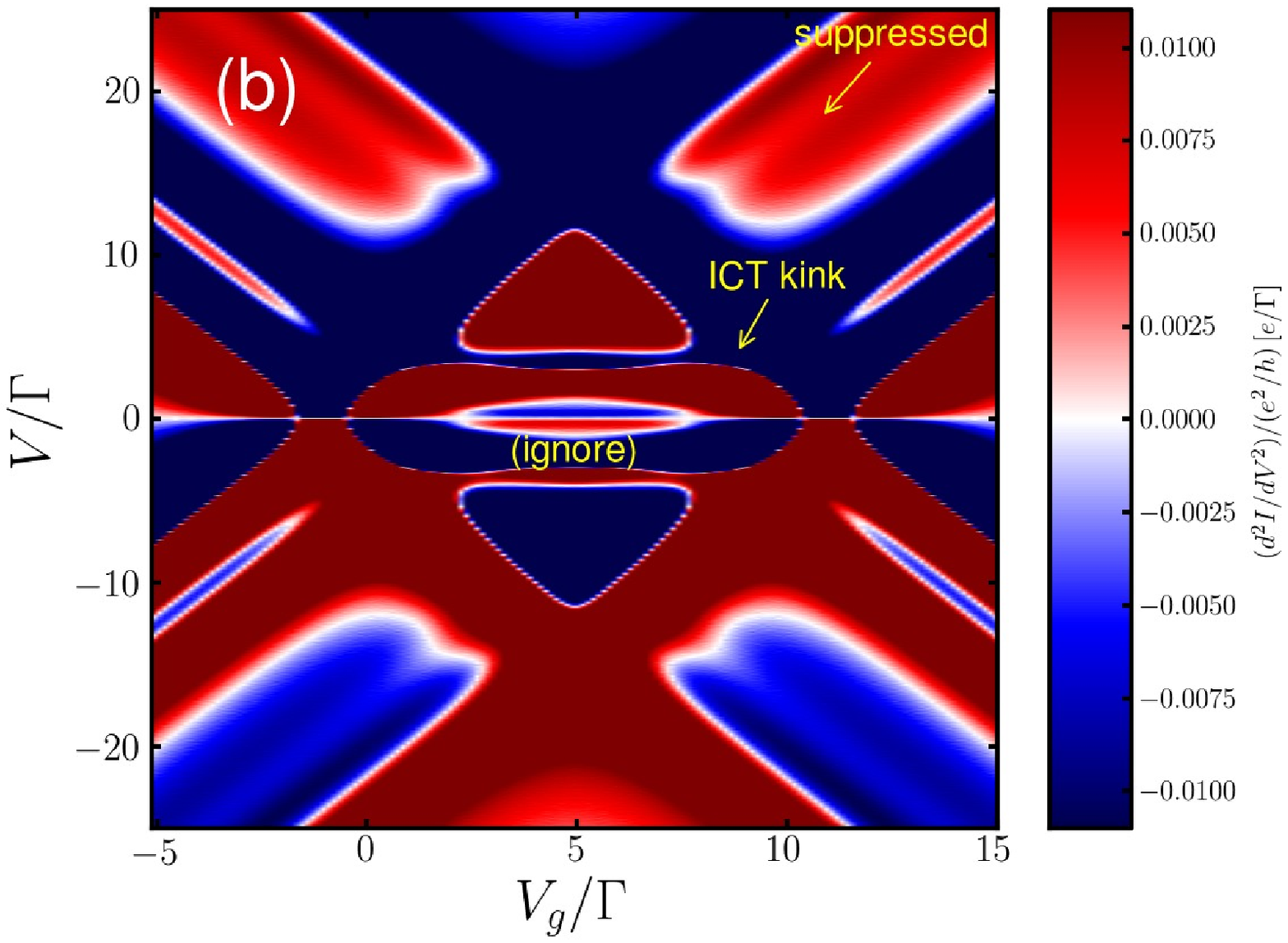}
  \caption{
    (a) Non-linear conductance, $dI/dV$, as in \Fig{fig:largeU}(b)
    but for reduced interaction $U=10\Gamma$ and magnetic field $B=3\Gamma$
    \hl{and a symmetric bias range.}
    (b) Derivative of (a),  $d^2I/dV^2$, highlighting the renormalization of \HL{the} SET peak position
    (see caption \Fig{fig:largeU_kink}(b)), in particular \HL{near} the onset of ICT. 
  }
  \label{fig:smallU}
\end{figure}

\subsection{Single electron resonance: Level renormalization and
broadening\label{sec:set}}

\subsubsection{Kinks}

Careful inspection reveals that the SET resonance lines,
in fact, change their slope when crossing $V=0$ and $U$.
This can already be seen for small $\Gamma$ \hl{in \Fig{fig:largeU_kink}(a)}, where we plot the $V$-derivative of \Fig{fig:largeU}(a) to follow the peak positions, 
adding a linear extrapolation.
The SET resonance lines of the inner diamond change in such a way that 
the charge gap hardly renormalizes for $U\gg \Gamma$,  although diamond distortions are visible by kinks in the linear extrapolations.
The charge gap can both be determined from the height of the diamond (non-linear response) or from the width of the diamond (linear response) and no 
significant deviation is found in this limit.

A related effect arises in a magnetic field:
the SET slope below (above) the ICT threshold is smaller (larger)
than the slope of the bare resonance line.
As a result, the SET lines now show a kink at finite $V=\pm B$.
We note that no such kink is seen at $V=U-B$:
this indicates that indeed the ICT is responsible for this effect
since at $V=U-B$, in contrast to $V=B$ there is \emph{no} ICT excitation.
In (a) at $V=0$ there is some non-linearity around $V=\Gamma$ (small $V$ should be ignored, see above), which persists in (b) around $V=B$ in a magnetic field.

Upon increasing $\Gamma$ relative to $U$, these effects are enhanced as shown in 
\Fig{fig:smallU}(a).
The edges of the $N=1$ Coulomb-blockade regime tend to bend inwards, towards the diamond center.
Notably, above the onset of \hl{ICT} the slope is slightly larger than that of the bare resonance line.
We furthermore observe that this also leads to the suppression of the excitation
$\ket{\downarrow} \rightarrow \ket{2}$ at $\mu_L - \mu_R = \epsilon_\uparrow + U$
in \Fig{fig:smallU} (see arrow), which is still clearly visible in \Fig{fig:largeU}, see arrows in \Fig{fig:smallU}(a) and (b).

\subsubsection{Analysis}

\begin{figure}[tbp]
  \includegraphics[width=0.99\linewidth]{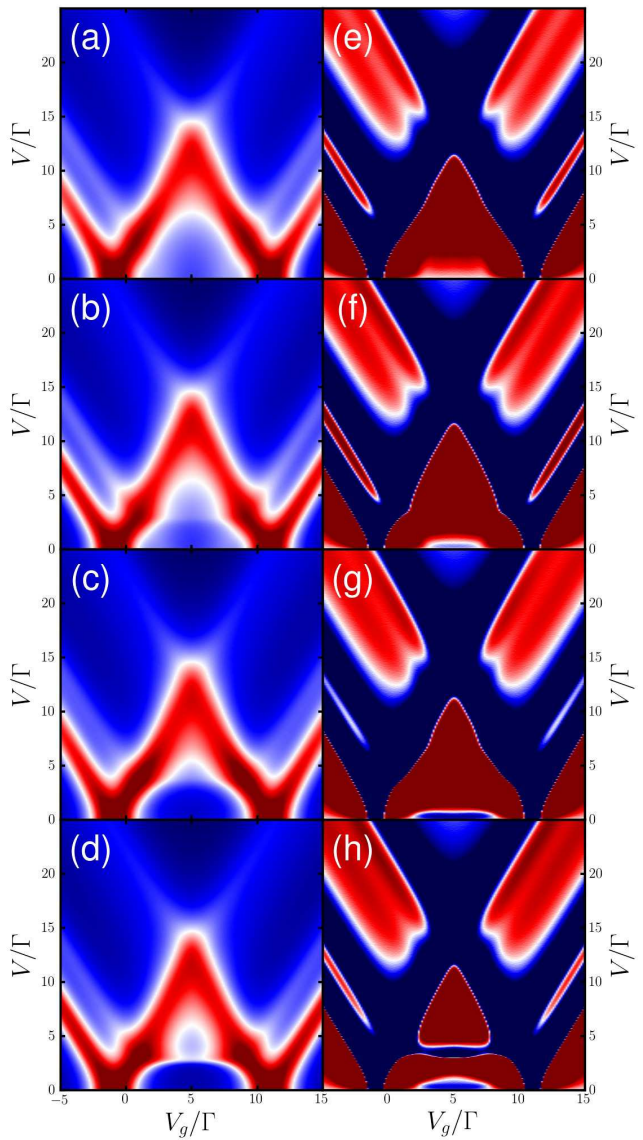}
  \caption{
    Comparison of $dI/dV$ (same color scale as \Fig{fig:smallU})
    calculated in the 1-loop (a,b) and 2-loop approximation (c,d),
    neglecting all Matsubara axes (a,c) and
    fully converging using $15$ frequency axes (b,d).
    The \HL{second} column shows the corresponding $dI^2/dV^2$ maps, allowing
    the $dI/dV$ peak positions to be followed (see caption \Fig{fig:largeU_kink}(b)).
    In particular, the inelastic cotunneling excitation at $V=B$ evolves from a step in (a)-(c) into a peak in (d).
    Clearly, the non-equilibrium Matsubara frequency dependence is responsible for the ``kinks'' in
    the SET resonance (they already appear in 1-loop with quantitative modifications in 2-loop).
  }
\label{fig:kink}
\end{figure}

It seems not possible to analytically extract a simple physical picture explaining the above non-linearities. The following analysis aims to indicate
 why this is the case: we
trace back at which stage of the 2-loop RG scheme the various effects are generated taking the parameter set of \Fig{fig:smallU}(b) as a starting point.
In \Fig{fig:kink} we show the conductance calculated both in 1- and 2- loop RG, both with and without converging the calculations with respect to the non-equilibrium Matsubara axes.

\begin{figure}[tbp]
  \includegraphics[width=0.99\linewidth]{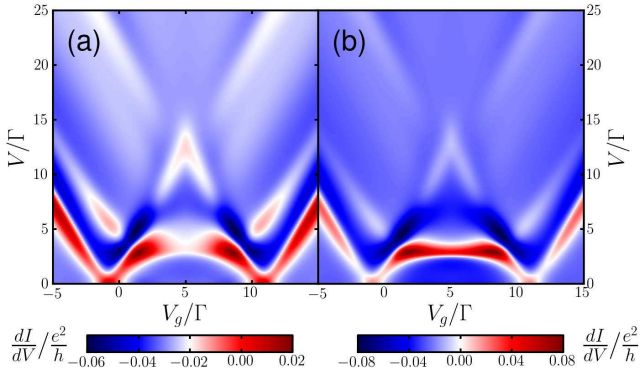}
  \caption{
    Effect of Matsubara axes convergence, \HL{i.e,}
    the converged result minus the result neglecting all Matsubara axes:
    \HL{(a) 1-loop RG, \Fig{fig:kink}(b)$-$(a) and
    (b) 2-loop RG, \Fig{fig:kink}(d)$-$(c).}
    Adjacent red and dark blue regions indicate that the correction is
    an S-shaped curve, which, when added to a peaked curve,
    results in a shift the peak position.
    \key{Clearly, the Matsubara frequency dependence
      has an impact on the \emph{positions} of all resonances
    and should be fully accounted for.}
  }
\label{fig:diffmatsubara}
\end{figure}

Clearly, the different slopes and non-linearities  already arise in the one-loop RG: \HL{this} is visible \HL{from} \Fig{fig:kink}(b) \HL{and (f)}.  Their strength correlates with that of
the signatures of ICT appearing in the stability map in the various
approximations.
However, \HL{this effect only arises when} the Matsubara axes are accounted for:
this is seen by directly comparing \Fig{fig:kink}(a) \HL{and (b)} and is
confirmed by \Fig{fig:diffmatsubara}, where we explicitly plot the
difference of former two figures.

\begin{figure}[tbp]
  \includegraphics[width=0.99\linewidth]{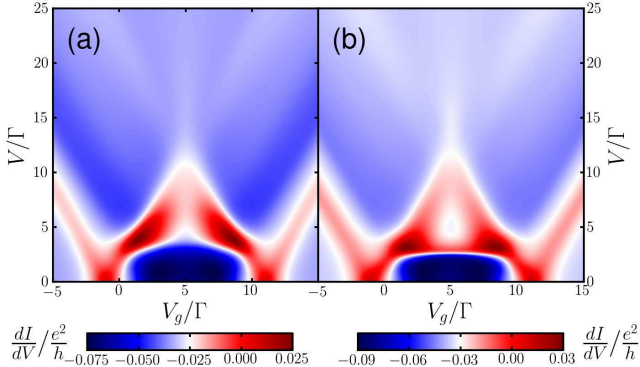}
  \caption{
    Effect of 2-loop corrections, \HL{i.e., the 2-loop result minus the 1-loop result
    (a) without Matsubara frequency dependence,
    \Fig{fig:kink}(c)$-$(a);
    (b) with converged Matsubara frequency dependence,
    \Fig{fig:kink}(d)$-$(b).}
    Note the positive corrections to \HL{the} \emph{magnitude} of the inelastic cotunneling in (b).
  }
\label{fig:diff12}
\end{figure}

Overall, the two-loop corrections are most pronounced along the
SET-regime boundaries and the ICT
threshold as comparison of \Fig{fig:kink}(b) and \Fig{fig:kink}(d) and
the plot of their difference in \Fig{fig:diff12} shows.  Also, in 2-loop order,
the kinks are most pronounced in the Matsubara-converged result.
We conclude that the 2-loop fluctuation effects result in a non-trivial energy dependence of the vertices and Liouvillian, which shows up in anomalous features of the measurable stability diagram, 
even for such a simple Anderson model of  a quantum dot.
We emphasize that these features are not related to renormalization processes that generate the Kondo effect (3-loop, not included here)
and have an effect at $V \sim B\gtrsim \Gamma$.

\subsubsection{Experimental implications}

Having traced the origin of the change of the slopes and the non-linearities of the SET
resonances,
we \HL{now} discuss their relevance to experiments.  In fact, kinks
in SET resonances are often observed in various type of quantum dot systems.~\cite{Goss11,Eliasen10,Tans98,Oosterkamp99}
Our calculations indicate that tunnel-induced renormalization
is a possible mechanism for their occurrence, but other
(e.g., electrostatic) mechanisms~\cite{Reimann99,Reimann02} should not be ruled out in an experimental situation.  However, for strong coupling, it is physically
not unexpected that when ICT sets on, the level renormalization
significantly changes, resulting in such a kink.

\begin{figure}[tbp]
 \includegraphics[width=1.0\linewidth]{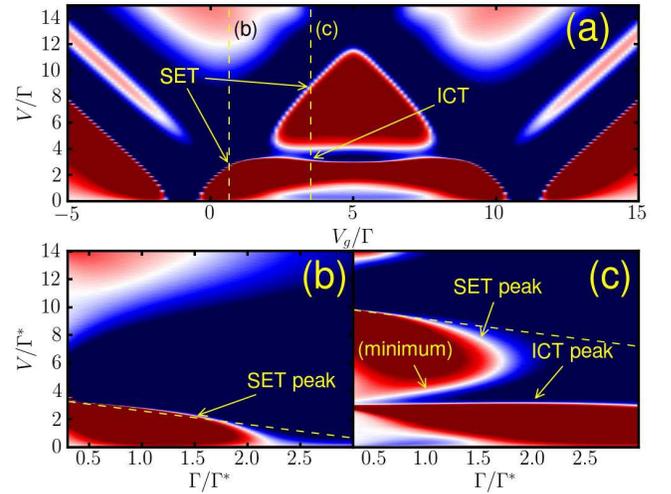}
  \caption{
    Distortion of the stability diagram with increasing tunnel coupling $\Gamma$.
    (a) Zoom in of \Fig{fig:smallU} of  $dI^2/dV^2$ (same scales and units),
    highlighting the gate-voltage dependence of the SET and ICT bias-thresholds generated by tunnel renormalization.
    Here, $U=10.0\Gamma^{*}$, $B=3.0\Gamma^{*}$ and $\Gamma^{*}$ is the reference value of $\Gamma$.
    In panel (b) and (c) we show the evolution of the zeros of $dI^2/dV^2$ as the tunnel coupling is increased \HL{for fixed $U$ and $B$.
    Here, $\Gamma$ varies from} $0.3\Gamma^{*}$ to $3.0\Gamma^{*}$ along a fixed gate-voltage cut in (a), marked the vertical dashed line
    $V_{g}= -\epsilon = 0.65\Gamma^{*}$ for (b) and
    $V_{g} = -\epsilon = 3.50\Gamma^{*}$ for (c).
    The dashed linear approximation to the renormalized SET positions in (c) is copied with a vertical offset to (b), showing that the renormalization is \emph{non-uniform} in the gate-voltage.
    This signals a distorted stability diagram.
    The ICT peak clearly has a weaker $\Gamma$ dependence
    \HL{which, moreover, can be seen to be non-monotonic when calculated for a larger $\Gamma$ range.} 
  }
  \label{fig:kink_gamma}
\end{figure}

A direct test of this assumption would be to track the Coulomb diamond as a
function of the coupling strength $\Gamma$.
\hl{In \Fig{fig:kink_gamma} (b) and (c)} we show predictions for the
evolution of the SET resonance point for two fixed gate voltages, one below  and one above the ICT threshold, respectively.
The main observation from such
a plot is that the peaks evolve along curves that are \emph{not} simply offset by a constant bias. This indicates that the renormalization of SET resonance becomes increasingly nonlinear and a kink must develop.  (Note that experimentally $\Gamma$ may change non-linearly with control voltages but this does not spoil the argument.)
It is important to properly choose the point above the ICT threshold: depending on the gate voltage position the peak may renormalize
stronger or weaker than the peak below the threshold.
We are aware that experimentally such tuning of $\Gamma$ with gate voltages
 may lead to other side effects \HL{that} may be hard to distinguish
from the effect.
Here the different renormalization of the ICT resonance can be of use, which is discussed next.

\begin{figure}[tbp]
  \includegraphics[width=0.99\linewidth]{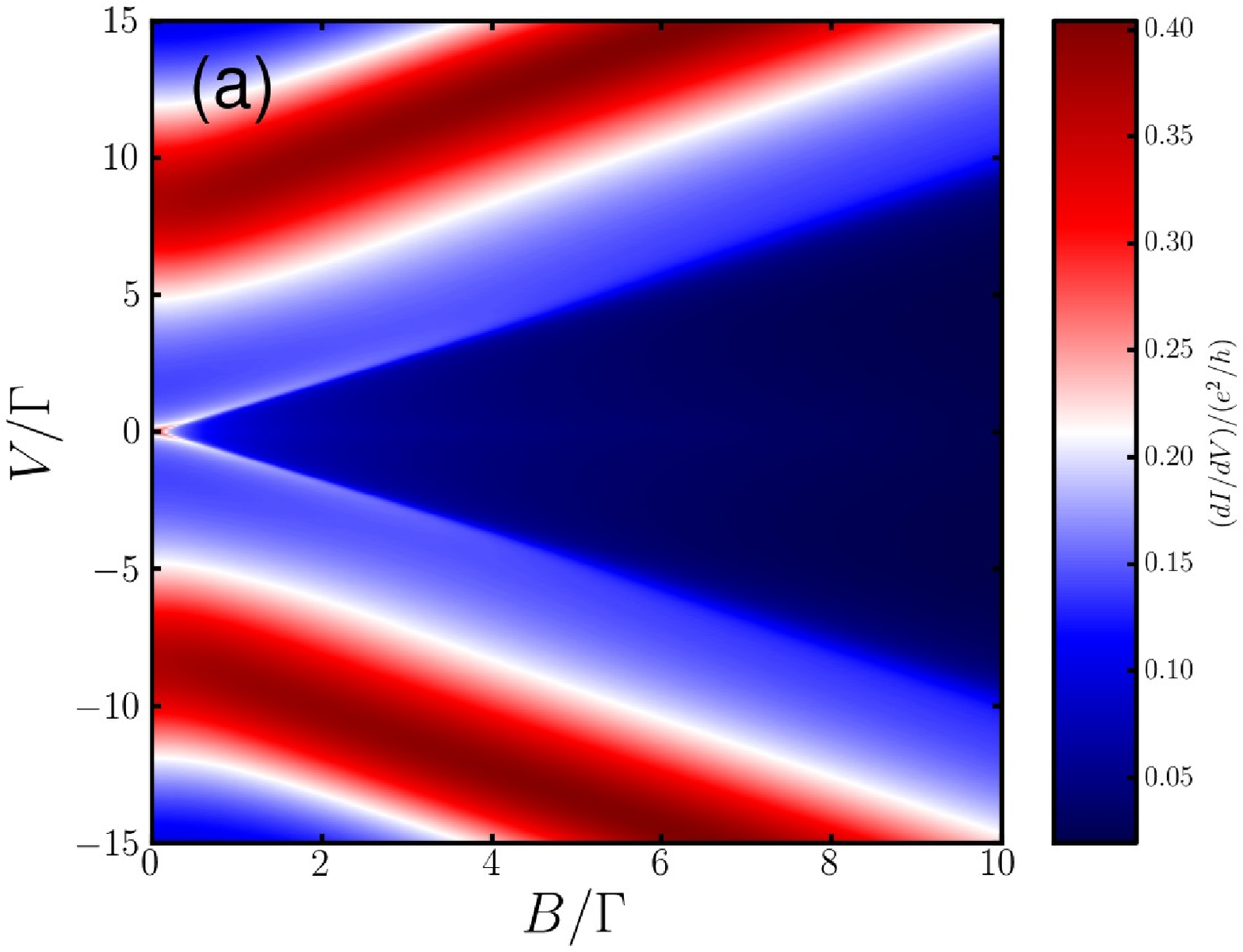}
  \includegraphics[width=0.99\linewidth]{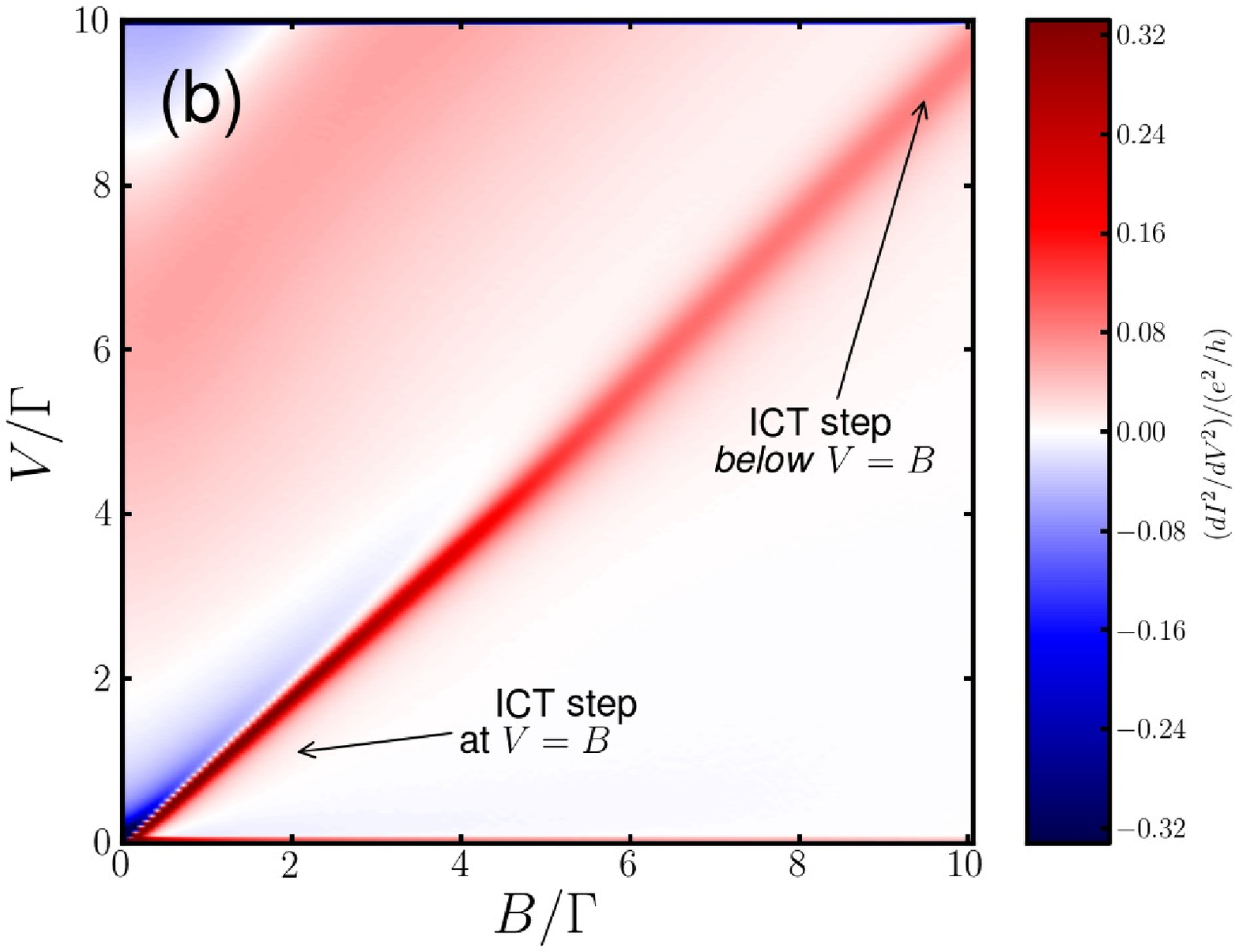}
  \includegraphics[width=0.99\linewidth]{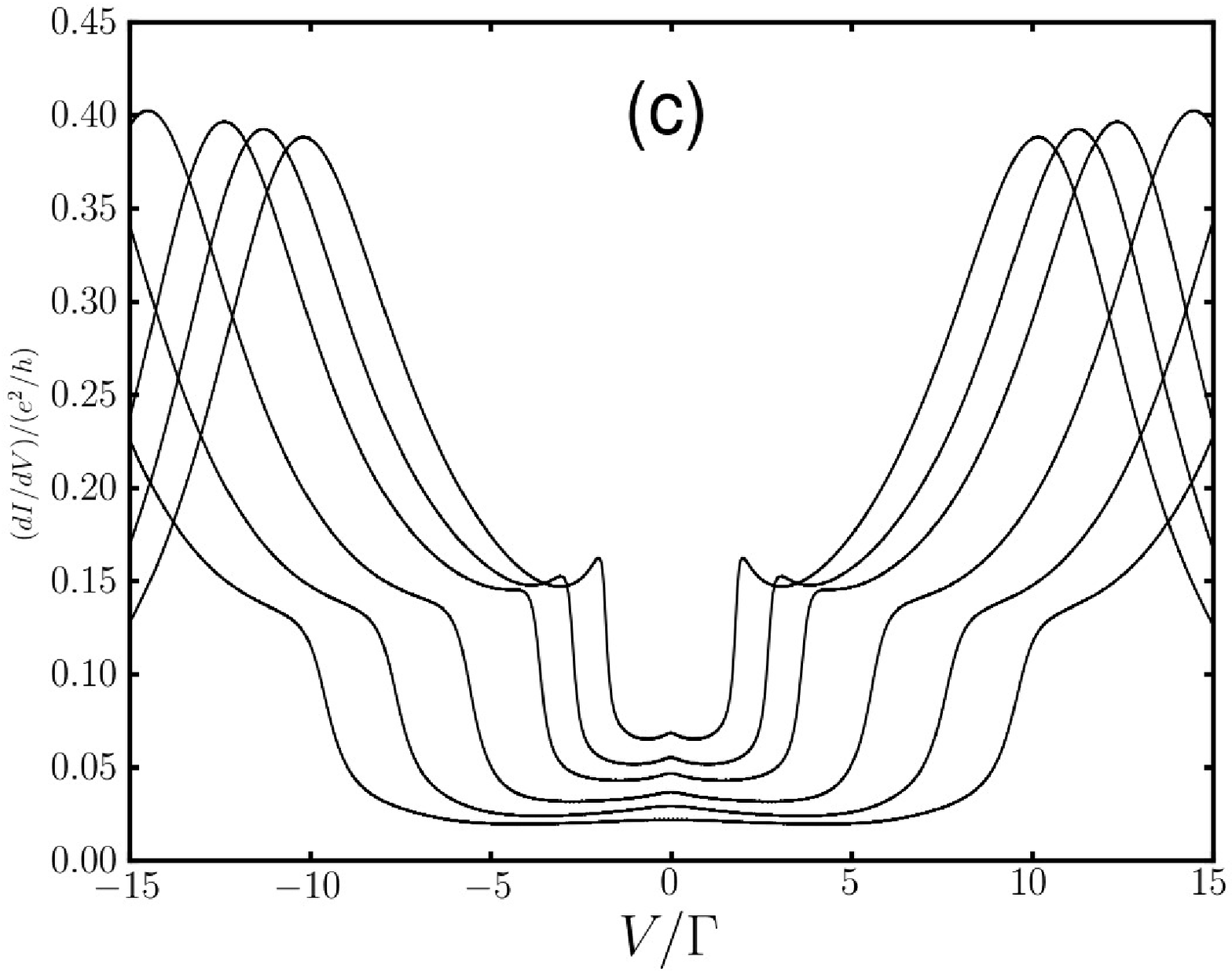}
  \caption{\hl{
    (a)
    Zeeman splitting of the cotunneling resonance (change in blue color)
    and shift of the SET peaks (red peak)
    for $U=30\Gamma$ and $V_g = - \epsilon = U/6 = 5\Gamma $.
    (b) Zoom in of the bias-derivative of (a), i.e., $dI^2/dV^2$
    showing the resonance position.
    The sharp peak in $dI/dV$ in (a), starting from the lower left corner, closely follows the diagonal line $V=B$.
    The step in $dI/dV$ in (a), starting from the upper right corner,
    instead follows a line that is parallel to the diagonal $V=B$ but offset by a constant, negative magnetic field (in this case $\approx 0.45 \Gamma$).
    (c) $dI/dV$ traces of (a) for $B/\Gamma = 2.0, 3.0, 4.0, 6.0, 8.0, 10.0$
  }
  }
  \label{fig:zeeman}
\end{figure}

\subsection{Cotunneling resonance: Gap renormalization
 and reduced broadening \label{sec:}}

Having discussed the effect of the ICT on the SET resonances, we now study
the ICT features themselves in more detail.
\Fig{fig:zeeman} shows how the inelastic cotunneling resonance exhibits a Zeeman splitting with increasing magnetic field.
Despite the zero temperature, the width of the inelastic cotunneling feature is finite and \HL{can be} clearly seen to depend on the magnetic field, and thereby, on the voltage at which this resonance occurs.
This is in contrast to high-temperature 2-loop perturbation theory~\cite{Leijnse08a,Koller10} where this resonance appears as a thermally broadened feature at the unrenormalized excitation energy.
At larger voltage $V=B$, the resonance width increases, reflecting a decreasing life-time.
This energy-dependence of the width is generated by our 2-loop renormalization
since the initial Liouvillian of the RG flow, \Eq{initial}, has imaginary parts that are all $\sim \Gamma$.

Next, as the magnetic field is reduced, but still on the order of several times $\Gamma$, the differential conductance develops a pronounced peak on top of the inelastic tunneling step:
in \Fig{fig:zeeman}(b) this is signaled by the onset of a negative second-derivative  of the current (blue) and \HL{is clearly seen in the} conductance traces in \Fig{fig:zeeman}(c).
It is known that part of such a peak on top of the well-known inelastic tunneling step~\cite{Lambe68} is due to non-equilibrium occupations~\cite{Eto01,Paaske06,Leijnse08a,Zyazin10} that we also fully take into account.
The enhanced conductance at the cotunneling resonance is due to the 2-loop renormalization,
 including the frequency dependence.
Only after including both 2-loop corrections \emph{and} converging with respect to the Matsubara frequency axes
the ICT resonance evolves from a $dI/dV$ step into a peak.
This was illustrated (for smaller $U$) in \Fig{fig:kink}: (a)-(c) show no ICT peak (since their derivatives (e)-(g) have no zero at the ICT threshold $V \approx B$)
in contrast to \Fig{fig:kink}(d), see also \Fig{fig:kink_gamma}.

This enhancement is not related to Kondo-exchange tunneling, which is known to lead to additional logarithmic enhancements~\cite{Paaske06,Schoeller09b}: their renormalization is not included in our 2-loop calculations and is expected to be of limited importance at this large magnetic fields (several times $\Gamma$).
The qualitative change of the inelastic conductance feature from a step to a peak in \Fig{fig:zeeman}(b) implies some ambiguity in the extraction of the excitation energy, either \HL{from} the inflection point of the 
step (upper right corner) or \HL{from} the peak position (lower left corner).

Finally, close inspection \Fig{fig:kink_gamma}(a) shows that the ICT bias peak position has a weak gate voltage dependence.
In contrast, in the high-temperature limit this resonance lies at the unrenormalized cotunneling excitation $V=B$.
Strong tunneling thus leads to an apparent renormalization in the cotunneling peak position, even for this simple model, cf.~\cite{Holm08}.
Although our effective Liouvillian contains parameters $E_\sigma$ and $F_{\eta\sigma}$ that relate to the magnetic field (compare \Eq{initial} and \Eq{obliv}),
these parameters depend on energy and \HL{our full result requires
their values at several different energy scales}
 since many non-equilibrium Matsubara frequency axes must be accounted for.

\section{Overview and outlook\label{summary}}

We have studied the standard model of an interacting quantum dot, 
an Anderson impurity, in the low-temperature, non-equilibrium limit.
We have calculated the effective time-evolution kernel $L(z)=L+\Sigma(z)$ for the kinetic equation of the reduced density operator, 
where $\Sigma$ is the non-trivial self-energy superoperator.
In contrast to many previous studies using such a generalized / quantum master equation approach, we have calculated \HL{the} time-evolution kernel using the 
real-time renormalization group (RG).
The equations for the effective kernel are integrated as a function of a cutoff parameter $\Lambda$:
as $\Lambda$ is reduced, the renormalized Liouvillian $\bar{L}_{\Lambda}$ in principle flows to the exact result, 
$\bar{L}_{\Lambda}|_{\Lambda=0}= L(z)$.
This RG calculation of the effective kernel involves a number of key elements:
\begin{itemize}
\item
  Transitions between \emph{all} \emph{Liouville space supervectors} need to be accounted for.
  This includes those elements of $\bar{L}_{\Lambda}$ and $L(z)$,
  which in a perturbative calculation of the kernel drop out
  due to conservation laws (charge, spin, and possibly particle-hole symmetry).
  The reason for this is that  as one integrates out energy scales,
  effectively higher-order diagrams are included into a renormalization of the kernel
  \HL{that} describe \emph{virtual} intermediate states,
  which are less restricted by conservation laws.
\item
  The dependence of the \HL{renormalized} kernel on the \emph{real QD frequency ($E$)}
  (Laplace variable conjugate to time)
  is important
  \emph{even in the stationary state}.
  During the RG flow the renormalized Liouvillian at frequency $E$ self-consistently couples
  to its action on virtual intermediate states at frequencies differing from $E$
  by multiples of the bias voltage $\mu_L-\mu_R$.
  We have shown that this non-equilibrium effect leads to significant quantitative corrections
  and may require tens of bias-multiples to achieve convergence.
\item
  The \emph{reservoir-frequency} ($i\omega$) dependence of both the kernel and the vertices
  becomes important when going beyond the leading, 1-loop approximation,
  in addition to the QD frequency $E$.
  This dependence is generated on the imaginary frequency axis when the reservoirs are integrated out
  and it may \emph{cancel} 2-loop corrections calculated without frequency dependence.
\end{itemize}
We have systematically accounted for the leading frequency correction
within a one-plus-two-loop approximation to the exact RT-RG equations
and derived an effective RG equation for the time-evolution kernel \emph{only}.
This includes the relevant vertex renormalization corrections.
Importantly, the leading frequency correction of the one-loop renormalization of the Liouvillian $\bar{L}_{\Lambda}$ was found to exactly cancel the two-loop zero frequency term.

For the non-interacting Anderson problem ($U=0$), we found that the current is exactly captured already in the one-loop approximation without any frequency dependence
even though the complete solution (i.e., including the density operator) is contained
 only within the one-plus-two-loop approximation.
For the strongly interacting case of interest the non-linear transport spectrum ($dI/dV$ stability diagram) was calculated for a wide range of 
parameters. The different, intrinsic broadening of the single-electron tunneling and cotunneling resonances was captured,
as well as the zero-temperature renormalization of their positions.
As emphasized throughout, due to the restriction to one- and two-loop diagrams, the \HL{small} Kondo regime cannot be addressed. 
This regime has been recently studied in detail using the RT-RG approach based on the mapping to a Kondo model. 
This allowed the entire crossover from weak to strong coupling to be described.~\cite{Pletyukhov12a}
Our study thus provides a starting point for a three-loop analysis of the non-equilibrium Anderson model in which the interplay of Kondo 
spin fluctuations and charge fluctuations can be described.

The RT-RG study benefited a lot from an extensive reformulation of the underlying real-time perturbation theory in terms of vertex superoperators 
$\bar{G}$ and $\tilde{G}$.
Although originally introduced in the \HL{context of the RT-RG~\cite{Schoeller09a}}, we have revealed their full significance as fermionic field superoperators that directly 
generate the Liouville Fock space in complete analogy to \emph{closed} quantum many-body systems:
\begin{itemize}
\item
  Field \HL{superoperators} obey definite anticommutation relations,
  and a simple fluctuation-dissipation relation similar to \HL{that of the} underlying \HL{usual} field operators.
  The Wick theorem in Liouville space at finite temperature can be obtained 
  \emph{algebraically} using relation \eq{FDT} \HL{in the usual way,~\cite{Gaudin60}
    avoiding the necessary careful explicit accounting of fermion-parity signs in other proofs.~\cite{Schoeller09a}
    This difficult aspect of Liouville-space fields was noted earlier.~\cite{Mukamel03,Mukamel08}}
\item
  The causal structure of the theory
  is reflected by the vanishing of 2 out of 4 reservoir correlation functions
  (rather \HL{than} 1 out of 4, as in the Keldysh Green-function technique).
   This results in an exponential reduction of the \HL{number of} terms contributing
  to the time-evolution kernel
  (additional to the reduction in the wide-band limit).
  The remaining contributing diagrams are easily identified by their topology.
\end{itemize}  
We have extended the use of these fermionic field superoperators to the perturbation theory underlying the RT-RG.
This resulting \emph{causal representation} of the perturbation theory has many advantages:
\begin{itemize}
\item
  Probability conservation of the kernel is manifest term-by-term,
  allowing non-conserving approximations to be easily spotted.
    In addition, other \emph{exact} eigenvectors and eigenvalues of the kernel were
  found, \HL{which} limit the form of the exact effective Liouvillian $L(z)$
  as expressed in our central result~\Eq{obliv}.
\item
  Term-by-term \emph{diagrammatic} evaluation of the wide-band limit.
  Diagrams that vanish in this limit can be identified by their topology
  and the remaining diagrams \HL{can be shown to be independent of the cutoff
  using} the fundamental fermionic algebra of the field superoperators.
  This results in a further strong reduction in the number of contributing terms
  as function of the perturbation order.
  Moreover, the advantage of working with the complete space of the QD states becomes explicit.
\item
  Finally, the fundamental importance of the infinite temperature limit
  becomes explicit.
  It defines the Liouville Fock-space vacuum
  and the perturbation theory can be explicitly decomposed into
  infinite temperature renormalization effects
  and the non-trivial finite-temperature contributions.
  This provides a natural starting point for the RT-RG,
  which readily suggests itself.
\end{itemize}
Aside from their application to the RT-RG, we have already found useful application of some of these points in perturbative studies, as discussed here in connection \HL{with}~\Cite{Contreras12} and \HL{in forthcoming works on time-dependence~\cite{Saptsov13a} and adiabatic driving.~\cite{Reckermann13}}

\acknowledgments
This work was initiated by intense discussions with H. Schoeller, which are gratefully acknowledged.
We thank D. DiVincenzo, T. Costi and J. Splettstoesser for valuable comments and A. Flesch, E. Gorelov and F. Reckermann for suggestions regarding the numerical calculations.
We are grateful to D. Kosov for pointing out Prosen's work and his kind help in establishing the relations between our and other
existing superfermion approaches. 
This work has also benefited from a preliminary study undertaken together with M. Kurz.
 \appendix
\section{Wick theorem for causal field superoperators of the reservoirs $J^{q}$\label{sec:Wick}}
Here, we show how the algebraic proof of the Wick theorem~\cite{Gaudin60} for standard field operators directly applies to the field superoperators $J$ in the causal representation.
In this proof, one considers the average of $n$ reservoir field superoperators $J_i^{q_i}$ defined \HL{by} \Eq{kr_el},
\begin{align}
  X=\Tr{\R} \left( J_1^{q_1} ... J_n^{q_n} \rho^R \right)
  ,
\end{align}
which \hl{can be} nonzero only for even $n$.
\HL{One then} commutes $J_n^{q_n}$ to the left-hand side, using that the field \HL{superoperators} obey the anticommutation relations \eq{commut}.
We first consider the case $q_n=-$ and make use of \HL{the} zero-trace property \eq{trJ} of the causal representation.
This reduces the average to expressions of the same form,
\begin{align}
  X  = &\sum \limits_{k\neq n}   (-1)^{N_{k,n}}
  \tilde{\gamma}_{k,n}
  X_{k,n}
  \label{Xrecursion}
  ,
\end{align}
but with averages over $n-2$ operators
\begin{align}
  X_{n,k} = &
  \Tr{\R} \left( J_1^{q_1} ... J_{k-1}^{q_{k-1}} J_{k+1}^{q_{k+1}} ... J_{n-1}^{q_{n-1}} 
    \rho_{res}\right)
  ,
\end{align}
weighted with \HL{the} contraction function
\begin{align}
  \tilde{\gamma}_{k,n} = \frac{\Gamma_0}{2\pi} \delta_{k,\bar{n}}  \delta_{+,q_k}
  .
\end{align}
Here, $N_{k,n}=n-k-1$ is the number of permutations to bring $J_k^{q_k}$ and  $J_n^{q_n}$ together.

For $q_n=+$ we proceed in the same way, except that \HL{before permuting $J_n^{q_n}$  to the far left} we apply the fluctuation-dissipation superoperator identity \eq{FDT}. 
This transforms the expression into that for the $q_n=-$ case multiplied with the Keldysh distribution function: \Eq{Xrecursion} applies again but with $\tilde{\gamma}_{k,n}$ replaced by
\begin{align}
  \bar{\gamma}_{k,n} =  \tanh(\eta_n\omega_n/2T_{r_n}) ~ \tilde{\gamma}_{k,n}
  ,
\end{align}
\HL{where} the multiindex $n$ of  $J_n^{q_n}$ reads \HL{$n=\eta_n, \sigma_n, r_n, \omega_n $.}

We thus find the standard Wick recursion relation
\begin{align}
  X  = &\sum \limits_{k\neq n}   (-1)^{N_{k,n}}
  \gamma_{k,n} 
  X_{k,n}
  \label{Xrecursion_f}
\end{align}
with $\gamma_{k,n} =\bar{\gamma}_{k,n} \delta_{+,q_n} + \tilde{\gamma}_{k,n} \delta_{-,q_n}$,
which by iteration gives the Wick theorem \eq{wick} with \HL{the} contractions \eq{ret-contr}-\eq{keld-contr}.
 
\HL{\section{Main properties of the causal vertex superoperators $G^q$\label{sec:operG}}}

\paragraph{``Bare'' vertices}
In \Sec{sec:causal} the main properties \eq{commutG} and \eq{Gconj}
\HL{of} the causal superoperators $\bar{G}$ and $\tilde{G}$, \HL{respectively}, were introduced,
namely, their anticommutation relations \eq{commutG} and their relation by Hermitian conjugation in Liouville space, \Eq{Gconj}. We now give the proof of the latter relation.
We note that both relations are fundamental as they imply a formal correspondence of causal $G^{\pm}=\bar{G},\tilde{G}$ operators with the usual fermionic field operators and allow us to develop the ``\HL{second} quantization'' technique for fermionic Liouville Fock-space.

Super-Hermitian conjugation is defined relative to the 
 scalar product in Liouville space, $(A|B)=\mathrm{Tr} A^\dagger B$, where $A$ and $B$ are dot operators.
To prove \Eq{Gconj} we first notice that the ``naive'' field superoperators \eq{keld_g} obey
\begin{align}
  \left(\mathscr{G}^p_1\right)^\dagger=\mathscr{G}^p_{\bar{1}}
  \label{keld-g-conj}
  ,
\end{align}
where $1=\eta\sigma$ and $\bar{1}=\bar{\eta}\sigma$.
Superoperators with the same Keldysh index $p$ are thus conjugated to each other in the usual way: Hermitian conjugation is equivalent to inverting 
the particle-hole index $\eta$.
This follows \HL{from} the cyclic property of the trace:
\HL{\begin{align}
  (A|\mathscr{G}^{+}_1|B)
  &=\Tr{D} A^\dagger d_{\eta, \sigma} B
  &=(\Tr{D} B^\dagger d_{\bar{\eta}, \sigma} A)^*
  &
  \nonumber
  \\
  &=(B|\mathscr{G}^{+}_{\bar{1}}|A)^*
  &=(B|(\mathscr{G}^{+}_1)^\sdagger|A)^*
  &
  ,
  \\
  (A|\mathscr{G}^{-}_1|B)
  &=\Tr{D} A^\dagger B d_{\eta, \sigma}
  &=(\Tr{D}d_{\bar{\eta},\sigma} B^\dagger  A)^*
  &=(\Tr{D}  B^\dagger  A d_{\bar{\eta},\sigma})^*
  \nonumber\\
  &=(B|\mathscr{G}^{-}_{\bar{1}}|A)^*
  &=(B|(\mathscr{G}^{-}_1)^\sdagger|A)^*
  &
  ,
\end{align}}where $*$ denotes complex-conjugation and $\dagger$ denotes either usual Hermitian-conjugation or super-Hermitian-conjugation depending on whether it acts on an operator or superoperator, respectively. 
These superoperators, however, have the disadvantage that they satisfy no definite fermionic or bosonic commutation relations [cf. \eq{keld_g-comm}].

The transformed field superoperators \eq{semi-naive_g}, which include the fermion-parity sign, \hl{obey the definite anticommutation relations \eq{semi-naive-comm}}. 
However, they are not related by super-Hermitian conjugation in the usual way:
\begin{align}
\label{semi-naive-conj}
  \left(\mathcal{G}_1^p\right)^\dagger = p~\mathcal{G}_{\bar{1}}^p
  ,
\end{align}
which follows from \Eq{keld-g-conj}
and the properties
$\left(L^n\right)^\sdagger=L^n$ and \hl{$[L^n, \mathscr{G}^p_1]_{-}=\eta \mathscr{G}^p_1$}
 of $L^n=[n,\bullet]_{-}$:
\begin{alignat}{2}
  \left(\mathcal{G}_1^p\right)^\dagger
  &= (p^{L^n}\mathscr{G}^p_1 )^\dagger
  &
  &= (\mathscr{G}^p_1 )^\dagger p^{L^n}
  \nonumber \\
  &= p^{L^n+1}\mathscr{G}^p_{\bar{1}}
  &&
  = p~\mathcal{G}_{\bar{1}}^p
  .
\end{alignat}

Finally, the causal field superoperators \eq{kr_d} obtained from $\mathcal{G}^p$ by a Keldysh-rotation
obey both definite commutation relations \eq{commutG} \emph{and} standard Hermitian-superconjugation relations \eq{Gconj}: using $p^2=1$ \HL{we obtain}
\begin{alignat}{2}
  (G^q_{1})^\sdagger 
  &= \tfrac{1}{\sqrt{2}} \sum_p p^{\hl{(1-q)/2}} (\mathcal{G}^{p}_{1})^\sdagger
  \nonumber \\
  &= \tfrac{1}{\sqrt{2}} \sum_p p^{\hl{(3-q)/2}} \mathcal{G}^{p}_{\bar{1}}
  & & = G^{\bar{q}}_{\bar{1}}
  .
\end{alignat}
This clearly demonstrates the fundamental advantage of the causal representation over the other ones.

Note that due to the \HL{$p$-dependence of the} superoperators entering \HL{into} the definition \eq{kr_d} of our superoperator $\bar{G}_{\eta\sigma}$,
the latter is either a commutator or \HL{an} anticommutator with the fermionic operator \HL{$d_{\eta,\sigma}$},
\emph{depending on the argument} on which it acts. For example 
 $\bar{G}_{\eta\sigma} A = [\HL{d_{\eta,\sigma}},A]_{-}$ if $A$ is a fermionic dot operator (odd in charge, see
\Eq{basis-fermion}) and $\bar{G}_{\eta\sigma} A = [\HL{d_{\eta,\sigma}},A]_{+}$ if $A$ is bosonic [even in charge, see Eqs. \eq{basis-boson}-\eq{zr}]. For the 
$\tilde{G}_{\eta\sigma}$ the opposite holds.
The crucial relation \HL{\eq{trG}, $\mathrm{Tr}_D \bar{G}_{\eta\sigma}=0$} nevertheless always holds: one obtains zero
in the first case as \HL{the} trace of a commutator and in the second \HL{case} as \HL{the} trace of an operator which is off-diagonal in charge.

The analogy to \HL{the} usual field operators extends also to the transformation behavior under spin-rotations.
\HL{The} usual field operators in Fock space, ${d}^\dagger_\sigma$ \HL{and} $\sigma{d}_{\bar{\sigma}}$, transform as irreducible tensor operators (ITOs) of rank $1/2$ 
and index $\sigma/2$ \HL{(note the $\sigma$-signs)}.
\HL{Similarly,  superoperators  corresponding to particle fields ($\eta=+$), $G^{q}_{+\sigma}$, and hole fields ($\eta=-$), $\sigma G^{q}_{-\bar{\sigma}}$,
are irreducible tensor \emph{super}operators (ITSOs) of rank ${1}/{2}$ with index $\sigma/2$.}
This applies to all the representations of the field superoperators that we used:
the same holds for
$\mathcal{G}^{p}_{+\sigma}$, $\sigma \mathcal{G}^{p}_{-\bar{\sigma}}$ and
$\mathscr{G}^{p}_{+\sigma}$, $\sigma \mathscr{G}^{p}_{-\bar{\sigma}}$.

To prove this, we note that for any two superoperators ${A}^{p}$ and ${B}^{p'}$ generated
in the same way as ${G}^{p}_1$ (cf. \Eq{keld_g}) by two operators $A$ and $B$,
\begin{align}
  A^{p} \bullet =
  \begin{cases}
      A\bullet   &  p=+\\
      \bullet A  & p=-
    \end{cases}
    ,
  \end{align}
the commutator of superoperators can be expressed in the superoperator corresponding to the commutator:
$
  [A^{p},B^{p'}]_-=p \HL{\delta_{p,p'}} \left([A,B]_-\right)^p
  .
$
This directly shows that the field superoperators $\mathscr{G}_1^p$ \eq{keld_g} transform in the same way as the field operators $d_1$ under spin-rotations:
$
  [L^{S_i},\mathscr{G}_1^p]_- =\left( [S_i,d_1]_-\right)^p
  .
$
The  superoperators \eq{semi-naive_g} and \eq{kr_d} simply inherit this property
since spin and charge superoperators commute, $[L^{S_i},L^n]_{-} =  0$.

\HL{\section{Prosen's fermionic superoperators\label{sec:Pros}}}
\HL{For open systems described by a Lindblad equation that is quadratic in fermion operators,
superoperators fields similar to ours [cf. \Eq{kr_d}],
were introduced by Prosen\cite{Prosen08}
with the aim of explicitly constructing the steady state in the Liouville Fock space.
These fields obey the anticommutation relations \eq{commutG} and the Hermitian-conjugation relations \eq{Gconj}.
Moreover,} the superoperators $c_j^\dagger$ introduced in that work also satisfy \Eq{trG} and thus possess the causal structure \HL{discussed here}.
\HL{The latter property is, however, not explicitly seen in the formulation of~\Cite{Prosen08}, and becomes clear only after setting up the
correspondence \eq{Prosen-our} between our, and Prosen's, superoperators.}

In \Cite{Prosen08}, the 
opposite order of construction was used, i.e., first a Fock space was constructed \HL{(cf. our \Eq{basis-boson}-\eq{basis-fermion})}
as an ordered product of the Majorana operators, which are linear combinations
of usual fermion creation and annihilation operators.
\HL{Then,} fermionic superoperators were defined on \HL{this space (cf. our \Eq{kr_d}).}
The fermion-parity superselection rule, \HL{which plays an explicit, key role in our construction,}
was implicitly taken into account by \HL{the} ordering of the Majorana operators and by a specific definition of the creation and annihilation superoperators. 
Careful comparison
of the Fock spaces shows that superoperators $c_j^\dagger$ of \Cite{Prosen08} are related to our $\bar{G}_{\eta,\sigma}$ by the following
unitary transformation:
\begin{equation}
  \begin{aligned}
    c_j^\dagger
    & =
    \frac 1 {\sqrt{2}} \left(\bar{G}_{+,m}^{'}+\bar{G}_{-,m}^{'}\right),
    & & j=2m-1,
    \\
    c_j^\dagger
    & =\frac i {\sqrt{2}} \left(\bar{G}_{-,m}^{'}-\bar{G}_{+,m}^{'}\right),
    & & j=2m.
  \end{aligned}
  \label{Prosen-our}
\end{equation}
Corresponding rrelations for the superoperators $c_j$ can be obtained from \Eq{Prosen-our} by super-Hermitian conjugation.
Here, $m=1,2,3,...$ \HL{enumerates the} fermionic channels 
and $\bar{G}_{\eta,m}^{'}$ are the $\bar{G}_1$ superoperators renumbered by the channel index.
For \HL{our} single level \HL{Anderson} model we have only spin 
channels, \HL{giving}
\begin{align}
 \bar{G}_{\eta,m}^{'} \equiv\left\{
\begin{array}{rl}
 \bar{G}_{\eta,\downarrow},~~ m=1\\
 \bar{G}_{\eta,\uparrow}, ~~m=2
\end{array}\right.
\end{align}
For \HL{a} multilevel model with \HL{several} discrete channels \HL{numbered} $k=1,2,3,...$:
\begin{equation}
  \begin{aligned}
    \bar{G}_{\eta,m}^{'}
    & :=
    \bar{G}_{\eta,\sigma}^k,
    & m=2k+(\sigma-1)/2 .
  \end{aligned}
\end{equation} \HL{It is not directly clear how the definitions in
\Cite{Prosen08} can be applied to the case of \HL{an} infinite and,
especially, a continuous number of channels.}  % how one should
\HL{In fact, the} relation
\eq{Prosen-our} can be used as a definition of the $c_j,c_j^\dagger$
superoperators \HL{in this case} since the superoperators
$\bar{G}_1,\tilde{G}_1$ have a proper definition also in this limit,
see \Eq{commut}.  In contrast to our case, the superoperators
$c_j^\dagger$ are not irreducible tensor superoperators (ITSOs)
\HL{with respect either spin- or charge-rotations}. This fact
\HL{would} complicate the \HL{symmetry-}classification \HL{similar
that} performed in Sec.\ref{sec:basis}.  \HL{Moreover, the} causal
structure of the kernel, which in general plays a crucial role as our
analysis shows, also remains implicit in the
representation.~\cite{Prosen08}
Finally,
we emphasize the different scope of application of \HL{the} field superoperators in our work: Whereas Prosen's approach was formulated
to calculate steady states of quadratic effective Louvillians,
we here extend it to the reservoirs with continuous fields \HL{and use it to} simplify the microscopic \emph{derivation} of the effective Liouvillians for \emph{non-quadratic} problems.
\section{Schmutz's fermionic superoperators\label{sec:Schmutz}}
\HL{In} \Cite{Kosov,Mukamel08} alternative field superoperators \HL{denoted by} $a_\sigma,a_\sigma^\dagger$ and $\tilde{a}_\sigma,\tilde{a}_\sigma^\dagger$ \HL{were discussed, which were first introduced by Schmutz~\cite{Schmutz}}. 
In our notation \eq{etadef}-\eq{bar1} and \eq{keld_j},\eq{keld_g}, they read as:
\begin{align}
  a_1^p=a_{\eta,\sigma}^p=
\HL{
  \begin{cases}
    a_{\eta,\sigma},         & p=+1\\
    \tilde{a}_{\eta,\sigma}, & p=-1.
  \end{cases}
}
\end{align}
They are related to our ``intermediate'' superoperators \eq{semi-naive_g} as follows
(the definition of \Cite{Mukamel08}, \HL{in fact}, contains an additional $p$ sign \HL{that is not relevant here}):
\begin{align}
  a_{\eta,\sigma}^p=p^{\frac{1-\eta}2}\mathcal{G}_{\eta,\sigma}^p
  .
\end{align}
The use of the additional $\eta$-dependent \HL{factor} $p^{\frac{1-\eta}2}$ allows one to compensate the inconvenient signs $p$ in the anticommutation
relations \eq{semi-naive-comm} and in the Hermitian-conjugation relation \eq{semi-naive-conj}, thus restoring the usual fermionic 
algebra \HL{\emph{without performing}} the Keldysh-rotation \eq{kr_d}:
\begin{equation}
  \begin{aligned}
    \label{a-prop}
    [ a_1^{p_1},a_{2}^{p_2}]_+ & =\delta_{1,\bar{2}}\delta_{p_1,p_2},
    & \left(a_1^p\right)^\dagger = a_{\bar{1}}^p
  \end{aligned}
  .
\end{equation}
The spin and particle-hole group transformations of these superoperators coincide with those introduced by us. However, they
do not reveal the important property \eq{trG} which is crucial in our formulation \HL{of the causal structure}. 
Also, the explicit $\eta$-dependence of the sign-prefactor does not allow one to perform a  simple Keldysh rotation of the $a_1^p$ superoperators to recover this property.
\HL{Finally, we mention that during the preparation of this paper, a work~\cite{Dzhioev} appeared, in which Schmutz's operators are modified by a
complex phase factor.}
\section{Fermion-parity operator and superoperator 
\label{sec:grassmann}}
In \Sec{sec:fock-liouville} the operator $Z_R$ \HL{was constructed as} the right eigen-supervector $|Z_{R})$ of causal field superoperator $\bar{G}_1$ (cf. \Eq{Gx1})
\HL{and} turned out to play an important role.
Here, we discuss its additional properties to further clarify its physical meaning.

\paragraph{Fermion parity}
The causal field superoperators \eq{kr_d}, written out in terms of field operators, read
\begin{align}
  G_1^q \bullet = \tfrac{1}{\sqrt{2}}
  \left\{ d_1\bullet + q (-1)^n \bullet d_1(-1)^n \right\}
  \label{Gexplicit}
  ,
\end{align}
where we used $(-1)^{L^n} \bullet  = (-1)^n \bullet (-1)^n$.
Note that $n=\sum_\sigma n_\sigma$ is the occupation \emph{operator} with $n_\sigma=d_\sigma^\dagger d_\sigma$.
The definition of this representation is based on the fermion-parity superselection rule \HL{of quantum mechanics}, cf. \Sec{sec:causal}.
The explicit form \eq{Gexplicit}  makes clear that $G^{-}_1 =\bar{G}_1$ has \HL{the operator}
\begin{align}
  Z_{R} =  \tfrac{1}{2} (-1)^n
  \label{ZR-exponent}
\end{align}
as \HL{its} right zero eigen-supervector:
upon substitution the two terms in \Eq{Gexplicit} simply cancel since $(-1)^{2n}=1$.
Noting that $(-1)^n = \prod_\sigma e^{i\pi n_\sigma} = \prod_\sigma\left( 1-2n_\sigma\right)$ we recover the result \eq{ZR-product} in the main text:
\begin{align}
  Z_{R} 
  =\tfrac{1}{2}(2n_\uparrow-1)(2n_\downarrow-1)
  = 2\hat{n}_\uparrow\hat{n}_\downarrow-\hat{n}+\frac{1}{2}
  ,
  \label{ZRdefapp}
  .
\end{align}
Clearly, from \Eq{ZR-exponent} \HL{it follows that}
\begin{align}
  Z_{R}^2 =  \tfrac{1}{4} \unit{}
  \label{ZRsq}
  ,
\end{align}
which implies the normalization \HL{$(Z_R|Z_R)=1$.}
The eigenvalue equation \eq{Gx1},
$
  \bar{G}_1|Z_{R})=[d_1,Z_{R}]_{+}=0
  \label{ZR-Grassmann}
$
requires $Z_{R}$ to \emph{anticommute} with all the fermionic fields:
$Z_{R}$ is therefore the unique operator (up to normalization and a phase) 
that (anti)commutes with all QD bosonic (fermionic) operators in the QD Liouville space,
in close analogy to Grassmann numbers used in functional integral approaches~\cite{Kamenev09}.
\Eq{ZR-exponent} most clearly illustrates the physical meaning of the operator $Z_R$:
we identify $Z_R$ as the \emph{fermion-parity operator} (up to a constant),
\HL{
\begin{align}
  2 Z_R|x\rangle = \pm |x\rangle
  .
  \label{ZRparity}
\end{align}
when the state $|x\rangle$ has a well-defined, even / odd fermion number.}
Finally, it follows from
$(-1)^n d_1(-1)^n=-d_1$
that $|Z_L)=\tfrac{1}{2} \unit$ is a zero eigen-supervector of $G^{-}_1$.
The eigenvalue equation \eq{x0G}, $\tilde{G}_1|Z_{L})=[d_1,Z_{L}]_{-}=0$, 
requires $Z_{L}$, considered as an operator, to commute with the QD field operators and 
therefore with all QD operators. \HL{This implies} that indeed $Z_{L}$ is proportional to the unit operator.

\paragraph{Spin and charge rotations}
By construction the two independent operators $Z_{L}$ and $Z_R$ transform as
scalars under both spin and charge rotations (cf. \Tab{tab:itos}).
They must therefore be related \HL{to} the two scalar \HL{operators with respect to} these groups, \HL{the Casimir operators of their Lie-algebras 
$S^2=\sum_i S_i^2=3/4 \sum\limits_\sigma \ket{\sigma}\bra{\sigma}$ and
$T^2=\sum_i T_i^2=3/4 (\ket{0}\bra{0}+\ket{2}\bra{2} )$:}
\begin{align}
  \begin{aligned}
    Z_{L} & = \tfrac{1}{2}\unit = \tfrac{2}{3}(T^2+S^2),
    &
    Z_{R} &=\tfrac{2}{3}(T^2-S^2)
    .
  \end{aligned}
\end{align}
\HL{The first relation follows from the completeness relation and second one from \Eq{ZRparity}.}

\paragraph{Fermion-parity superoperator}
We next consider the fermion-parity \emph{super}operator,
defined naturally by the right action of the fermion-parity operator on an operator (identical results follow for the left action):
\begin{align}
  \begin{aligned}
    U \bullet & :=   \bullet ~2Z_R  & = \bullet (-1)^n 
  \end{aligned}
  .
\end{align}
This unitary and Hermitian superoperator  ($U^\sdagger = U, U^2 =\unit$)
transforms the two types of \HL{causal} field superoperators into each other,
\begin{align}
  U G_1^q U = G_1^{\bar{q}}
  .
  \label{mapG_G}
\end{align}
We can thus interchange the role of creation and annihilation superoperators
in  Liouville Fock-space by a linear transformation $U$.
This is similar to the field operators $d_\sigma$ and $d_\sigma^\dagger$ that generate the standard Fock-space: the $\eta$ index distinguishing can be inverted by a unitary transformation
\HL{\begin{align}
  e^{i\pi( L_{T_y}-L_{S_y})}d_{\eta,\sigma} 
  &
  = e^{i\pi(T_y-S_y) }d_{\eta,\sigma}  e^{-i\pi(T_y-S_y) }
  \\ \nonumber
  &=d_{\bar{\eta},\sigma}
  ,
\end{align}}cf. \Eq{SU2spin} and \Eq{SU2p-h}.
Note that $Kd_{\eta\sigma} =d_{\bar{\eta}\sigma}$ as well, cf. \Eq{Kdef}, but this is an antiunitary transformation.

\HL{The result \eq{mapG_G}} follows by considering an arbitrary fermionic operator ${F}$ for which $(-1)^{L^n}F= (-1)^n{F} (-1)^n=-F$. The superoperator $U$ transforms a commutator of ${F}$ with \emph{any} operator to an anticommutator and \emph{vise versa}:
defining $L_F^{\pm} \bullet=[F,\bullet]_{\pm} = F\bullet \pm \bullet F$:
\begin{align}
  \label{U}
  U L_F^\pm U \bullet & = 
  {F}\bullet (-1)^{2n}\pm  \bullet (-1)^n{F} (-1)^n
  \\
  & = {F}\bullet \mp  \bullet {F} =   L_F^\mp \bullet
  .
\end{align}
Commutators with any bosonic operator $B$ remain unaffected.
Since the superoperator $\bar{G}_\sigma$ is a (anti)-commutator when acting on a fermionic (bosonic) operator and \emph{vice versa} for $\tilde{G}$,
the superoperator $U$ interchanges these two:
$U \bar{G}_1 U = \tilde{G}_1
$
and-
$
U \tilde{G}_1 U = \bar{G}_1
$.

\paragraph{Multiorbital Anderson models}
Finally, we indicate how the operators $|Z_L)$ and $|Z_R)$ can be constructed for more general multiorbital Anderson-type models.
The super vacuum state is
$$
|Z_L)=\tfrac {1}{2^N}\unit
,
$$
where $N$ is the number of orbitals and the prefactor takes into account normalization $(Z_L|Z_L)=1$.
Equation \eq{ZR-constr} is then simply extended to
the maximally occupied state with respect to this vacuum
\begin{align}
  |Z_{R})
  =
  \prod\limits_{k=1}^N \prod_\sigma \left( \prod_{\eta} \bar{G}_{\eta\sigma}^k \right)
  |Z_{L})
  \label{ZR-constr-multi}
  .
\end{align}
%Using \Eq{ZR-constr-multi} and \Eq{chi-creation} in is straightforward to generalize \Eq{ZR-product} for the multiorbital case:
Using \Eq{ZR-constr-multi} and \eq{chi-creation}, \HL{we can generalize \Eq{ZR-product} to}
\begin{align}
  |Z_{R}) = \frac{1}{{2^N}} \prod_k \prod_\sigma (2n_{\sigma}^k - \unit)
  =\frac{1}{{2^N}} e^{i\pi n}
  \label{ZR-product-multi}
  ,
\end{align}
implying $Z_R^2= \frac{1}{4^N} \unit$.
Here, $n=\sum_{k,\sigma} n_\sigma^k$ is the total dot particle-number operator.
All properties of \HL{the} single-orbital $Z_R$ operator, i.e, (anti)commutation relations with fermionic (bosonic) operators, 
transformation properties under charge and spin rotations, etc.,
\HL{also} hold for the multiorbital case.
\section{Symmetry of the self-energy\label{sec:selrule}}

In this Appendix we derive the symmetry for QD \Eq{selrule} in contact with the reservoirs
from the global symmetries \eq{chgcons} and \eq{spincons}.
Quite generally, a quantity with QD and reservoir contributions, $A^{\tot} = A + A^{\R}$, 
is globally conserved when  $[A^\tot, H^\tot]_{-}=0$.
The corresponding Liouville superoperators $L^{A^\tot }=[A^\tot, \bullet]_{-}$
and $L^\tot=[ H^\tot,\bullet]_{-}$, then also commute:
\begin{align}
  [L^{A^\R }, L^\R ]_{-} =0,
  \label{Arescons}
  \\
  [L^{A^\tot }, L^\tot ]_{-} =0
  .
  \label{Atotcons}
\end{align}
The commutator of the local part $L^{A}=[A, \bullet]_{-}$ with the Laplace transformed evolution superoperator
$\Pi(z) =  \mathrm{Tr}_{\R}\, i (z-L^\tot)^{-1} \rho^\R$ of the reduced density operator (cf. \Eq{rhot}) must then vanish:
\begin{align}
  L^{A} \Pi(z)
  & = L^{A} \Tr{\R} \frac{i}{z-L^\tot} \rho^\R
  &
  & =  \Tr{\R} L^{A^\tot}\frac{i}{z-L^\tot} \rho^\R
  \nonumber \\
  &  =  \Tr{\R} \frac{i}{z-L^\tot} L^{A} \rho^\R
  &
  &  = \Pi L^{A}
\end{align}
using subsequently
$\mathrm{Tr}_{\R} L^{A^\R}\bullet =\mathrm{Tr}_{\R} [H^\R,\bullet]_{-}=0$,
\Eq{Atotcons},
and finally $L^{A^\R} \rho^\R= 0$.
\hl{We note that this last result follows from \Eq{grand_canon} under the assumption that $A^\R$ conserves both the reservoir energy and \HL{the} particle number, $[A^\R,H^\R]_{-}=[A^\R,n^\R]_{-}=0$, which is the case in our applications of the result.}
Clearly, this proof applies also \HL{to} the time-evolution superoperator $\Pi_{0}$ without any QD-reservoir interaction, i.e., for $L^V=0$.
Then, using $[L^{A},\Pi(z)]_{-}=[L^{A},\Pi_{0}(z)]_{-} =0$
and taking the commutator of $L^A$ with the Dyson equation that defines the self-energy $\Sigma(z)$,
$\Pi(z) = \Pi_{0}(z) -i \Pi_{0}(z) \Sigma(z) \Pi(z)$,
we find that $[L^{A},\Sigma(z)]_{-}=0$.
%%% Local Variables: 
%%% mode: latex
%%% TeX-master: "paper"
%%% End: 
\section{Hermiticity\label{sec:K-pr}}
\paragraph{Hermitian conjugation superoperator}
The density operator is restricted to be invariant under Hermitian conjugation in Hilbert space.
When considering density operators as supervectors in Liouville space,
this Hermitian conjugation of an \emph{operator} then corresponds to a superoperator that we denote by $K$,
\begin{align}
  K|A) := |A^{\dagger} )
  .
  \label{Kdef}
\end{align}
It is to be distinguished from the Hermitian conjugation of a superoperator.
$K$ is antilinear and satisfies
\begin{align}
  \label{antiunit}
  K^2 = \mathcal{I}
\end{align} 
where $\mathcal{I}$ is the unit superoperator.
Changing the basis in Liouville space by the superoperator  $K$, we effect an antilinear transformation of a superoperator,
$
S \rightarrow K S K
$, \HL{referred to as} ``$c$-transform'' in \Cite{Schoeller09a}.
In the time representation the density operator is invariant under this transformation:
$
K\rho (t) =\rho (t),
$
implying for the Laplace-transformed density matrix \eq{Laplace-rho}: 
$
K {\rho} (z) = {\rho} (-z^* )
$.
Applying $K$ to the kinetic equation \eq{kineq}, we obtain a conjugation relation
restricting the kernel $\Sigma(z)$:
\begin{align}
  \label{c-opeation}
  K \Sigma(z) K = -\Sigma (-z^*)
\end{align}
\HL{Of course, this property also} holds for \HL{the} initial Liouvillian:
$K L K \bullet = [{H},{\bullet}^\dagger]^\dagger_{-} = -[H^\dagger,\bullet]_{-}=-[H,\bullet]_{-}=-L\bullet$.

The transformation of the fields $\mathscr{G}^p$ follows by applying
$K$ to \Eq{keld_g} \HL{and} using $
({d}^\dagger \bullet )^\dagger = \bullet^\dagger {d} = (K\bullet) d
$,
\begin{align}
  K\mathscr{G}_1^{p}K=\mathscr{G}_{\bar{1}}^{\bar{p}}
  ,
\end{align}
giving with \Eq{semi-naive_g}, \eq{antiunit} and $KL^nK = -L^n$:
\begin{align}
  K \mathcal{G}^{p}_1 K 
  =  p^{-L^n} \mathscr{G}_{\bar{1}}^{\bar{p}}
  = (-1)^{L^n} \mathcal{G}^{\bar{p}}_{\bar{1}}
  .
\end{align}
The transformation of the causal field superoperators $G^q$ [\Eq{GbarK} in the main text] \HL{now follows by applying} \Eq{kr_d}:
\begin{align}
  \label{conjug_g}
  K G^{q}_1 K = q (-1)^{L^n} G^{q}_{\bar{1}}.
\end{align}

\paragraph{Simplifications using conjugation}

The transformation behavior of the kernel $\Sigma(z)$ under $K$-conjugation of the basis vectors,
$K\bar{\Sigma}(z)K=-\bar{\Sigma}(-z^*)$,
restricts the structure of the contributions to $\bar{\Sigma}(z)$
in the renormalized perturbation theory \Eq{barSigma}.
The RG-equations for the Liouvillian \eq{RG-L}, \eq{1loop} \HL{and} \eq{final} have a similar structure (since we can eliminate the renormalized vertices, even in the 2-loop RG approximation, we can restrict our considerations to the bare vertices $\bar{G}$ as in \Eq{barSigma}).
To make use of this, we decompose \HL{$\bar{\Sigma}(z)$} into conjugate pairs \HL{of contributions}.
To illustrate the idea, consider first the 1-loop \HL{approximation to the perturbation theory \HL{\eq{barSigma}} for $\bar{\Sigma}(z)$:}
\begin{align}
  \bar{\Sigma}(z) = \sum_{1} \bar{\Sigma}_{1\bar{1}}(z)
\end{align}
\HL{Here, we} write the sum over $1$ explicitly and
let $\bar{\Sigma}_{1\bar{1}}(z)
:=\bar{\gamma}(x_1) \bar{G}_1\Pi_{1}\bar{G}_{\bar{1}}
=-K \bar{\Sigma}_{\bar{1}1}(-z^*) K
$ denote a term in which the multiindex $1$ has a \emph{fixed} value
and $\Pi_1 = (z-\bar{L}-x_1)^{-1}$.
\HL{Furthermore, noting} that $1$ is dummy \HL{multiindex}, we can restrict the summation \HL{to} one \emph{fixed} $\eta$ configuration, e.g., $\eta_1=+$, while manifestly preserving the structure $K\bar{\Sigma}(z)K=-\bar{\Sigma}(-z^*)$:
\begin{align}
  \bar{\Sigma}(z) =
  \sum_{1}     \delta_{\eta_1+}
  \left( \bar{\Sigma}_{1\bar{1}}(z) -K\bar{\Sigma}_{1\bar{1}}(-z^{*})K \right)
  \label{barSigmaKdecomp}
\end{align}
The calculation of the supermatrix elements is now simplified: using the notation of \Sec{sec:rg}
and the antilinearity of $K$, $(A|KSK|B)=(KA|S|KB)^{*}$ we obtain:
\begin{align}
  &(\kappa_3| \bar{\Sigma}(z)|\kappa_0)  =
  \label{Sigma_eta_simple}
  \\
  & \sum_{1}   \delta_{\eta_1+}
  \left[
    (\kappa_3| \bar{G}_1 |\kappa_2)      (\kappa_1| \bar{G}_{\bar{1}} |\kappa_0)
    \HL{(\kappa_2| \Pi_{1}(z) |\kappa_1)}
  \right.
  \nonumber
  \\
  &
  \left. 
    \HL{-}
    (K\kappa_3| \bar{G}_1 |\kappa_2)^{*} (\kappa_1| \bar{G}_{\bar{1}} |K\kappa_0)^{*}
    \HL{(\kappa_2| \Pi_{1}(-z^{*}) |\kappa_1)^{*}}
  \right]
  \nonumber
\end{align}
In our Liouville-Fock basis the matrices representing $G^{q}$ are \HL{real, just as those of the field operators $d_1$, cf. \Eq{Gexplicit}.}
When the basis supervectors $|\kappa_3),|\kappa_0)$ correspond to diagonal operators ($Z_i,\chi_\sigma,S_0,T_0$), \HL{then} the second term in \Eq{Sigma_eta_simple} simply relates to the first one
since these supervectors are mapped onto themselves by $K$, cf. \Eq{basisKb}. \Eq{Sigma_eta_simple} simplifies to
\begin{align}
   (\kappa_3| \bar{\Sigma}(z)|\kappa_0)  &=
   \sum_{1}   \delta_{\eta_1+}
   (\kappa_3| \bar{G}_1 |\kappa_2)      (\kappa_1| \bar{G}_{\bar{1}} |\kappa_0)
   \\
   & ~
   \HL{\times
   \left[ (\kappa_2| \Pi_{1}(z) |\kappa_1)
     -
   (\kappa_2| \Pi_{1}(-z^{*}) |\kappa_1)^{*}
   \right]
     \nonumber }
\end{align}
which is explicitly imaginary at $z=i0$ as it should be.
Supervectors corresponding to non-diagonal operators ($S_\sigma, T_\eta,\alpha^\nu_{\eta\sigma}$) come in pairs related by inversion of both indices $\eta$ and $\sigma$.
The superoperator $K$ maps these pairs onto each other \HL{(possibly with a sign change, see \Eq{basisKoff})}.
Noting also that $\bar{G}$ only has non-zero matrix elements for one superket of each pair, cf. \Eq{g-b+}-\eq{g-b-},
we see that in \eq{barSigmaKdecomp} either the first or second term contributes or neither when $\kappa_3$ and / or $\kappa_0$ is a non-diagonal operator.
In this way we have effectively eliminated the need to evaluate the terms for $\eta_1=-$ using the conjugation relations.
This consideration is generalized to 2-loop expressions by adding to \Eq{barSigmaKdecomp}
\begin{align}
  \sum_{12}     \delta_{\eta_1+}\delta_{\eta_2+}
  \left( \bar{\Sigma}_{1\bar{1}2\bar{2}}(z) -K\bar{\Sigma}_{1\bar{1}2\bar{2}}(-z^{*})K \right)
\end{align}
where $\bar{\Sigma}_{1\bar{1}2\bar{2}}(z)$ collects all 2-loop terms with a fixed multiindices $1$ and $2$.
The same analysis applies to terms in the RG equations \Eq{RG-L},\eq{1loop} \HL{and} \eq{final}.

\paragraph{Conjugation properties under RG flow}

In contrast to the charge and spin transformation properties,
the conjugation property \eq{conjug_g} -- associated with the fundamental Hermicity of the density operator -- 
is exactly preserved under the RG flow.
We first note that the vertices \HL{$\tilde{G}_1=G^{+}_1$} only determine the initial value of the Liouvillian $\bar{L}$ and therefore are not affected by the ensuing RG flow: they therefore simply obey \Eq{conjug_g}.
However, the vertices \HL{$\bar{G}_1=G^{-}_1$} flow together with the Liouvillian $\bar{L}$ and thereby acquire a dependence on the dot frequency $z$.
These vertices obey the following generalization of \Eq{conjug_g} to non-zero $z$:
\begin{align}
  \label{c-g-general}
  K G_1^{q} (z)K =q(-1)^{L^{n}} G_{\bar{1}}^{q} (-z^{*} )
  .
\end{align}
To prove \Eq{c-g-general} we use that the bare vertices, providing the initial values of the RG flow, \HL{has} the property \eq{c-g-general}.
It remains to show that for each scale $\Lambda$ the infinitesimal correction to the $\bar{G}$ generated by the RG flow also \HL{has} this property.
We therefore apply $K \bullet K$ to the RG \Eq{RG-G} for \HL{the} vertex $\bar{G}_\alpha$, insert \Eq{antiunit}, \HL{use} \Eq{c-opeation} and
assume that the property \eq{c-g-general} holds:
\begin{align}
  \label{KtrG}
  &  K\frac{d\bar{G}_\alpha}{d\Lambda}K
  =
  (-1)^{(k+1) L^n + k(k+1)/2}
  \left(
    \frac{d\bar{\gamma}}{d\Lambda}\prod\limits_i \bar{\gamma}_i (\bar{\omega}_i)\right)_\text{irr}\times
  \nonumber
  \\
  & \bar{G}_{\bar{1}} \frac{1}{z_1^{*}+\bar{L}(-z_1^{*})}
  \bar{G}_{\bar{2}}  ...\bar{G}_{\bar{\alpha}}... \bar{G}_{\overline{k-1}}~
  \frac{1}{z_l^{*}+\bar{L}(-z_l^*)}\bar{G}_{\bar{k}}
  \\ \nonumber
  & = (-1)^{L^n  }\left(\frac{d\bar{\gamma}}{d\Lambda}\prod\limits_i \bar{\gamma}_i (\bar{\omega}_i)\right)_{\mathrm{irr}} 
  \\
  &  \times\bar{G}_{{1}} \frac{\bar{\gamma} (\bar{\omega}_1)}{-z_1^{*}-\bar{L}(-z_1^{*})}  
  \bar{G}_{{2}} ...\bar{G}_{\alpha}  ...\bar{G}_{{k-1}}
  \frac{\bar{\gamma} (\bar{\omega}_l)}{-z_l^{*}-\bar{L}(-z_l^*)}\bar{G}_{{k}}
  \nonumber
  \\
  &=-(-1)^{L^{n}} \frac{d\bar{G}_\alpha}{d\Lambda} (-z^{*} )
  .
\end{align}
At the first equality a sign factor arises when commuting all $(-1)^{L^n}$ factors to the left, using the fermion-parity property $(-1)^{L^n} \HL{\bar{G}_1} =\HL{\bar{G}_1} (-1)^{L^n+1}$ which is preserved under the RG as well (since both sides of \Eq{RG-G} have an odd number of $\bar{G}$'s).
\HL{After} the second equality we inverted all dummy multiindices $i \rightarrow \bar{i}$ and inverted the integration variable $\bar{\omega}_i\rightarrow -\bar{\omega}_i$, giving a sign $(-1)^{k/2}$ due to the $k/2$ antisymmetric contraction functions (counting $\bar{\gamma}$'s and $d\bar{\gamma}/d\Lambda$), where $k$ is the even number of vertices other than $\bar{G}_\alpha$.
Since \HL{for even $k$ the integers} $k/2$ and $k(k+1)/2$ have opposite parity the result follows.

 \bibliographystyle{apsrev}
\end{document}